\documentclass[journal]{IEEEtran}
\UseRawInputEncoding

\usepackage{cite}

\ifCLASSINFOpdf
\else
   \usepackage[dvips]{graphicx}
\fi
\usepackage[cmex10]{amsmath}
\usepackage{algorithmic}

\usepackage{array}

  \usepackage[caption=false,font=footnotesize]{subfig}
\usepackage{fixltx2e}
\usepackage{url}

\begin{document}
%
\title{The First Airborne Experiment of Sparse Microwave Imaging: Prototype System Design and Result Analysis}
%
%
%

\author{Zhe Zhang,~\IEEEmembership{Student Member,~IEEE,}
        Bingchen Zhang,
        Chenglong Jiang,~\IEEEmembership{Student Member,~IEEE, }
        Xingdong Liang,~\IEEEmembership{Member,~IEEE, }
        Longyong Chen,~\IEEEmembership{Member,~IEEE, }
        Wen Hong,~\IEEEmembership{Member,~IEEE, }
        Yirong Wu,~\IEEEmembership{Member,~IEEE}
\thanks{Z. Zhang is with Science and Technology on Microwave Imaging Laboratory, Beijing, 100190, China; Institute of Electronics, Chinese Academy of Sciences,  Beijing, 100190, China and University of Chinese Academy of Sciences, Beijing, 100190, China. E-mail: pzhgrsrs@gmail.com.}
\thanks{B. Zhang, X. Liang, L. Chen, W. Hong and Y. Wu are with  Science and Technology on Microwave Imaging Laboratory, Beijing, 100190, China and Institute of Electronics, Chinese Academy of Sciences, Beijing, 100190, China}
\thanks{Manuscript received April 19, 2005; revised December 27, 2012.}}

%
%

\markboth{IEEE Journal of Selected Topics in Applied Earth Observations and Remote Sensings,~Vol.~xx, No.~x, xxxxx}%
{Z. Zhang \MakeLowercase{\textit{et al.}}: The First Airborne Experiments of Sparse Microwave Imaging: Prototype System Design and Result Analysis}
%



\maketitle

\begin{abstract}
In this paper we report the first airborne experiments of sparse microwave imaging, conducted in September 2013 and May 2014, using our prototype sparse microwave imaging radar system. This is the first reported imaging radar system and airborne experiment that specially designed for sparse microwave imaging. Sparse microwave imaging is a novel concept of radar imaging, it is mainly the combination of traditional radar imaging technology and newly developed sparse signal processing theory, achieving benefits in both improving the imaging quality of current microwave imaging systems and designing optimized sparse microwave imaging radar system to reduce system sampling rate towards the sparse target scenes. During recent years, many researchers focus on related topics of sparse microwave imaging, but rarely few paid attention to prototype system design and experiment. Here we take our first step on it. We introduce our prototype sparse microwave imaging radar system, including its system design, hardware considerations and signal processing methods. Several design principles should be considered during the system designing, including the sampling scheme, antenna, SNR, waveform, resolution, etc. We select jittered sampling in azimuth and uniform sampling in range to balance the system complexity and performance. The imaging algorithm is accelerated $\ell_q$ regularization algorithm. To test the prototype radar system and verify the effectiveness of sparse microwave imaging framework, airborne experiments are carried out using our prototype system and we achieve the first sparse microwave image successfully. We analyze the imaging performance of prototype sparse microwave radar system with different sparsities, sampling rates, SNRs and sampling schemes, using three-dimensional phase transit diagram as the evaluation tool. As a conclusion, sparse microwave imaging theory is proved to be effective, and shows its advantages comparing with traditional SAR. The design and experiment of prototype system is showing us a promising future of sparse microwave imaging applications.

\end{abstract}


%
\IEEEpeerreviewmaketitle

\section{Introduction}
%
%
%
%
\IEEEPARstart{S}{parse} microwave imaging is a new concept, theory and methodology of microwave imaging, which is a framework of SAR imaging based on sparse constraints \cite{ender2010, stojanovic2009compressed, potter2010sparsity}. Real aperture radar and synthetic aperture radar are two main technologies of microwave imaging. Synthetic aperture radar, or SAR, attracts most of our attention because of its high resolution imaging ability, and has become one of the main technologies of modern remote sensing. Comparing with optical sensing technology, SAR has the all-time and all-weather imaging ability because it is an active sensing technology, and has been widely used in many fields such as agriculture, forestry monitoring, oceanic observation, topography mapping and military reconnaissance \cite{wiley1965, wiley1985, cummingbook, henderson1998principles}. Modern SAR systems commonly aim at two key features of system performance: wide mapping swath and high imaging resolution. According to the SAR theory, wider swath means a larger observing area i.e. more imaging data; higher resolution requires a wider signal bandwidth, which leads to a higher sampling rate and data amount because of the Nyquist theorem. Therefore, as the swath and resolution increase in modern SAR systems, the system complexity and dte and data amount because of the Nyquist theorem. Therefore, as the swath and resolution increase in modern SAR systems, the system complexity and data amount also increase remarkably, which brings difficulties to the hardware implementation. Furthermore, wider swath also requires a lower pulse repetition frequency (PRF) that means a lower azimuth sampling rate i.e. lower azimuth resolution \cite{cummingbook, curlander1991synthetic}. This is a bottleneck that limits the future evolution of SAR systems. As a solution, sparse microwave imaging is suggested to overcome these challenges.

The basic concept of sparse microwave imaging is simple: some features of radar imaging, e.g. the sparsity of target scene, could be used to deal with the problems. Though the target scene of SAR imaging is not always sparse, we notice that in many specific but important cases and applications e.g. oceanic ship imaging, the SAR target scene is sparse, i.e. there only exist a few strong targets in the whole scene, it makes us possible to introduce the sparse signal processing theory to the microwave imaging. Sparse signal processing is a mathematical tool that developed by mathematicians by the end of 20th century \cite{sparse}, which aims at efficiency processing of a sparse signal: if the processed signal is sparse, the signal can be measured with dramatically smaller data amount. A main achievement of sparse signal processing is named as compressive sensing or compressed sensing (CS) \cite{donoho2006, candes2006, candes2006stable}. According to CS, under some certain constraints, a sparse signal can be reconstructed with much less samples than that required by Nyquist theorem. This fact provides us potential possibility to simplify the SAR system design and develop a new microwave imaging concept.

During the last decade, many efforts have been done to introduce the sparse signal processing technology to SAR imaging, aiming at benefits including reducing the system complexity and improving the system performance. Before the trend of CS, many work has been done to introduce regularization method to SAR imaging to ehnance the image quality, reference \cite{ccetin2001feature} is a representative of it. In the reference, the authors proposed a regularization or Bayesian based SAR image reconstruction method which enhances features of the image. The first attempt of combining CS and radar imaging was suggested by \cite{baraniuk2007}.  In recent years, plenty of institutes in the world are working at the theory and application of radar imaging with sparse constraint. Reference \cite{stojanovic2009compressed} introduced a multi-static SAR system with distributed antenna array and solved it by decomposing the complex problem into the real and imaginary parts. Reference \cite{5673011} proposed a two-dimensional sparse SAR reconstruction framework and discussed the selection of sparse dictionary. References \cite{5966335, 6112799} paid special attention to sparse reconstruction based SAR tomography imaging and super-resolution discussion. Reference \cite{6857023} provided us a sparse SAR imaging result based on TerraSAR-X data. Reference \cite{ender2010}, \cite{6832835} and \cite{potter2010sparsity} made detailed overviews to applications of sparse constraints in radar and radar imaging, including its model, signal processing, performance evaluation, tomography imaging, moving target indication, error compensation, etc.   

The authors initiated our work on radar imaging framework with sparse constraints in 2009, naming \emph{sparse microwave imaging} which was detailed reported in reference \cite{overview, 6857026}. As a novel concept of radar imaging, sparse microwave imaging is mainly the combination of traditional radar imaging technology and newly developed sparse signal processing theory. In the signal processing stage of sparse microwave imaging systems, the sparse reconstruction algorithm such as $\ell_q$ ($0<q\leq 1$) regularization algorithm is utilized as the signal processing method instead of conventional matched filtering. As the result, it makes us possible to proceed the full-sampled and under-sampled data directly in both range and azimuth directions. Comparing with traditional radar imaging, sparse microwave imaging not only shows its advantages in improving the performance of current microwave imaging systems e.g. enhancing the image quality including better distinguishing ability, multilook application \cite{fangeusar} and reducing the sidelobes and ambiguity (in this way, the sparse microwave imaging only performs as a novel signal processing method), but also in possibility of reducing system sampling rate towards the sparse target scenes such as oceanic targets. In the latter way, as a systematic concept, we are able to design a optimized sparse microwave imaging system under the guidance of sparse microwave imaging theory, which allows us to achieve better performance e.g. a wider swath because of the down-sampling in azimuth direction. These results can be verified via theoretical analysis and empirical experiments. We need a prototype sparse microwave imaging radar system and enough experiments on it to test and prove the feasibility of our theories on sparse microwave imaging.

Among state-of-art researches on sparsity based SAR imaging, most of them focus on signal processing and imaging algorithm stage, with simulations or experiments with conventional SAR data. Rarely few of them focuses on the design and and implementation of a completely novel radar system under the sparse constraint and analysis of experiment results. In this paper, we will take our first step on it. Besides the system modeling and signal processing scheme, in this paper we will introduce our first prototype sparse microwave imaging radar system, including its system design, hardware considerations and signal processing methods. We also carry out airborne experiments using our prototype system and analyze its results.

The rest of this paper is organized as follows. In section II, we will introduce some theoretical considerations of sparse microwave imaging radar system design and experiment, including the model of sparse microwave imaging, system resolution, sampling schemes, performance evaluation methods and imaging algorithms. The system design of prototype radar system is introduced in section III. The airborne experiments using our prototype system is introduced in section IV, including the experiment setup and the signal processing. In section V we will provide the experiment results and analysis. Finally, the conclusions are given in section VI.

\section{Theoretical Considerations}

\subsection{Model}

We will provide a very brief introduction to the model and theory of sparse microwave imaging. Detailed discussion of sparse microwave imaging theory can be found in reference \cite{overview}.

\begin{figure}[!t]
	\centering
  \includegraphics[width=\columnwidth]{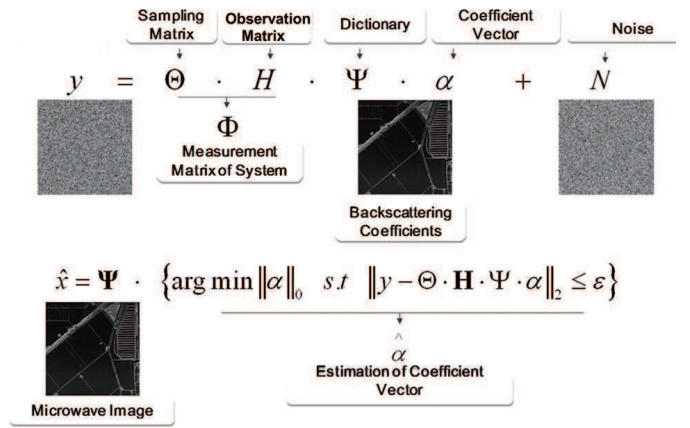}
	\caption{The diagram of sparse microwave imaging model \cite{overview}.}
	\label{fig:1}
\end{figure}

In a general form, The sparse microwave imaging  radar system model can be described as
\begin{equation}
	y=\mathbf{\Phi} x + N =\mathbf{\Theta} \mathbf{H} \mathbf{\Psi}\alpha + N,
		\label{eq:1}
\end{equation}
as shown in figure \ref{fig:1}

\begin{figure}[!t]
	\centering
  \includegraphics[width=\columnwidth]{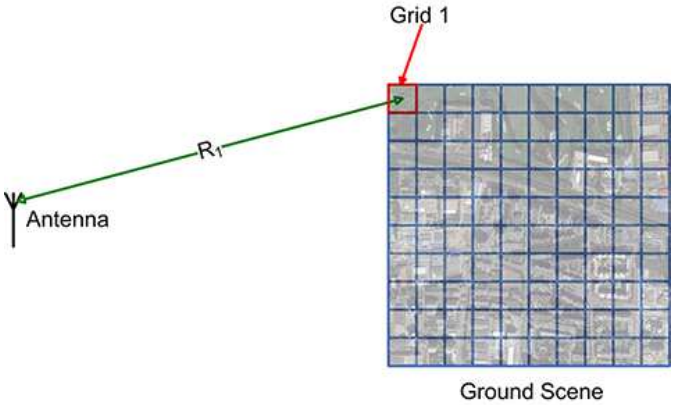}
	\caption{The target scene discretization.}
	\label{fig:2}
\end{figure}

\begin{itemize}
	\item $x$ is the back-scattering coefficients of targets in the scene. Usually the target scene is continuous, so we should do a discretization to the scene. A simple way of discretization is to divide the scene into multiple cells and then treat each cell as a point target which is indicated in figure \ref{fig:2}. The location of each point target is the center of corresponding cell, and the backscaterring coefficient is the equivalent backscattering coefficient of the cell. $x$ is a vector that consist of backscaterring coefficient of every target point one by one. The scene vector $x$ can also be written as $x=\mathbf{\Psi} \alpha$ with $\mathbf{\Psi}$ being a sparse dictionary/basis and $\alpha$ being the sparse expression of $x$. In most radar applications, it is difficult to find a universal-feasible dictionary. It means we use a identity matrix as $\mathbf{\Psi}$. 
	\item $y$ is the received echo signal. 
	\item $\mathbf{\Phi}$ is the measurement matrix of radar system. $\mathbf{\Phi}$ is defined as $\mathbf{\Phi}=\mathbf{\Theta H}$ with $\mathbf{\Theta}$ being the matrix representing the down-sampling of the radar system, and $\mathbf{H}$ being the the matrix describing the observation of radar. If the radar uses a full-sampling, the $\mathbf{\Theta}$ is an identity matrix $\mathbf{I}$; in the cases of under-sampling, $\mathbf{\Theta}$ will be some selected rows of $\mathbf{I}$.
	\item $\mathbf{H}$ is determined by the radar system. The creation of matrix $\mathbf{H}$ could be found in Appendix A.
	\item $N$ is the additive noise, and its noise-threshold is $\epsilon$.
\end{itemize}

$x$ can be reconstructed by a sparse reconstruction algorithm via the following optimization problem 
\begin{equation}
	\hat{x}= \mathbf{\Psi} \cdot \{ \arg \min_{\alpha}\|\alpha\|_{0} ~~\textrm{s.t.}~ \|y-\mathbf{\Theta} \mathbf{H \Psi} \alpha\|_{2} \le \epsilon \}.
	\label{eq:2}
\end{equation}

\subsection{Sampling Schemes}

Sampling includes the range sampling and azimuth sampling. In range direction, the radar signal is continuous analog signal, so the problem of sampling cannot be avoided. In azimuth direction, the sampling occurs on each pulse repetition point because of the operating principle of SAR. The traditional SAR system uses the Nyquist sampling strategy, which causes a huge data amount when signal bandwidth is wide. The sparse signal processing technology gives us the possibility to reduce the sampling rate in both directions. There exist several sampling schemes, different sampling schemes lead to different system performances.

The sparse signal processing technology usually samples a sparse signal with a linear incoherent measurement system. Such signal is usually sparse in a specific domain, has low information rate, and can be processed with a non-adaptive measurement system. But radar signal does not satisfy these conditions. Usually, radar signal has a high information entropy, so we are not able to apply a linear incoherent measurement directly. However, if the scene is sparse, we can directly treat the target scene as the signal that we would like to recover and the observation system as the measurement system, and then achieve some under-sampling schemes. A suitable sampling scheme should:
\begin{itemize}
	\item optimize the imaging performance of radar system; 
	\item easy to implement in hardware;
	\item and has as little side effect as possible. 
\end{itemize} 

The simplest sampling scheme is the uniform sampling. As the name suggests, a uniform sampling system has a definite sampling interval. It is quite simple in hardware implementation and can be utilized in both azimuth and range directions, but has a bad performance when under-sampling is applied because the sampling interval keeps invariant and brings high mutual coherence to the measurement matrix.

The opposite of uniform sampling is the random sampling, in which the sampling interval changes randomly from one sample to the next. Random sampling has a much better under-sampling performance, because the minimal sampling interval might be less than the Nyquist sampling interval though the average sampling rate is much less than the Nyquist rate. 

But random sampling has a side effect in azimuth direction. To a SAR system, its swath width is determined by its PRF, lower PRF allows wider swath. We want to reduce PRF to achieve a wider swath, but if the random sampling is applied to the azimuth direction, the minimal sampling interval might be even less than the Nyquist sampling interval so the swath might be limited even narrower. As a solution, we introduce the jittered sampling, a modified version of random sampling \cite{balakrishnan1962problem}. Jittered sampling is basically a random sampling scheme, but it ensures the minimal sampling interval must be more than a determined value.

\subsection{Performance Evaluation}

In sparse microwave imaging, most evaluation tools of traditional SAR imaging are no longer effective because of the totally different signal processing methods. In this paper, we use three-dimensional phase transit diagram as the performance evaluation tool of sparse microwave imaging \cite{tian2011}. Phase transit diagram \cite{stanley1971introduction} is originally a type of chart in physics to describe thermodynamics performance of a material. As a useful tool, Donoho et al. \cite{donoho2009observed, donoho2010precise, donoho2011noise} introduced it into the CS theory, to represent the system recovery abilities under different parameter sets of under-sampling ratio and sparsity in a visual intuit chart. Reference \cite{patel2010compressed} used two-dimensional phase transit diagram to analysis the performance of radar imaging system based on sparse constraint. Compared with two-dimensional phase transit diagram, three-dimensional phase transit diagram has three axes: sparsity, SNR and under-sampling ratio. By comparing the successfully recoverable area of phase transit diagram under different system conditions, we are able to evaluate and compare the performances of different system conditions. 

The advantage of phase diagram is that the performance of system could be displayed visibly and clearly, but its generation requires a large-scale calculation. As en empirical tool, the phase diagrams are not universal. It means, different system parameters and different imaging algorithm will make the shapes of phase diagrams different. 

\subsection{Imaging Algorithm}

Detailed discussion of imaging algorithm could be found in \cite{overview, jiang, 6522528}. Here we give a brief introduction to it.

The sparse signal processing method is initially applied to the range compressed data with few modifications to traditional SAR imaging framework. Because it requires preprocessing procedures (i.e., range compression), this kind of method may increase difficulties in system design. The sparse reconstruction algorithm that exploit SAR raw data with no preprocessing is practically feasible in simplifying the radar hardware \cite{overview}. Based on such consideration, we directly process the under-sampled two-dimensional raw data as the problem described in (\ref{eq:2}).

Many sparse reconstruction algorithms for radar imaging have been presented for the alternation of matched filtering based algorithms \cite{overview}. In this paper, we mainly use the $\ell_q, 0<q\leq 1$ regularization algorithm
\begin{equation}
	\label{eq:3}
	\arg\min_{\alpha} \left\{\|\mathbf{\Theta} \mathbf{H} \mathbf{\Psi}-\alpha\|_2^2+\lambda\|\alpha\|_q^q\right\},
\end{equation}
where $\lambda$ is the regularization parameter. We construct the sparse measurement matrix as the dictionary of the echo signal and reconstructs the SAR image using the iterative shrinkage-thresholding (IST) algorithm,
\begin{equation}\label{eq:4}
{x}^{(k+1)} = f\left({x}^{(k)}+\mu\mathbf{\Phi}^{H}\left({r}-\mathbf{\Phi} x^{(k)}\right), \lambda\right),
\end{equation}
Regularization parameter $\lambda$ is selected mainly based on experience. In our applications, we choose $\lambda$ based on the sparsity of target scene. Usually it is selected to make most clutters are lower than the threshold and most targets are above the threshold. Parameter $\mu$ controls the convergence, it is selected based on the trade-off of converging speed and reconstruction precision. $q$ could be selected in range $0<q\leq 1$. The $\ell_1$-norm is most widely used and discussed, while other $q$ e.g. $\ell_{1/2}$ \cite{zeng2012sparse} could also be applied in our framework. $\left(\cdot\right)^{H}$ denotes the conjugate transpose operation, and $f\left(\cdot\right)$ is the shrinkage function.

We choose IST here mainly because in the trade-off between efficiency and accuracy, we prefer the accuracy. Comparing with other sparse reconstruction algorithm categories e.g. greedy algorithms and non-convex algorithms, IST has a provably accurate recovery when signal is approximately sparse and are robust to the observational noise, while the disadvantage is its computation cost \cite{eldar2012compressed}. As an optimization algorithm, IST has relatively good imaging performance among the sparse reconstruction algorithm, but it is very computational-complex. IST has high complexity in both spacial and temporal domains because sparse microwave imaging applications usually comes with dramatically large matrix $\mathbf{H}$ scale. For example, in our estimation, to a scene with size $1024\times8192$, classic IST will take more than 64TB of memory and more than three months of calculation on a typical workstation.

In order to process the real SAR data, the accelerated sparse microwave imaging algorithms is necessary. The computational efficiency of $\ell_q$ regularization method can be uplifted by decoupling the azimuth and range of the raw data. The concept of decoupling lies in the formation of SAR signal from the reflectivity image of the observation area \cite{6522528}. The decoupling operator can be implemented by using the principle of conventional matched filtering algorithms, e.g. range Doppler algorithm~(RDA) and chirp scaling algorithm~(CSA).  Here we take CSA as the example. The measurement matrix is viewed as the azimuth convolution ${\mathcal H}_{a}^{H}$ imposed on the reflectivity image to obtain doppler history, the range-azimuth migration operator ${\mathcal H}_{sc}^{H}$ to generate different curves, and the operator ${\mathcal H}_{bulk\_pc}^{H}$ to complete the azimuth-range coupling and form the echo signal \cite{jiang}.
\begin{equation}
\mathbf{\Phi}\cdot{x} = \mathbf{\Theta}\odot {\mathcal H}_{sc}^{H}\left\{{\mathcal H}_{bulk\_pc}^{H}\left\{{\mathcal H}_{a}^{H}\left\{x\right\}\right\}\right\},
\end{equation}
where $\odot$ defines the Hadamard matrix product. ${\mathcal H}_{sc}$, ${\mathcal H}_{bulk\_pc}$ and ${\mathcal H}_{a}$ are corresponding to the phase functions in matched filtering algorithm. 

In the accelerated algorithm, The large scale matrix $\mathbf{H}$ has been decomposed into three (approx-)linear operators $\mathcal{H}_{a}^{H}$，$\mathcal{H}_{sc}^{H}$ and $\mathcal{H}_{bulk\_pc}^{H}$. With breaking down a large matrix into small matrices, the computation complexity could remarkably reduced from $O\left(N^2\right)$ to $O\left(N{\log}\left(N\right)\right)$ compared with the non-accelerated $\ell_q$ regularization method. The accelerated $\ell_q$ regularization method is effective in the reconstruction for sparse scenes with sub-sampling data. 
In additional, it can be also applied to process the non-sparse scene with Nyquist-sampling data.

\subsection{System Resolution}

We compare the signal processing method of sparse microwave imaging with traditional SAR in another aspect. Traditional SAR imaging algorithm is matched filtering, this is a signal recovery algorithm. Meanwhile, the imaging algorithm of sparse microwave imaging is sparse reconstruction algorithm, which is a signal detection algorithm. Ideally, the point spread function of sparse reconstruction algorithm has no side-lobe. So many references suggest that the sparse construction algorithm has the ability of super resolution \cite{5966335, 6112799}.

In our point of view, because the sparse microwave imaging is in fact signal detection rather than recovery, the terminology ``resolution'' is no longer accurate here.  Basically, the sparse microwave is not resolving, but ``distinguishing'' the targets in the scene. As the result, we suggest another terminology ``distinguishing ability'' in sparse microwave imaging. This is based on the following considerations:
\begin{itemize}
	\item Traditional SAR resolution is based on the matched filtering background. It is mainly a Rayleigh resolution, based on that fact that point spread function of SAR is a sinc pulse. In sparse microwave imaging, the point spread function is a Dirac pulse. It is difficult to define a Rayleigh resolution here.
	\item The Rayleigh resolution is a intrinsic property of SAR system, it is only determined by the system bandwidth. But in fact the ability of distinguishing two neighbouring points is infected by multiple factors. Especially in sparse microwave imaging, beside the radar system itself, the distinguishing ability is also influenced by the target scene, sparsity, imaging algorithm, SNR, etc. 
\end{itemize}

In one word, it is difficult to define a definite resolution of a specific sparse microwave system. We pay more attention to the ability of distinguishing two neighboring targets of a system. 

\section{System Design}

Of course, we can directly inherit a current conventional SAR hardware system and work along with the sparse signal processing methods. While, we more likely to design an optimized sparse microwave imaging system under the guidance of sparse microwave imaging theory. In this section, we will discuss some design principles, and try to achieve a design of the prototype sparse microwave imaging radar system.

\begin{figure}[!t]
	\centering
  \includegraphics[width=\columnwidth]{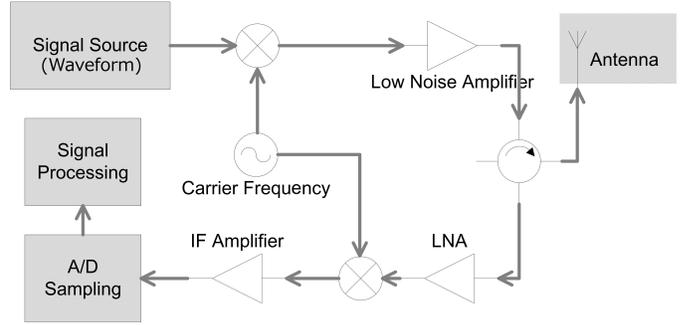}
	\caption{Diagram of components of a radar system. The grayed blocks are potentially modifiable \cite{overview}.}
	\label{fig:3}
\end{figure}

Figure \ref{fig:3} provides a diagram of components of a typical radar system. In this diagram, we can find some potential modifiable blocks which are shown in gray.
\begin{itemize}
	\item The waveform. We can use an alternative radar waveform to achieve optimized measurement matrix. Current widely used chirp signal can still be used in the sparse microwave imaging, while some other waveforms e.g. orthogonal coded signals might lead to a better system performance.
	\item The sampling. We can apply different sampling strategies e.g. the non-uniform sampling, aiming at a better measuring effectiveness. In our application, we believe that a pseudo-random sampling with definite interval is a reasonable choice.
	\item The antenna spacial location. We can modify the location of antenna or install antenna array to achieve spatial diversity. We believe that in the DPCA mode, we can at least loose the constraints to the antenna position and platform velocity.
	\item The antenna footprint. Besides, the platform motion and beam scanning might provide us better sparse reconstructing performance and wider swath. The sparse microwave imaging framework works for most beam scanning modes including ScanSAR and TOPSAR, and the combining of fast beam scanning and under-sampling in azimuth direction can provide us both wide swath and reduction of scalloping effect. Furthermore, with antenna array and orthogonal waveforms, the platform could even be stationary during imaging \cite{zhang2012influence}.
\end{itemize}

\subsection{Sampling}

To reduce the data amount and enlarge the swath, we want to reduce the sampling rate in both directions. In the range direction, we simply use a random under-sampling scheme, while in the azimuth direction, as we have mentioned before, the jittered sampling is best choice.

\begin{figure}[!t]
\centering
\includegraphics[width=\columnwidth]{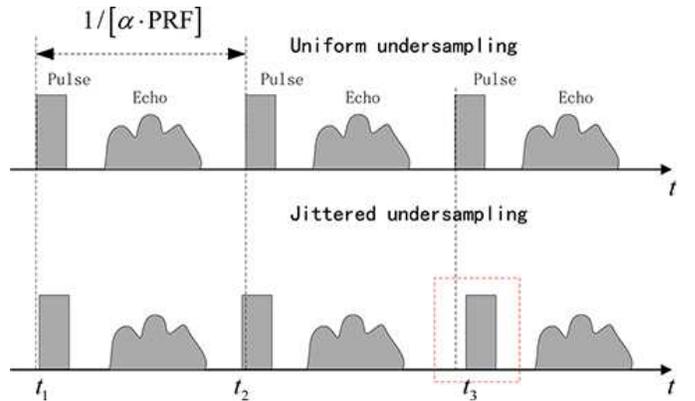}
\caption{Jittered under-sampling Scheme. Compared with uniform undersampling scheme, the sampling time of jittered under-sampling scheme is perturbed by additional factor.}
\label{fig:4}
\end{figure}

Implementation of such scheme is very simple and requires a minor change to the PRF scheduling / azimuth timing component of the radar since it is based on modifying the PRF of the radar. The under-sampling scheme is based on a regular under-sampling which is perturbed slightly by random noise. It is shown in Figure \ref{fig:4}. As the radar moves along its path, the $n$-th pulse is transmitted at $t_n$, which is expressed as:

\begin{equation}
t_n = n/\left[\alpha\cdot{\rm PRF}\right]+\epsilon_n
\label{eq:samp}
\end{equation}

where ${\rm PRF}$ is the traditional pulse repetition frequency, $\alpha\in(0,1]$ is the under-sampling factor, $\epsilon_n$ is the perturbing factor. Assume that $\epsilon_n$ is a uniform random variable which distributes in $\left[-t_p/2, t_p/2\right]$. Then the sampling interval $\Delta t_n = t_n - t_{n-1}$ is between $\left[\alpha\cdot{\rm PRF}-t_p, \alpha\cdot{\rm PRF}+t_p\right]$.

The jittered sampling scheme serves two major purposes. Firstly, by under-sampling in the azimuth direction and reconstruction through the means of sparse microwave imaging method, it have the ability to enlarge the swath. Secondly, the random changes imposed on the regular PRT can break up the periodicity of the aliasing artifacts and preserve the high frequency energy \cite{balakrishnan1962problem}.

There are several issues should be discussed about the designing of sampling scheme.
\begin{itemize}
	\item Design and implementation of under-sampling hardware
	
	The implementation of under-sampling is intuitive. In range direction, we setup a uniform sampling clock and (pseudo-)randomly select some among the uniformly sampled records. No modification to the range sampling instruments besides a pseudo random selector. 
	
	In azimuth direction, we generate a sampling interval sequence based on equation (\ref{eq:samp}) and write it into the ROM of azimuth timing instruments. 
	
	\item Under-sampling rates in each directions
	
	To achieve a specific under-sampling rate, the under-sampling could be done in either range or azimuth, or both domains. The proportion of under-sampling in both domains is not universal, but varies in each application. In fact, it is determined by the sparsity of both directions of the specific scene. 

	In the hardware, the maximal sampling rates on both directions are set more than the Nyquist rate. Under-sampling rate in both directions could be modified by randomly selecting among original sampled records. Such selection could be done on board.
	
	\item Imaging performance
	
	The influence of under-sampling in imaging performance is complicated. In a brief word, because our accelerated algorithm is 2D decoupled, the tolerable under-sampling rates in both directions are determined by the sparsity in each direction respectively. As the random under-sampling is applied in range and jittered under-sampling is used in azimuth, it is reasonable to expect a higher acceptable under-sampling rate in range direction.
	
	Detailed analysis of imaging performance with respect to the under-sampling rate in azimuth direction is given in the experiment section.
	
\end{itemize}

\subsection{Antenna and Sensor}

The sparse microwave imaging also has the potential to ease constraints on the selection of PRF for dual/multi-channel imaging radar system, or the displaced phase center antenna (DPCA) mode \cite{fang2012efficient}. As a prototype system, we still use a single-channel radar system. The sensor setup is identical to traditional signal-channel stripmap SAR.


\subsection{SNR}

In the sparse microwave imaging theory, a minimal system SNR is required. If the SNR is too low, the sparse reconstruction might fail. So the system SNR must be guaranteed. Simulations and experiments show that the SNR at the same level of traditional SAR is enough for sparse reconstruction. This requirement could be satisfied via enhancing the transmitting power, expand the pulse duration time and enlarging the antenna area. 


\subsection{Waveform}

Traditional SAR usually uses Chirp signal as its waveform. To a sparse microwave imaging system, we might have some alternative selections \cite{overview}. They lead to
different system performances.

\begin{itemize}
	\item Chirp Signal
\end{itemize}
Chirp signal is widely used in radar systems for its good matched filtering performance and low Doppler sensitivity. It has been broadly applied in traditional radar systems, and could also be applied to sparse microwave imaging systems.

\begin{itemize}
	\item Orthogonal Signals
\end{itemize}
Ideal orthogonal signals refer to a signal family $p_i(\tau)$ whose members are orthogonal. The orthogonal signals come in form of a signal family, meaning that the waveform varies in azimuth direction. Expression of orthogonal signals is
\begin{equation}
    \label{eq:5}
    \int_0^T p_i(\tau) p_j^\ast (\tau) d\tau= \left \{
        \begin{array}{ll}
            1, &~~i=j \\
            0, &~~i\neq j
        \end{array}
    \right .,
\end{equation}
where $i$ and $j$ are signal indexes in the family, and $[0, T]$ is the definition domain of signal family. Random noise signals family is one representative of orthogonal signals.  

Many researches find that comparing with traditional Chirp, orthogonal signals have better system performance, but the difference is not significant \cite{ripless}. Considering the system implementation complexity, we still use Chirp signal in our prototype system.

\subsection{Bandwidth}

SAR resolution is determined by its signal bandwidth. A random noise signal has infinite bandwidth, while Chirp not. Both theoretical researches and experiments show that the bandwidth requirement of sparse microwave imaging system is at the same level as traditional imaging radar system \cite{overview, ripless}. It means, the signal bandwidth is select under the same constraint as the traditional SAR according to the required system distinguishing ability. While, if the scene is sparse, the distinguishing ability of imaging system could be enhanced.

\section{Airborne Experiments}

In this section, we will introduce our prototype sparse microwave imaging radar system and the experiment setup.

\begin{figure}[!t]
	\centering
	\subfloat[]{
		\includegraphics[width=0.4\columnwidth]{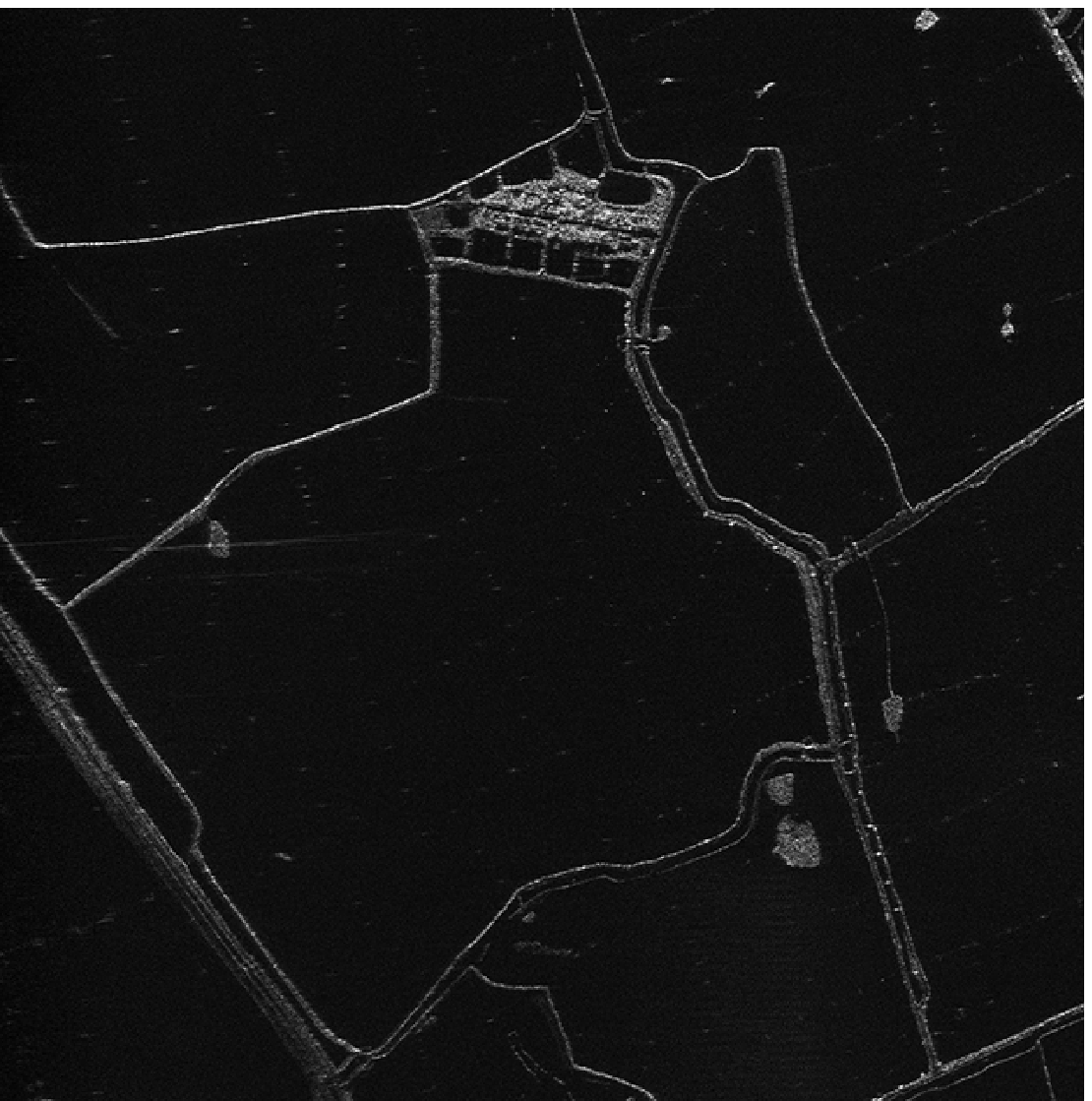}%
		\label{fig:51a}
	}
	\hfil
	\subfloat[]{
		\includegraphics[width=0.4\columnwidth]{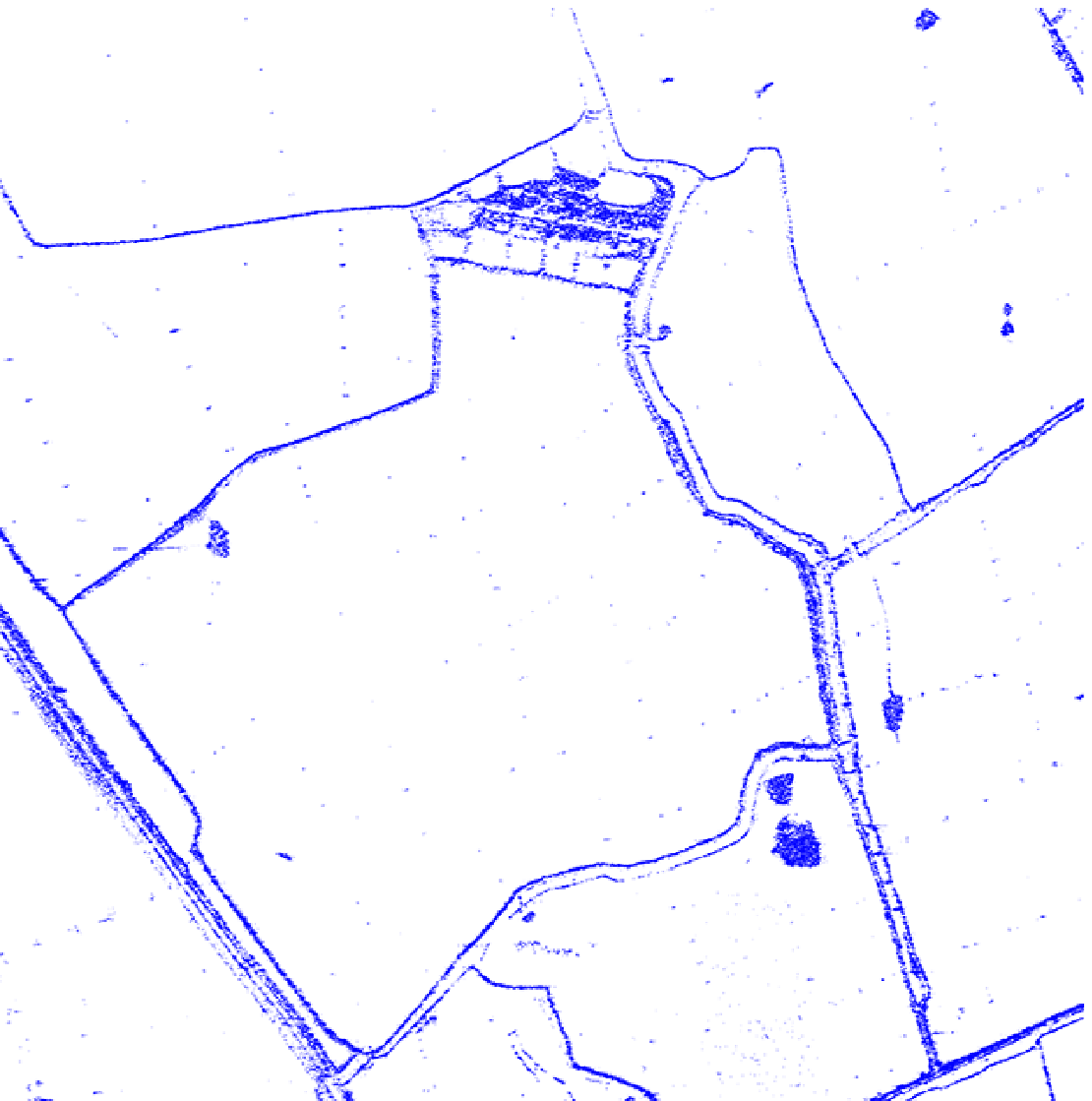}%
		\label{fig:51b}
	}
	\caption{Example image of a strong targets in salt pans. (a) The scene image and (b) strong targets are marked as blue.}
	\label{fig:51}
\end{figure}

\begin{figure*}[!t]
	\centering
	\subfloat[]{
		\includegraphics[width=2.5in]{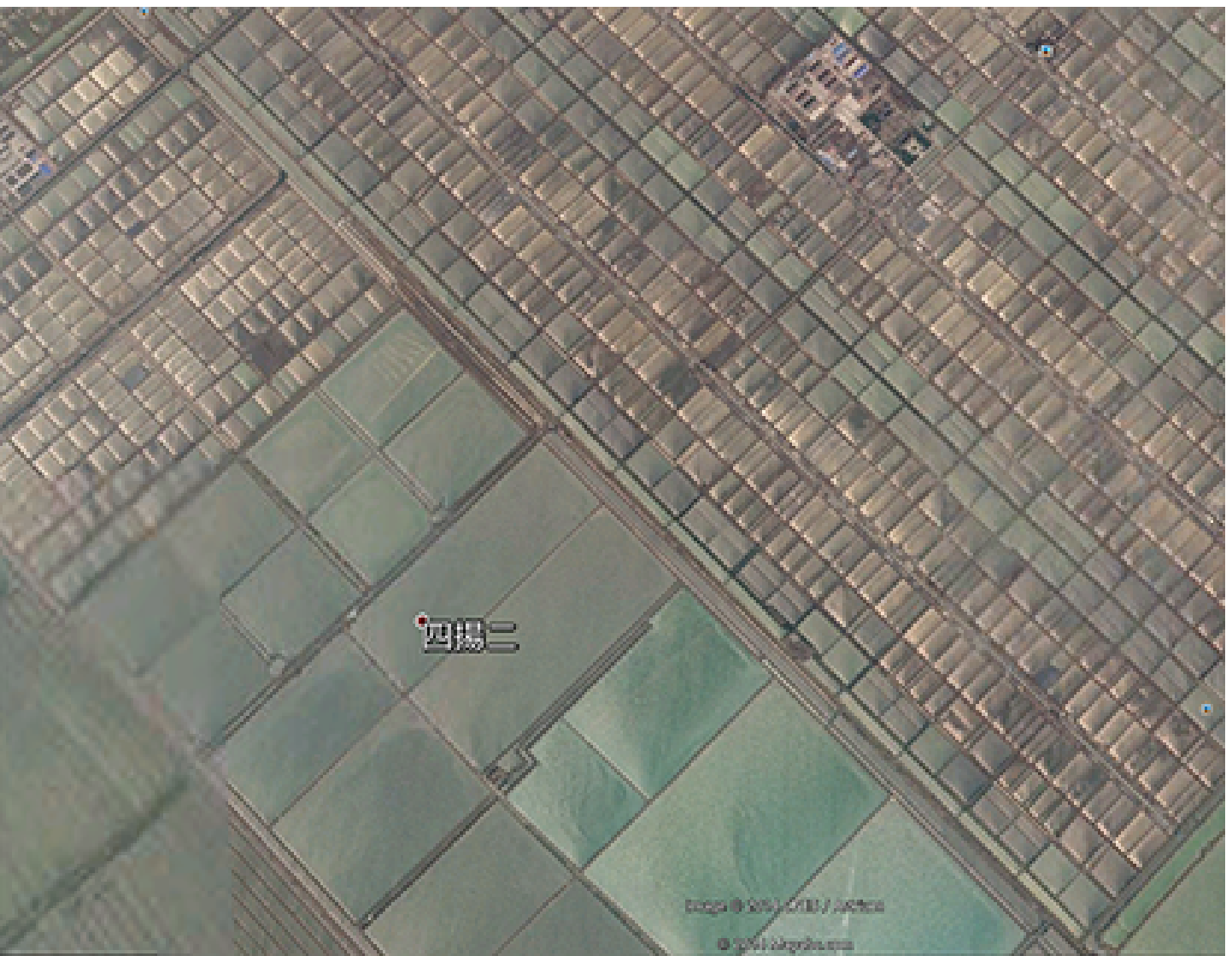}%
		\label{fig:5a}
	}
	\hfil
	\subfloat[]{
		\includegraphics[width=2.5in]{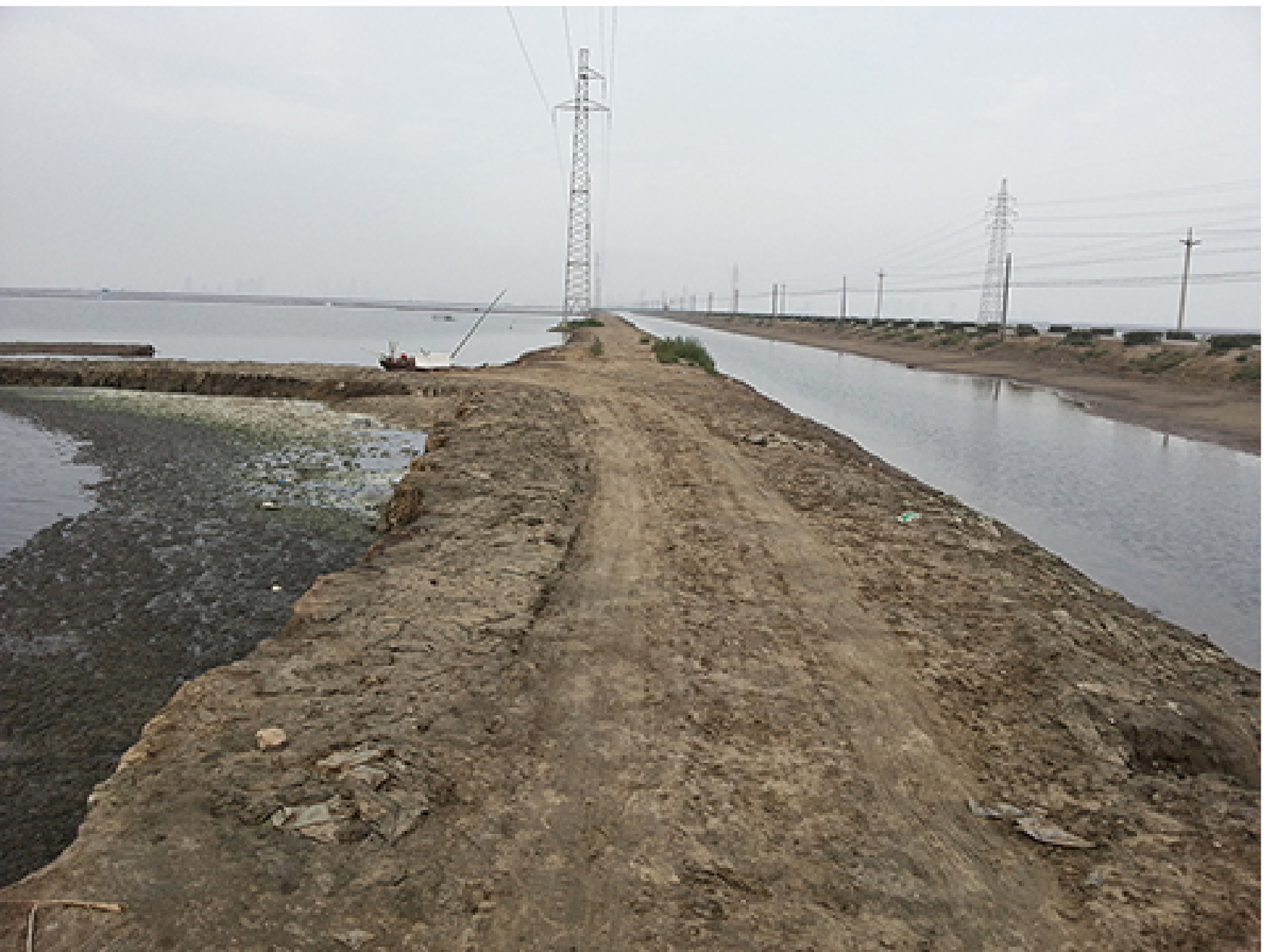}%
		\label{fig:5b}
	}
	\caption{Optical image of salt pans. (a) Optical image from Google Earth and (b) photo taken in the scene.}
	\label{fig:5}
\end{figure*}

\begin{figure*}[!t]
	\centering
	\subfloat[]{
		\includegraphics[width=2in]{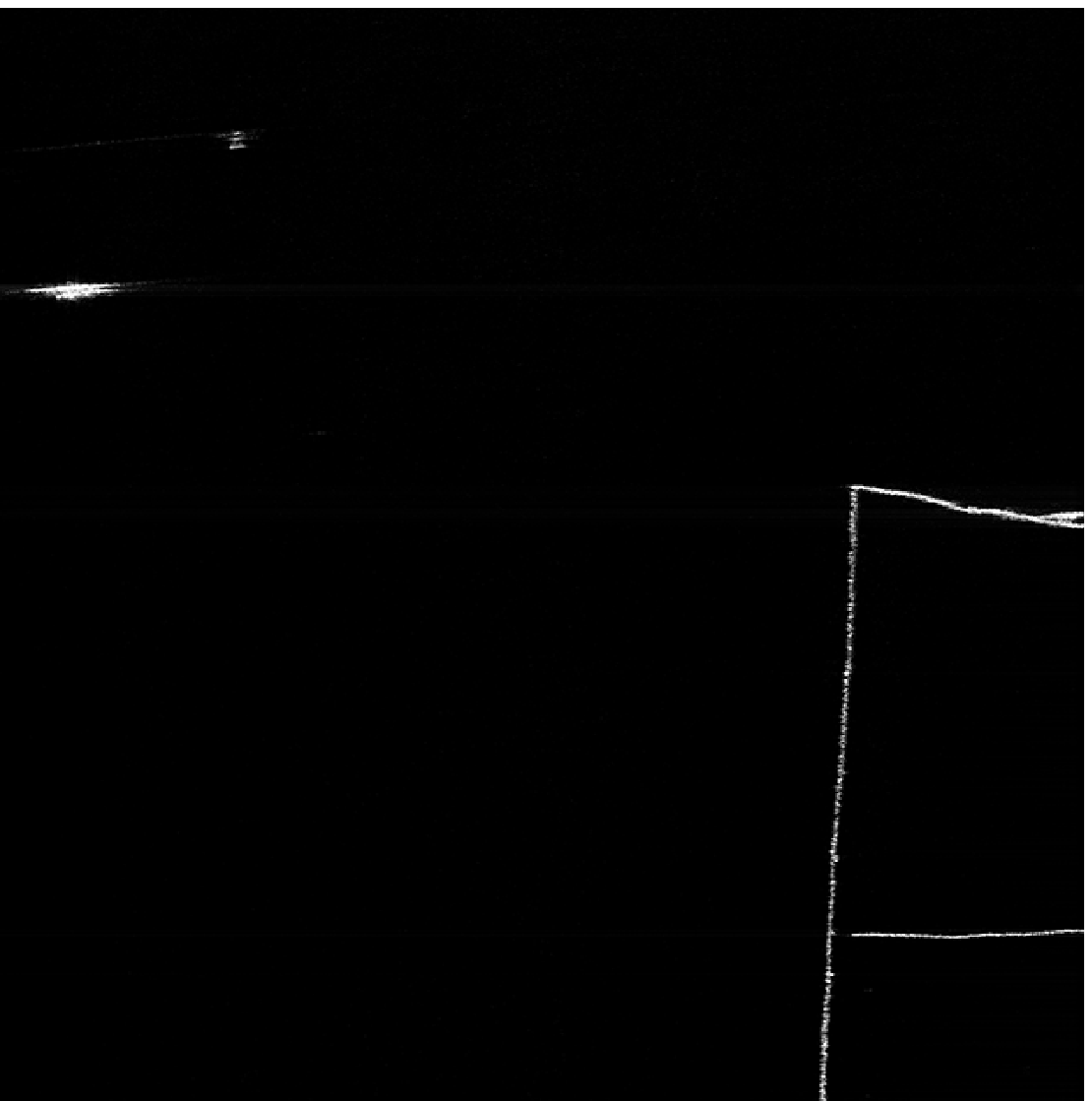}%
		\label{fig:6a}
	}~
	\subfloat[]{
		\includegraphics[width=2in]{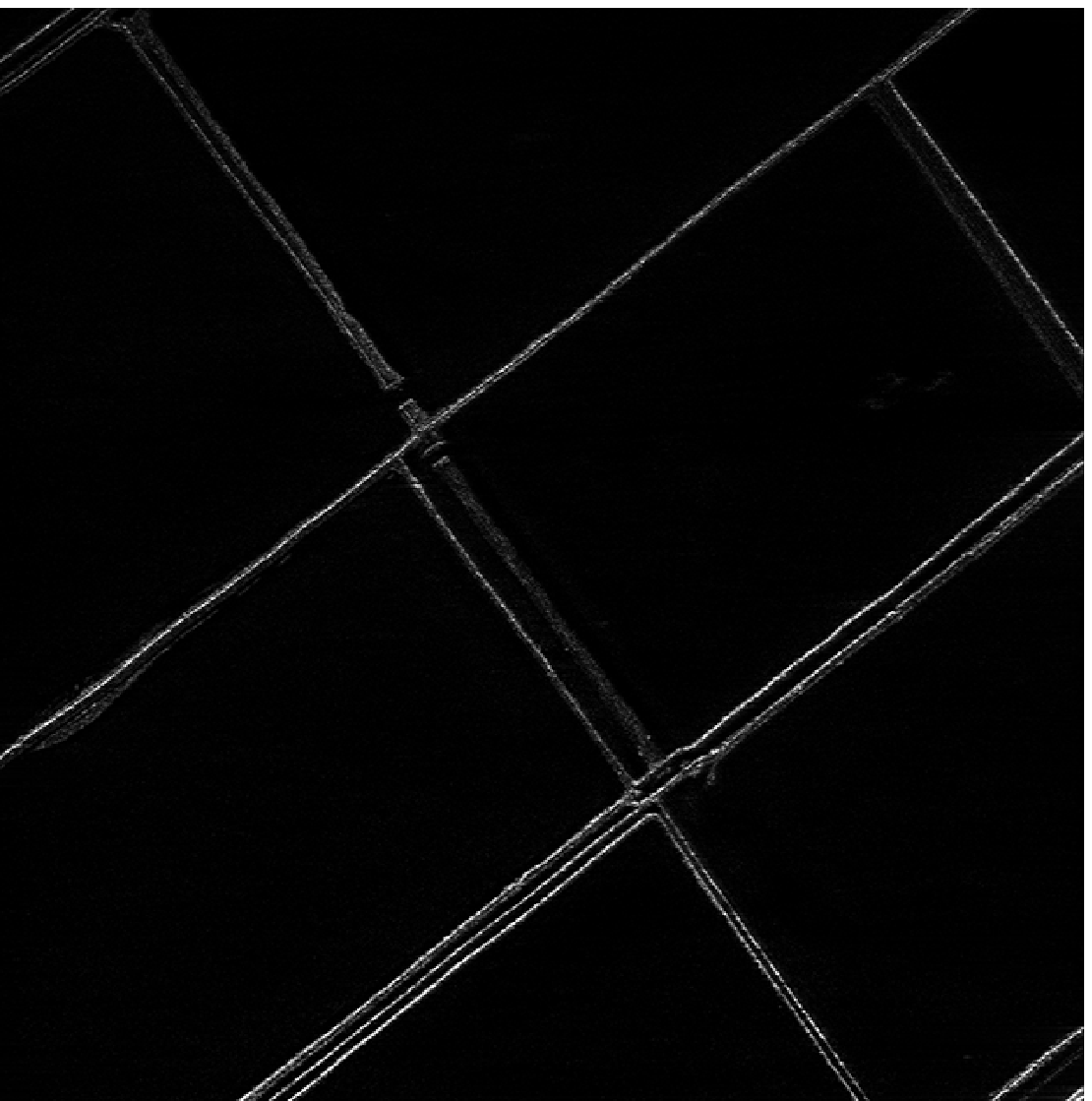}%
		\label{fig:6b}
	}~
	\subfloat[]{
		\includegraphics[width=2in]{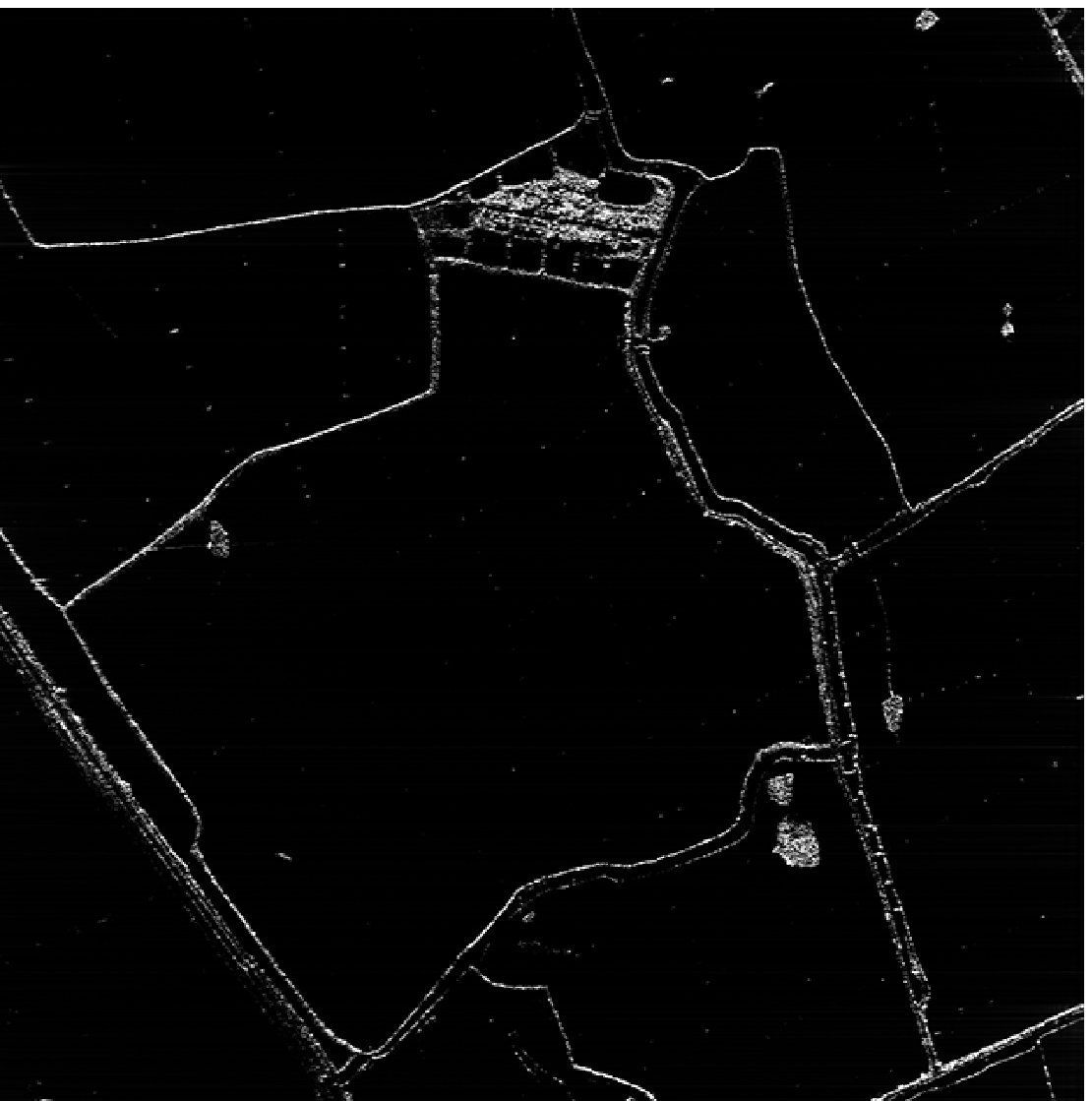}%
		\label{fig:6c}
	}\\
	\subfloat[]{
		\includegraphics[width=2in]{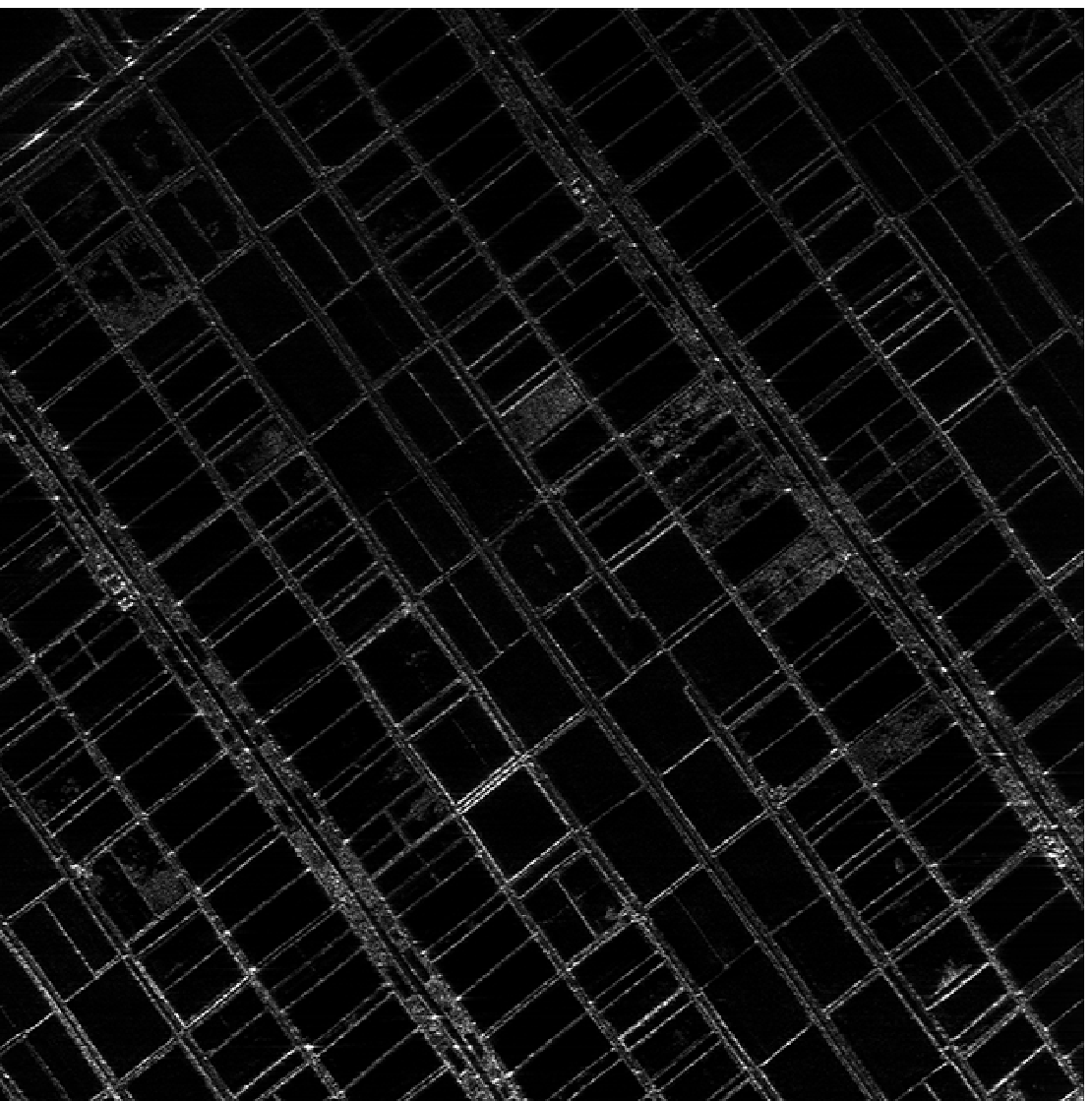}%
		\label{fig:6d}
	}~
	\subfloat[]{
		\includegraphics[width=2in]{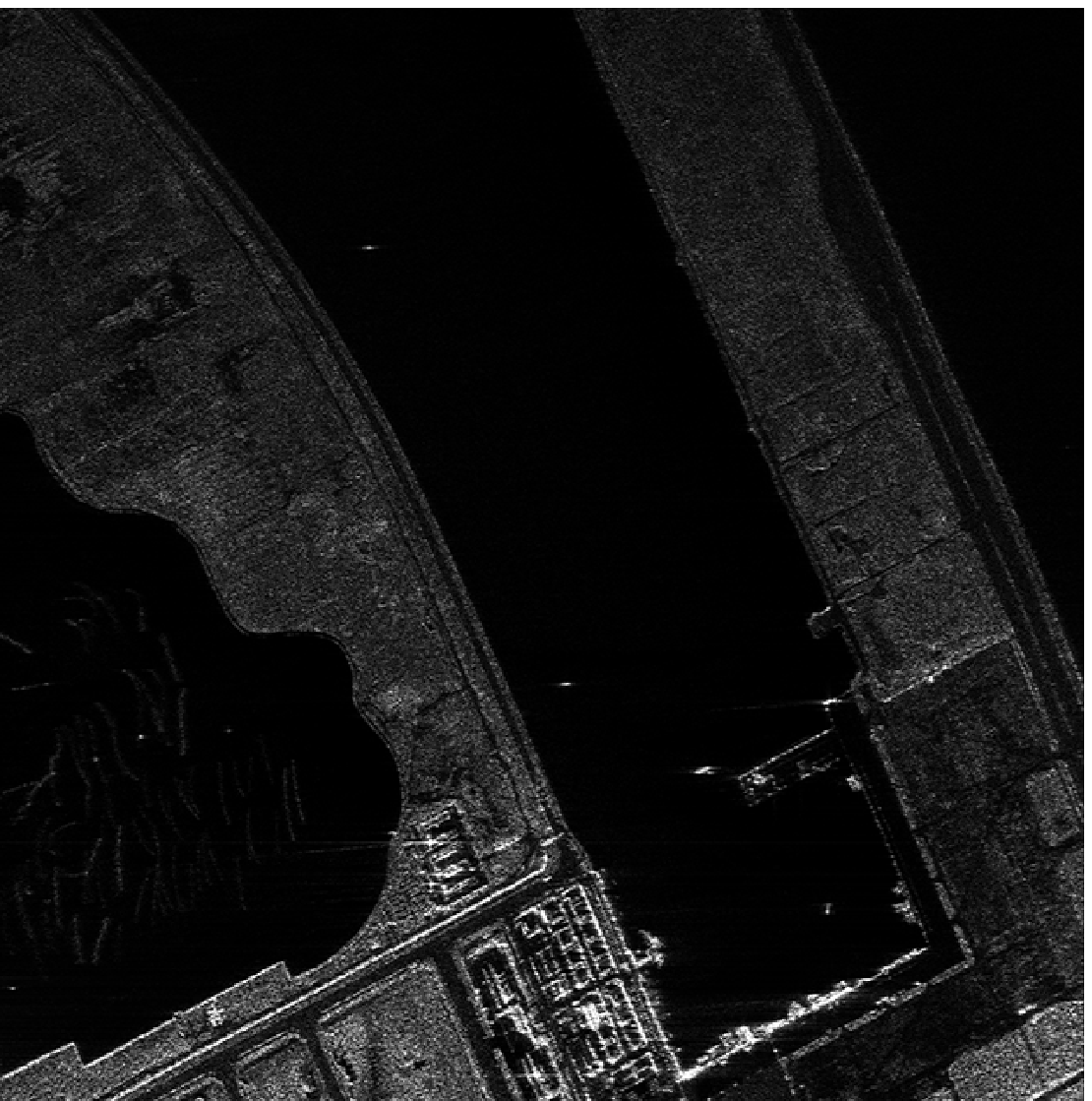}%
		\label{fig:6e}
	}~
	\subfloat[]{
		\includegraphics[width=2in]{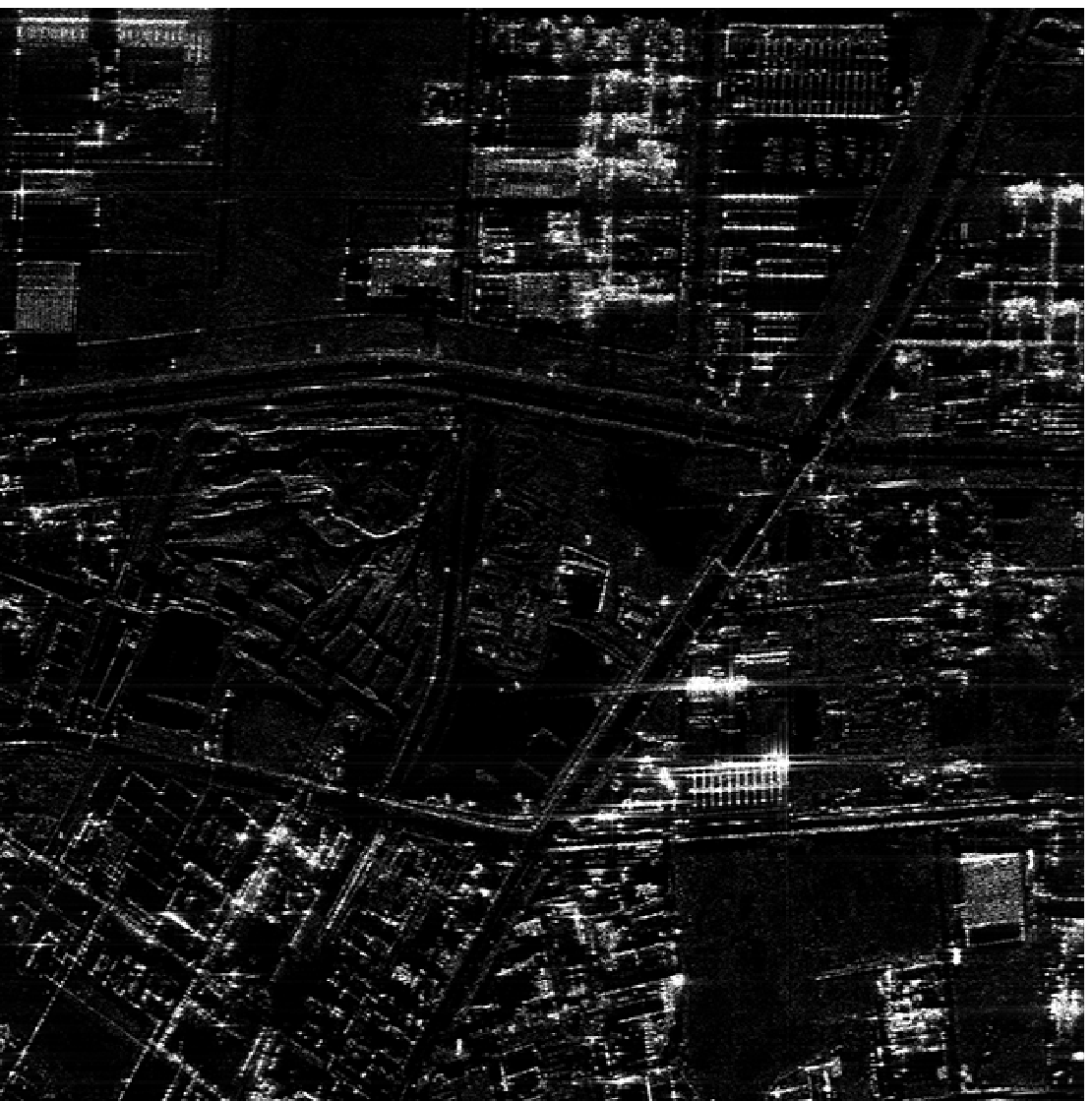}%
		\label{fig:6f}
	}
	\caption{Imaging results of different scene using $\ell_q$ regularization algorithm with full-sampled data. (a) Sea and ships; (b) salt pans; (c) salt pans; (d) salt pans; (e) habour and (f) urban area.}
	\label{fig:6}
\end{figure*}

\begin{figure*}[!t]
	\centering
	\subfloat[]{
		\includegraphics[width=2in]{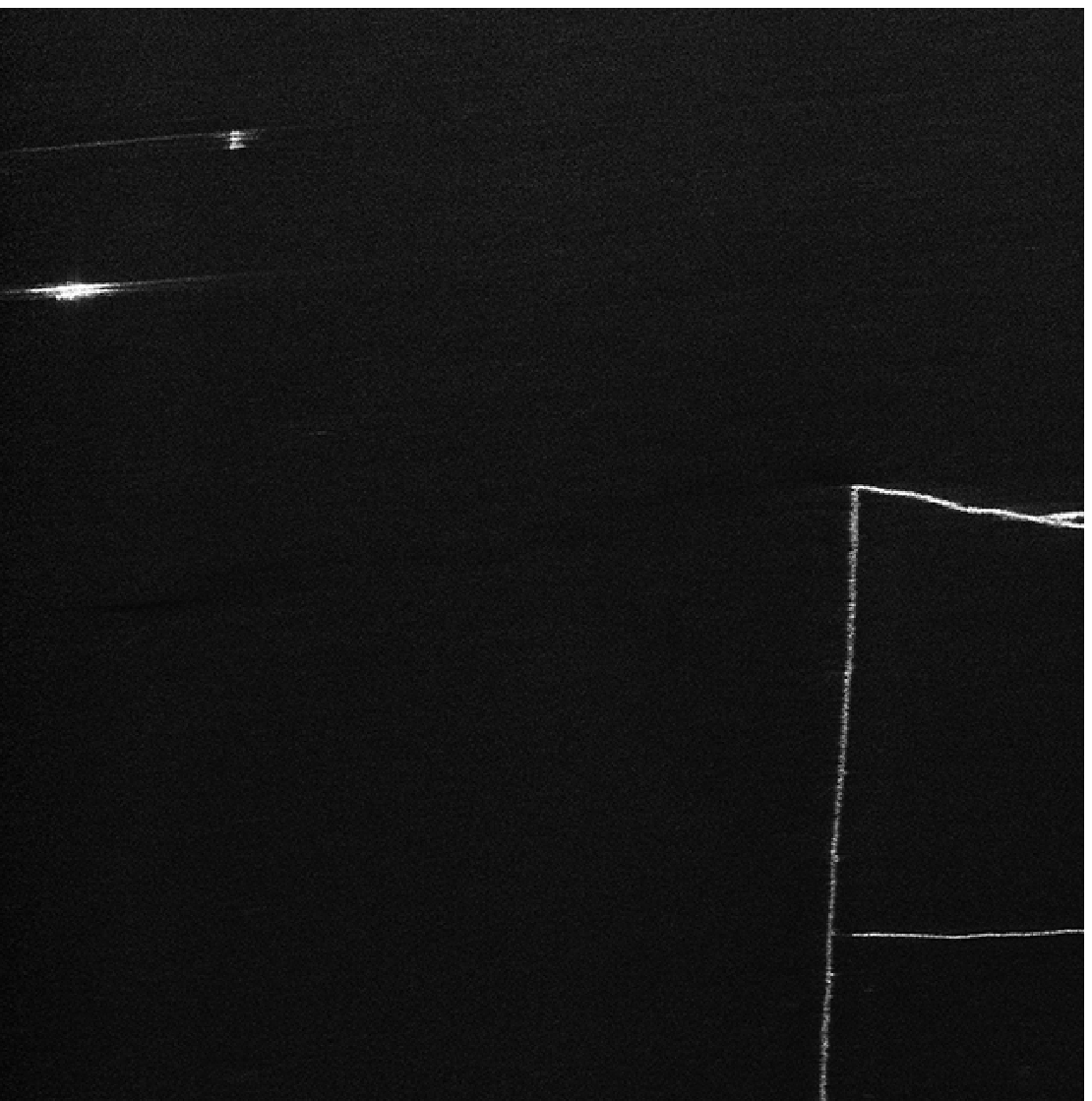}%
		\label{fig:61a}
	}~
	\subfloat[]{
		\includegraphics[width=2in]{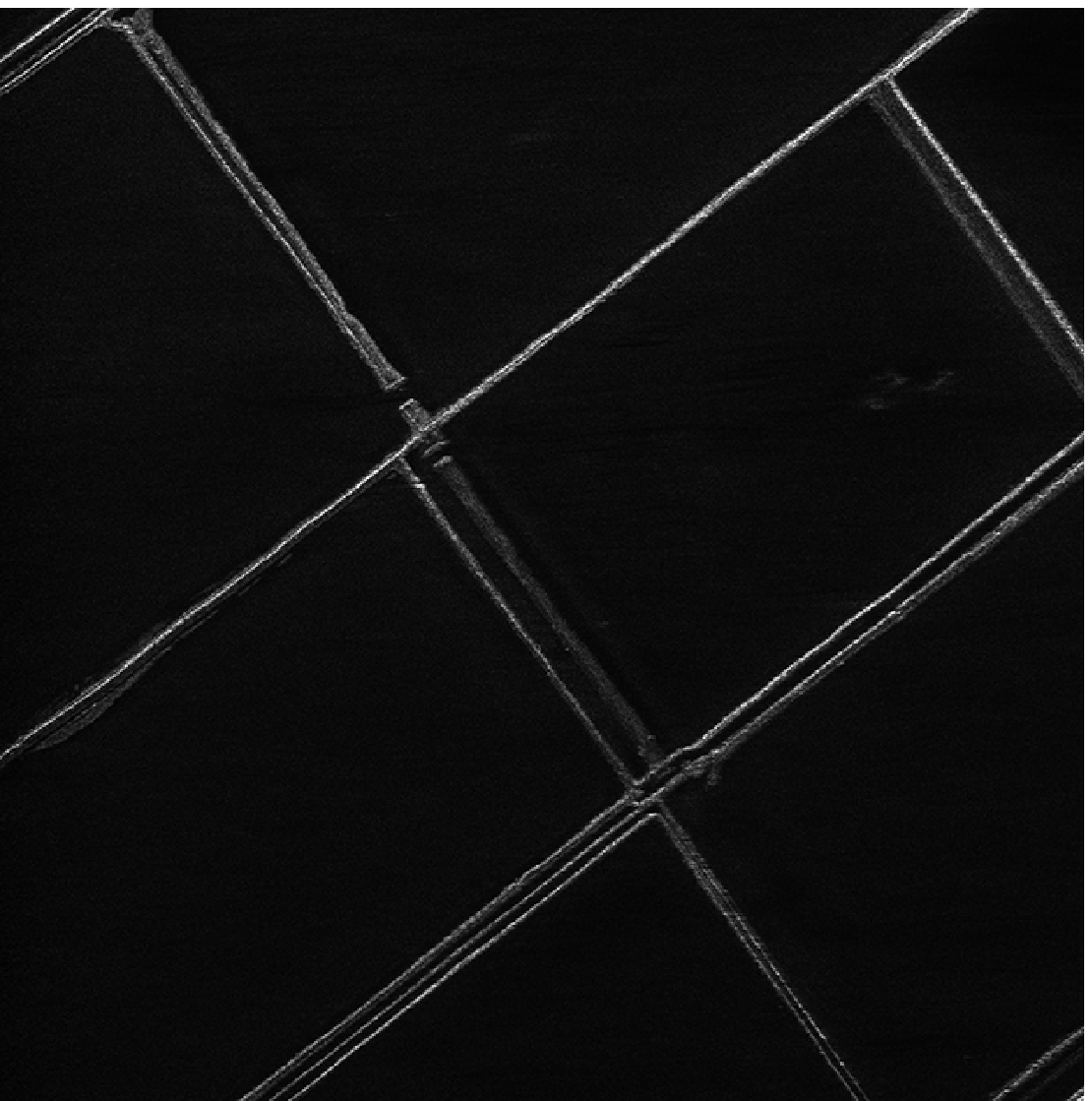}%
		\label{fig:61b}
	}~
	\subfloat[]{
		\includegraphics[width=2in]{s613.eps}%
		\label{fig:61c}
	}\\
	\subfloat[]{
		\includegraphics[width=2in]{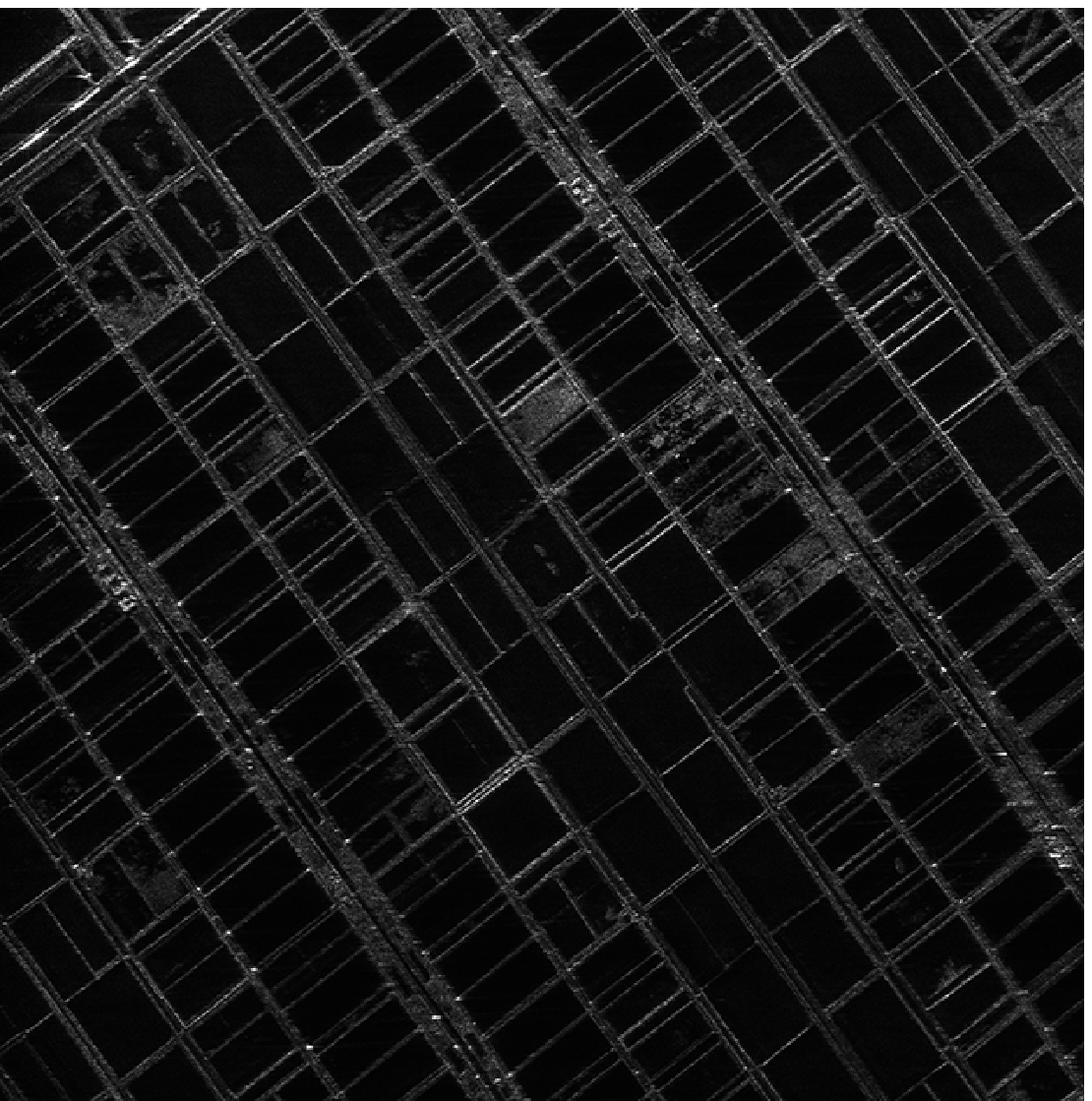}%
		\label{fig:61d}
	}~
	\subfloat[]{
		\includegraphics[width=2in]{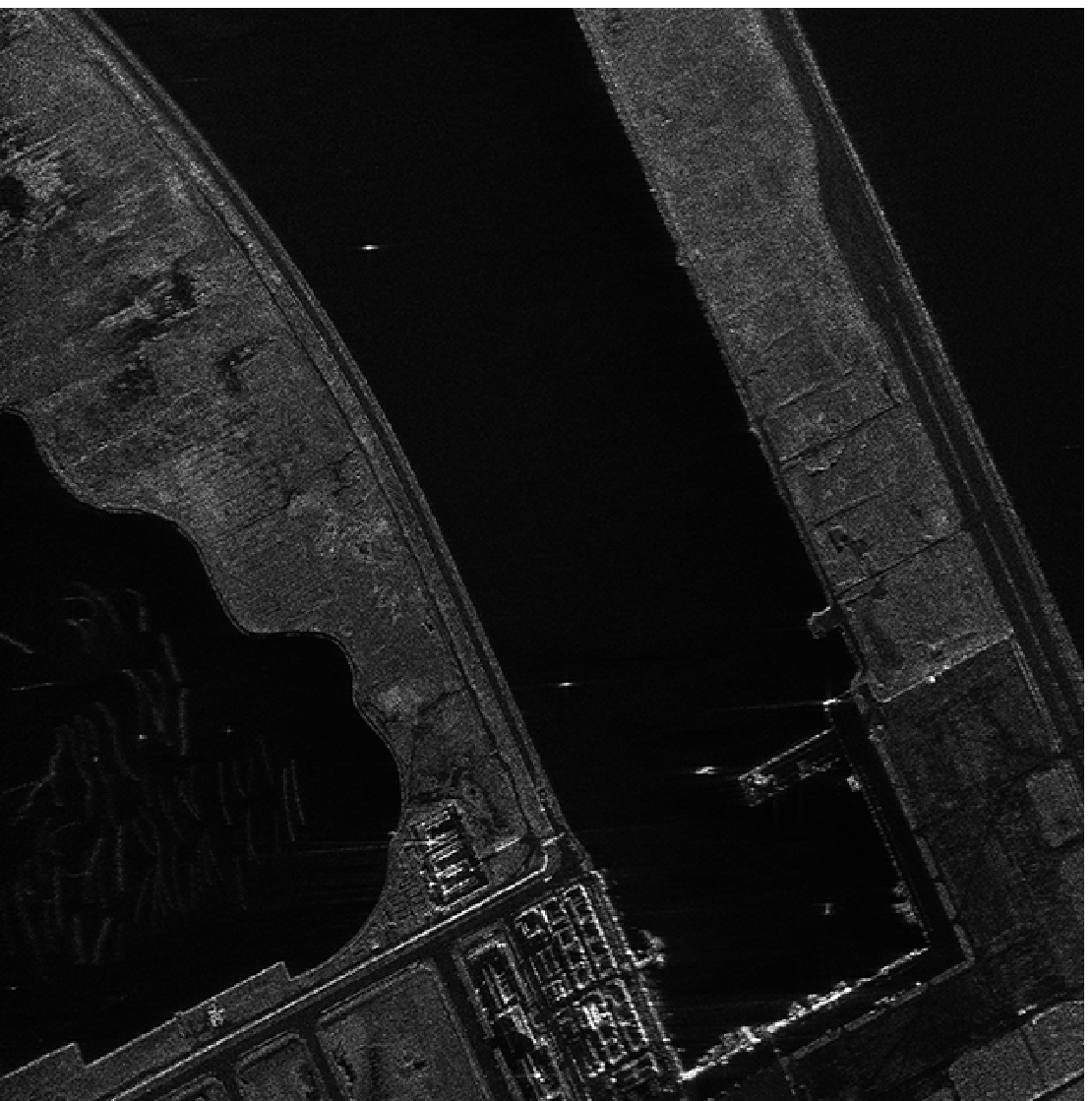}%
		\label{fig:61e}
	}~
	\subfloat[]{
		\includegraphics[width=2in]{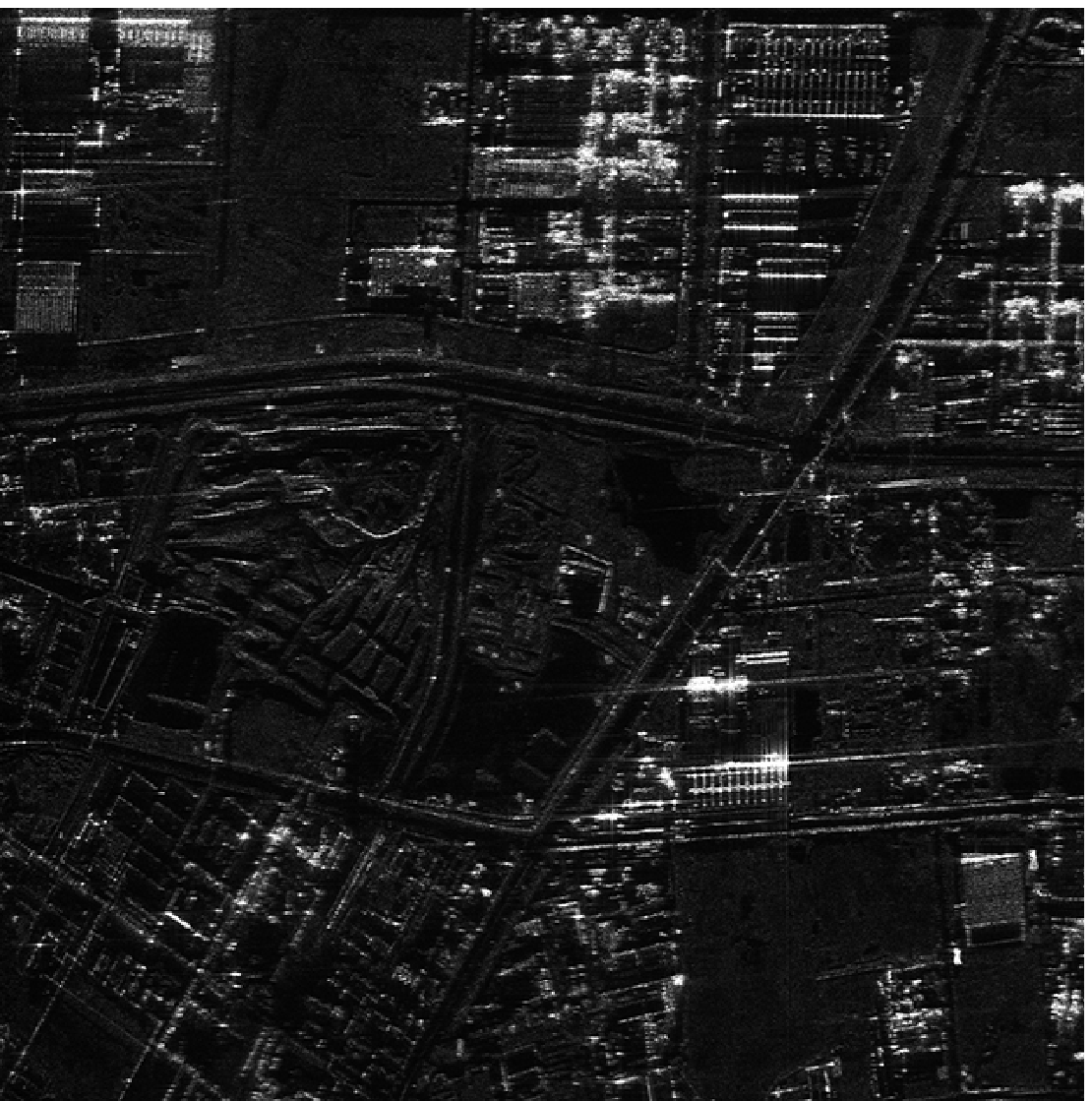}%
		\label{fig:61f}
	}
	\caption{Imaging results of different scene using traditional range-Doppler algorithm with full-sampled data. (a) Sea and ships; (b) salt pans; (c) salt pans; (d) salt pans (e) habour and (f) urban area.}
	\label{fig:61}
\end{figure*}

\begin{figure*}[!t]
	\centering
	\subfloat[]{
		\includegraphics[width=2in]{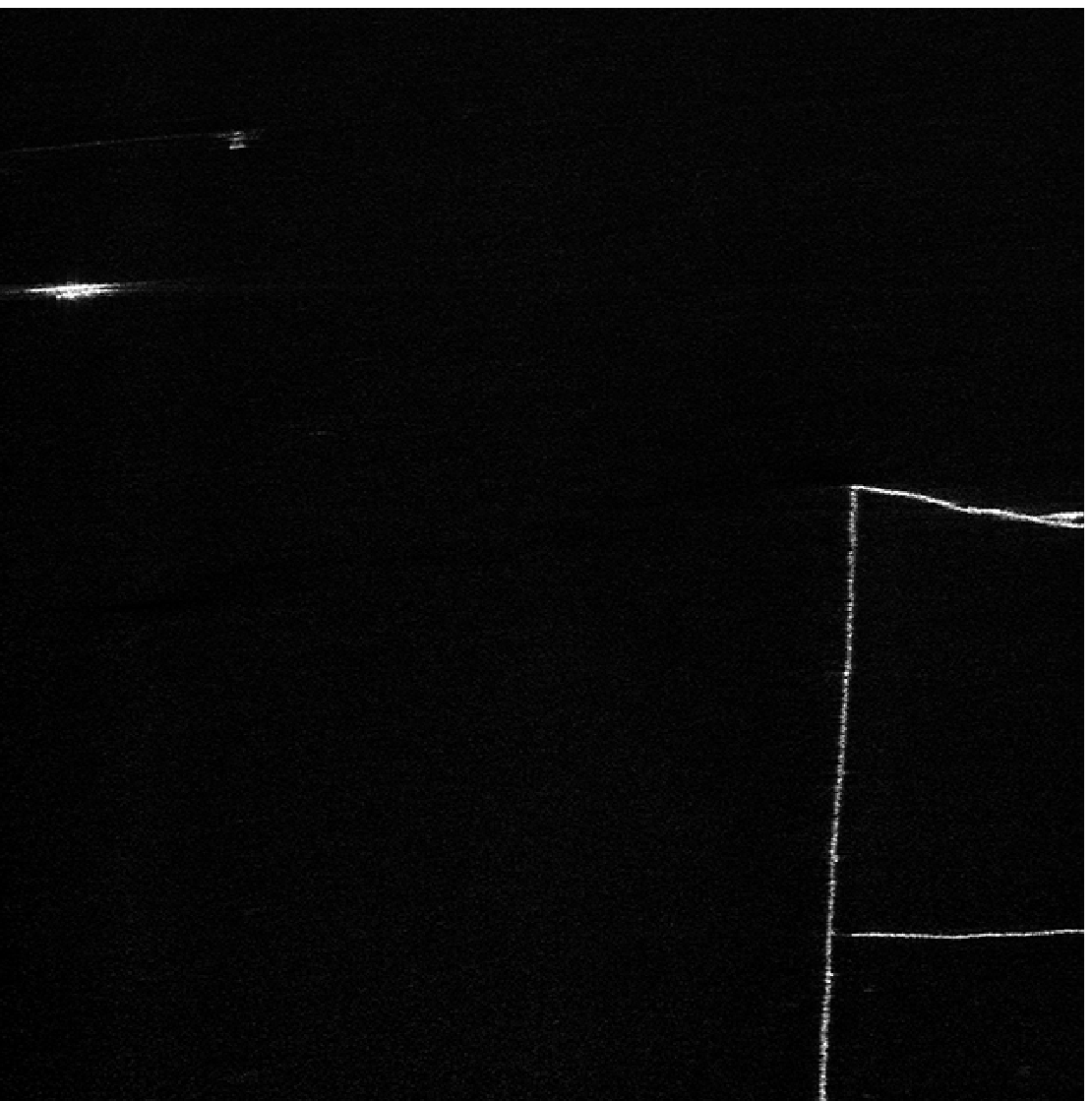}%
		\label{fig:9a}
	}~
	\subfloat[]{
		\includegraphics[width=2in]{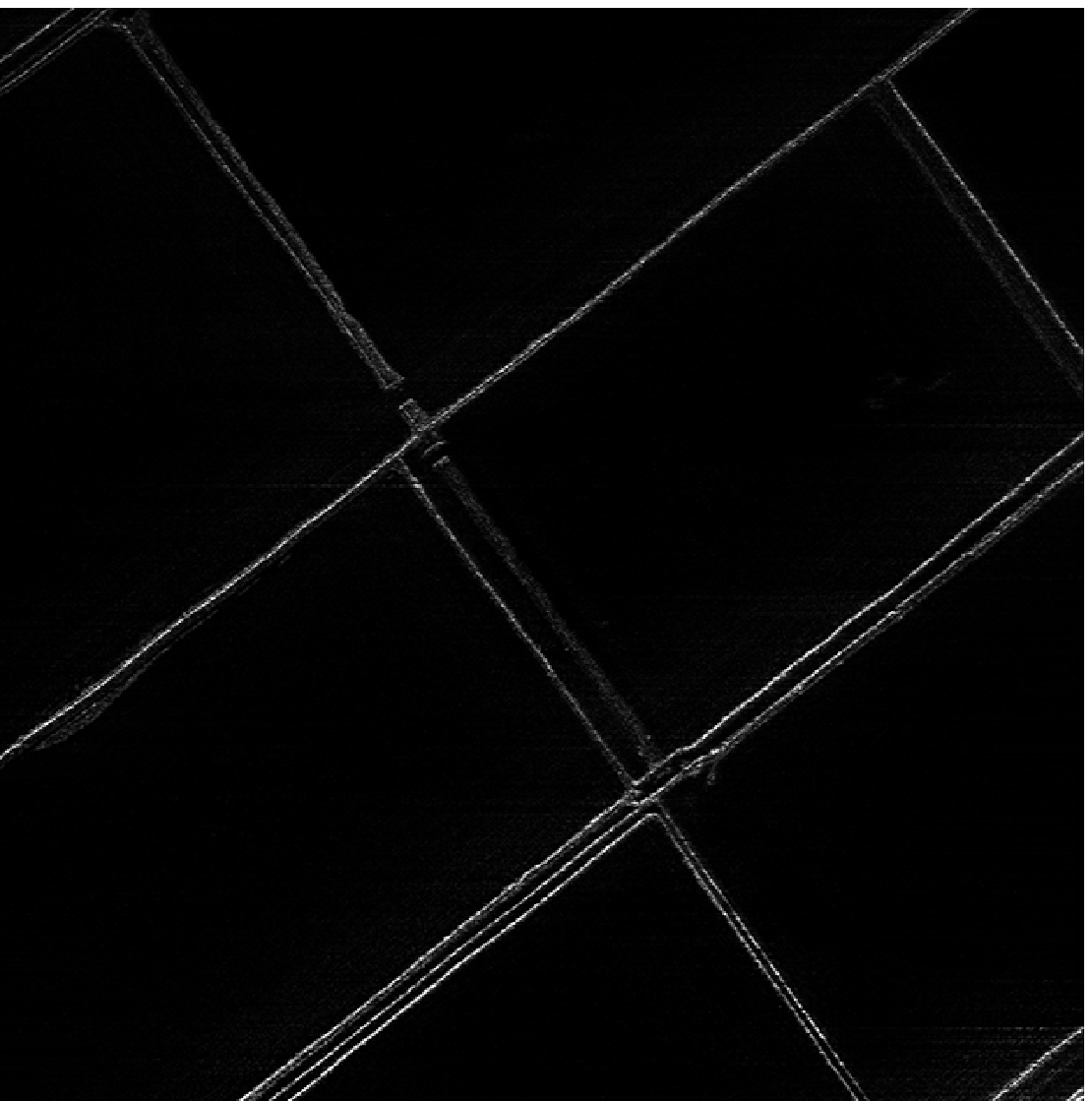}%
		\label{fig:9b}
	}~
	\subfloat[]{
		\includegraphics[width=2in]{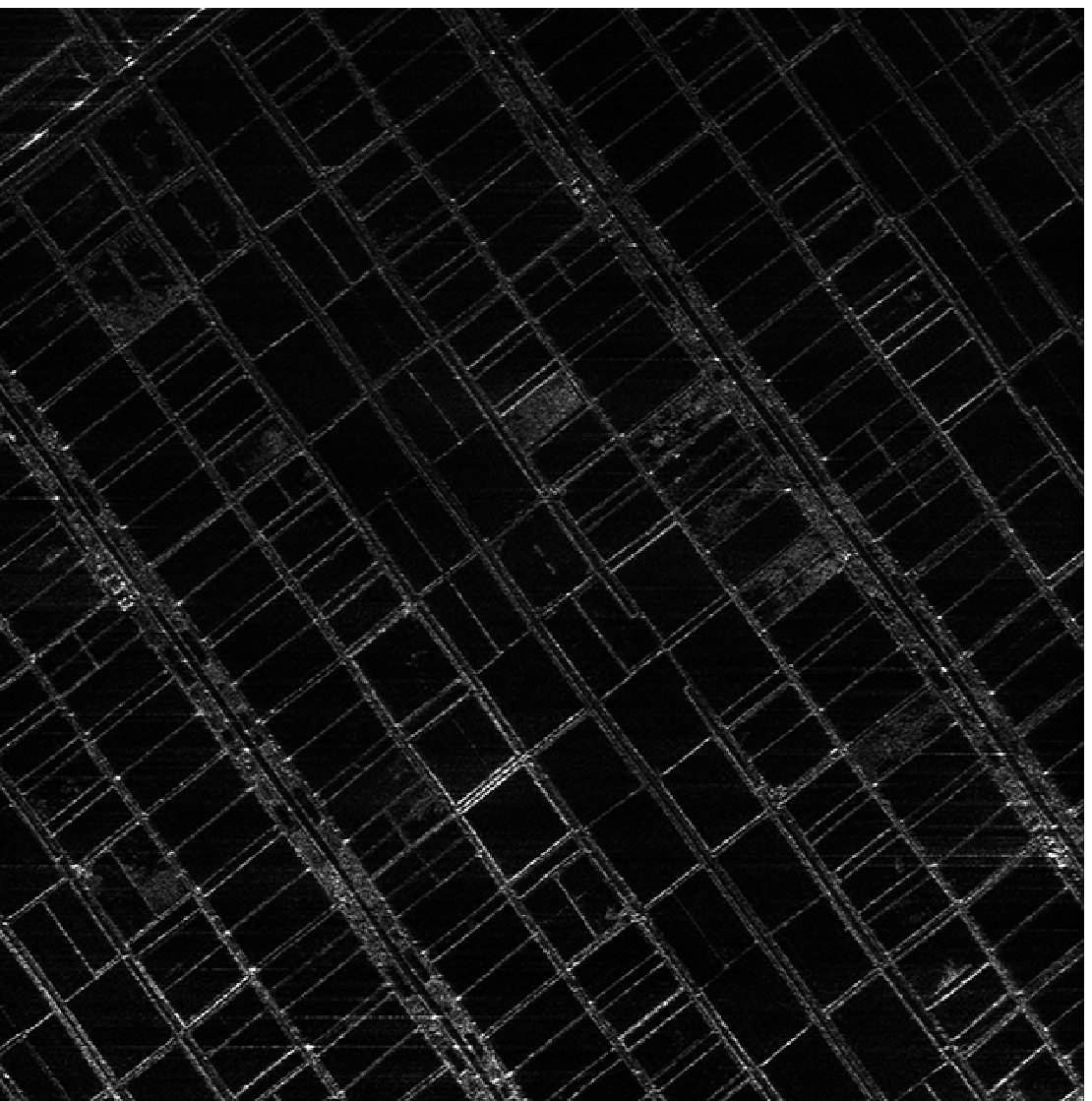}%
		\label{fig:9c}
	}\\
	\subfloat[]{
		\includegraphics[width=2in]{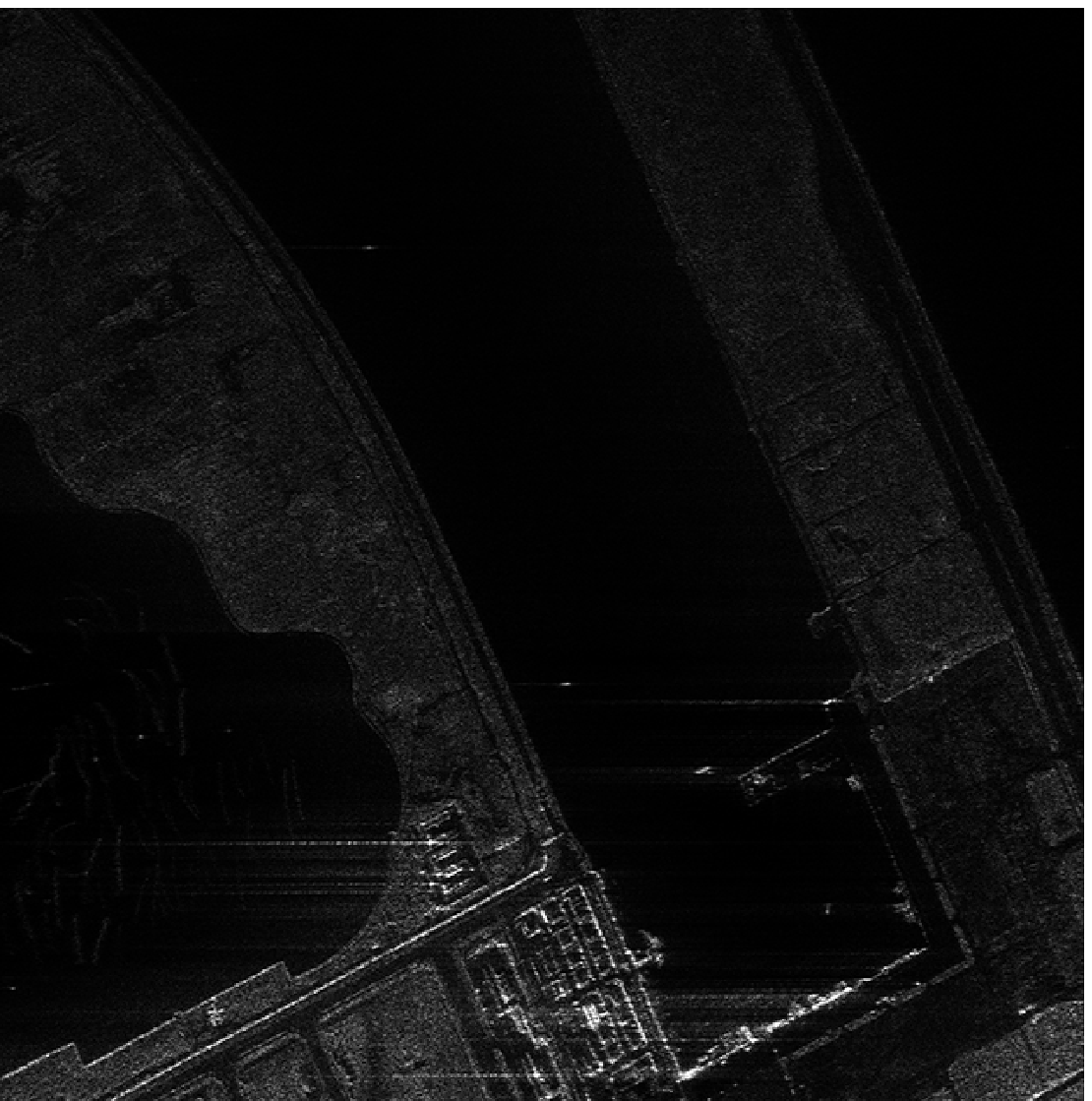}%
		\label{fig:9d}
	}~
	\subfloat[]{
		\includegraphics[width=2in]{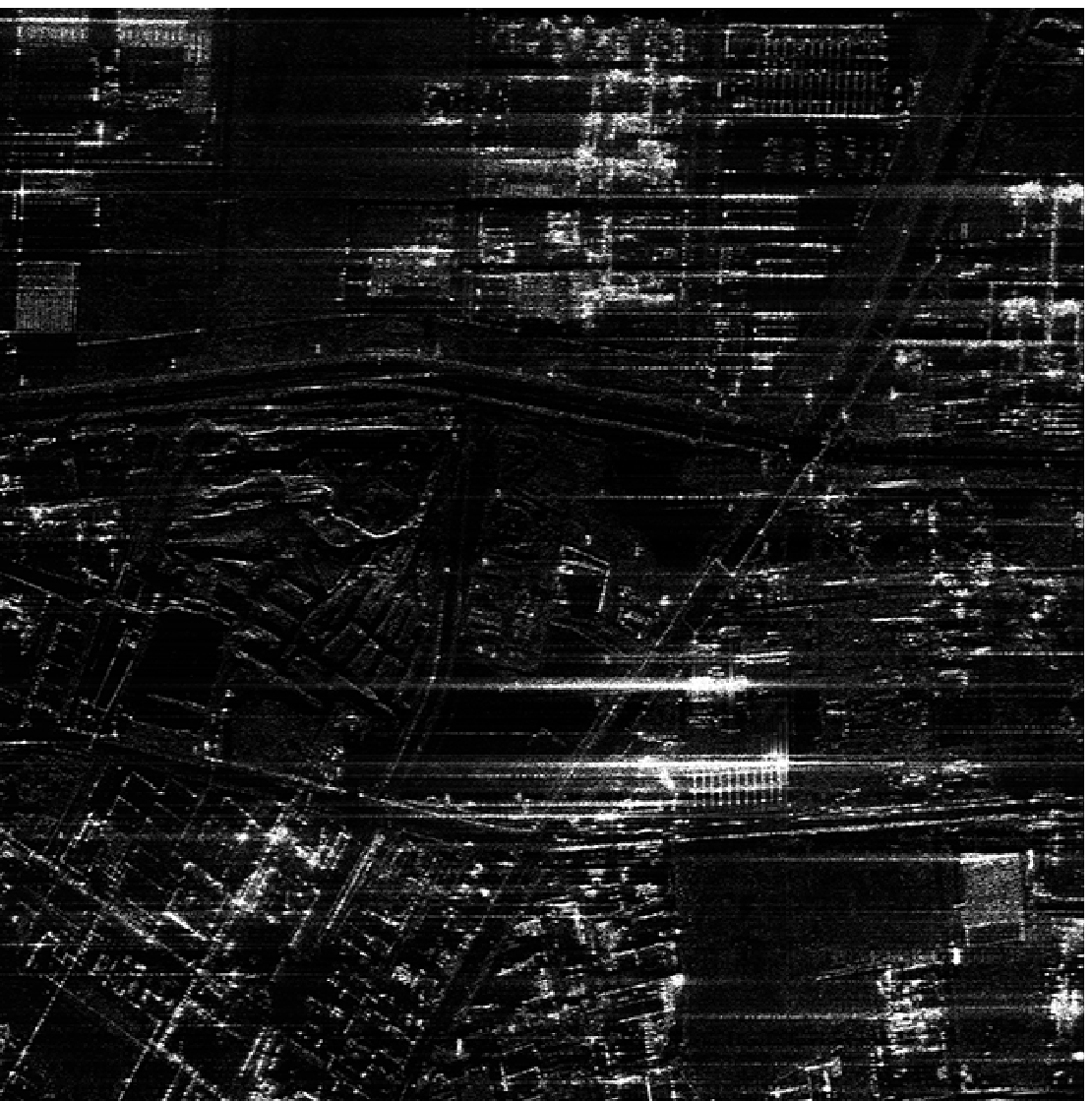}%
		\label{fig:9e}
	}
	\caption{Imaging results of different sparsity with 70\% down-sampled data. (a) scene sparsity 0.5\%; (b) scene sparsity 3\%; (c) scene sparsity 9.5\%; (d) scene sparsity 50\% and (e) scene sparsity over 90\%.}
	\label{fig:9}
\end{figure*}

\begin{figure}[!t]
    \centering
  \includegraphics[width=\columnwidth]{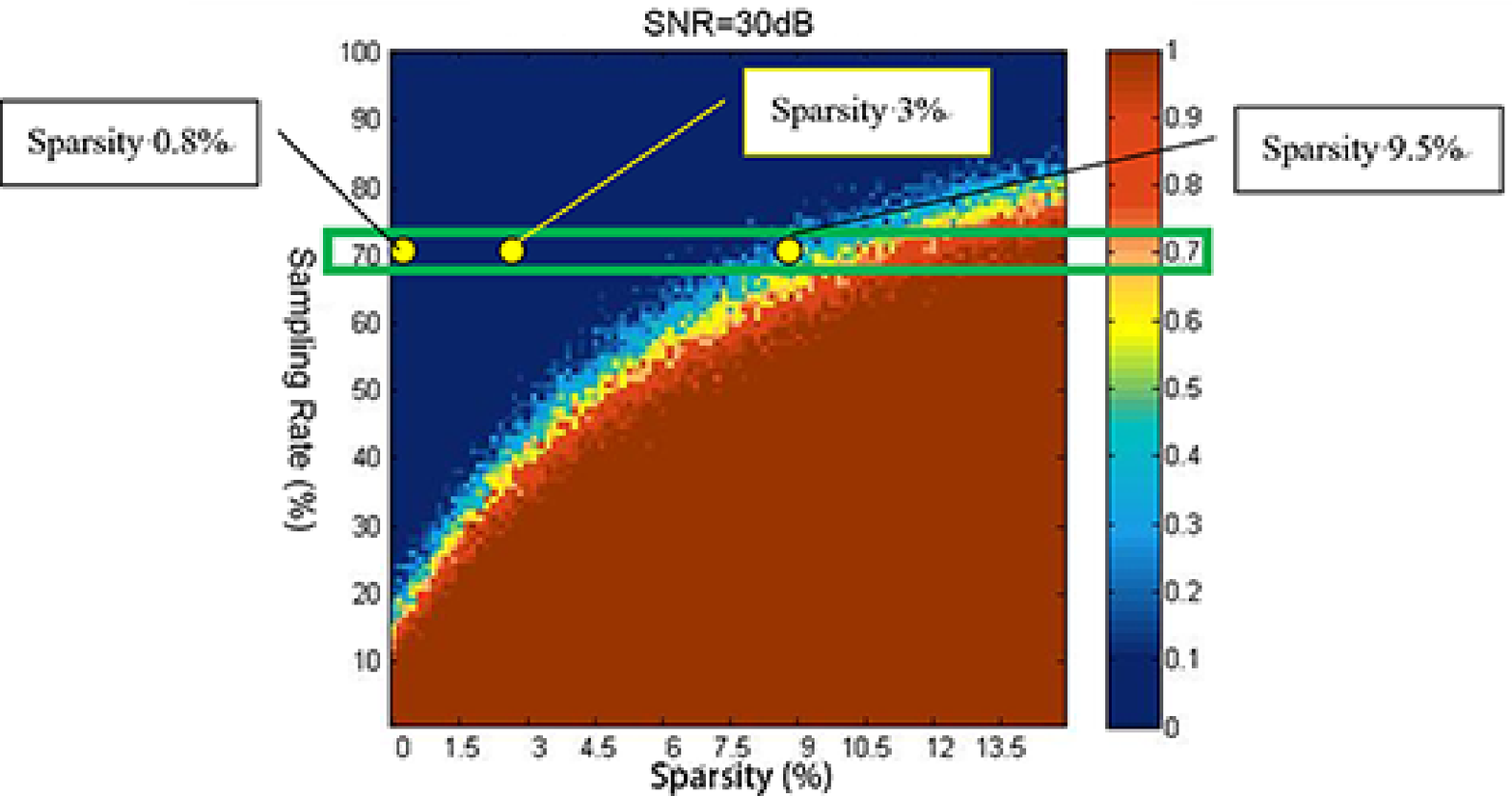}
  \caption{Phase transit diagram of system, red area means the failure area.}
  \centering
  \label{fig:10}
\end{figure}

\begin{figure*}[!t]
	\centering
	\subfloat[]{
		\includegraphics[width=2in]{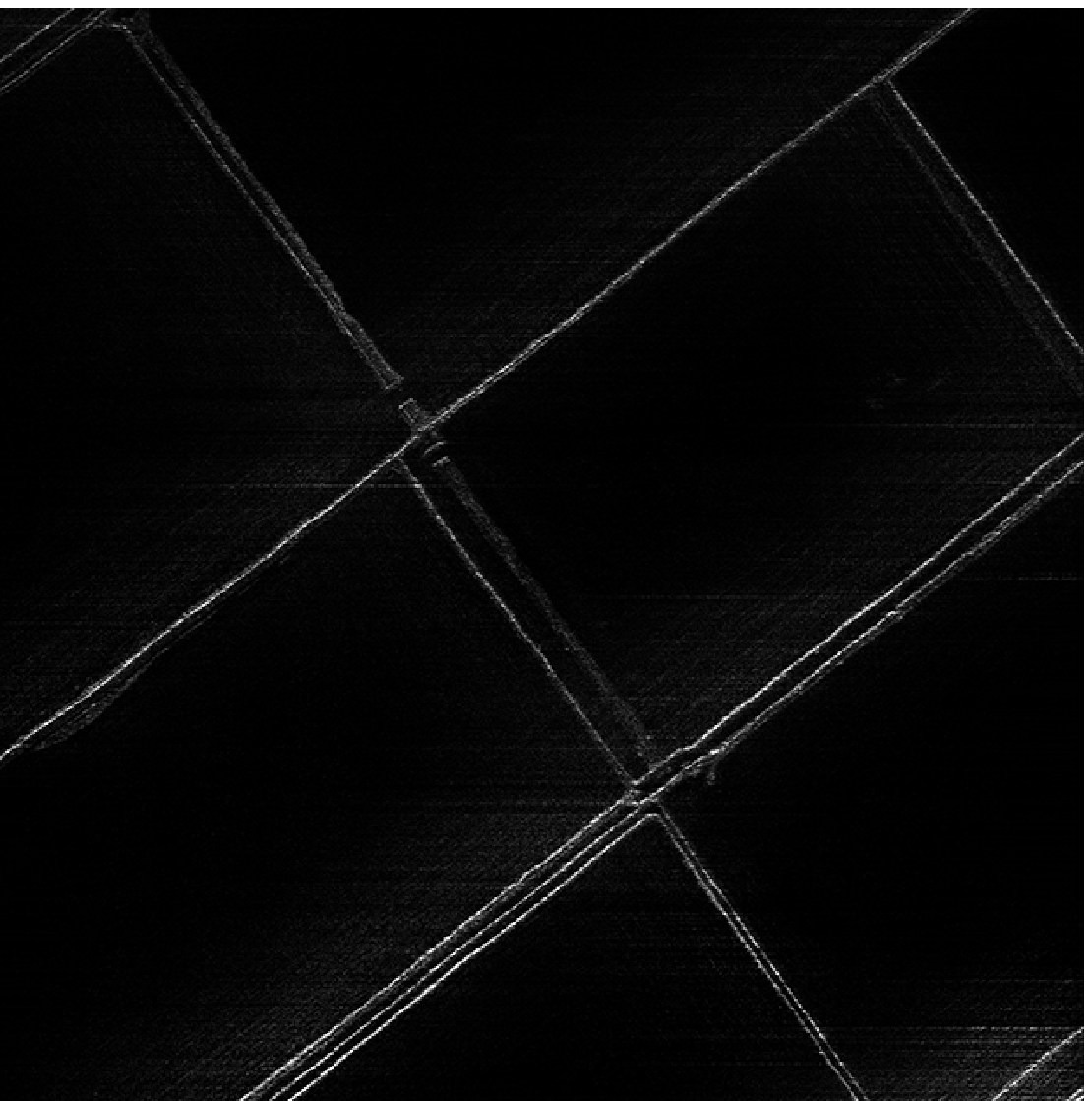}%
		\label{fig:11a}
	}\\
	\subfloat[]{
		\includegraphics[width=2in]{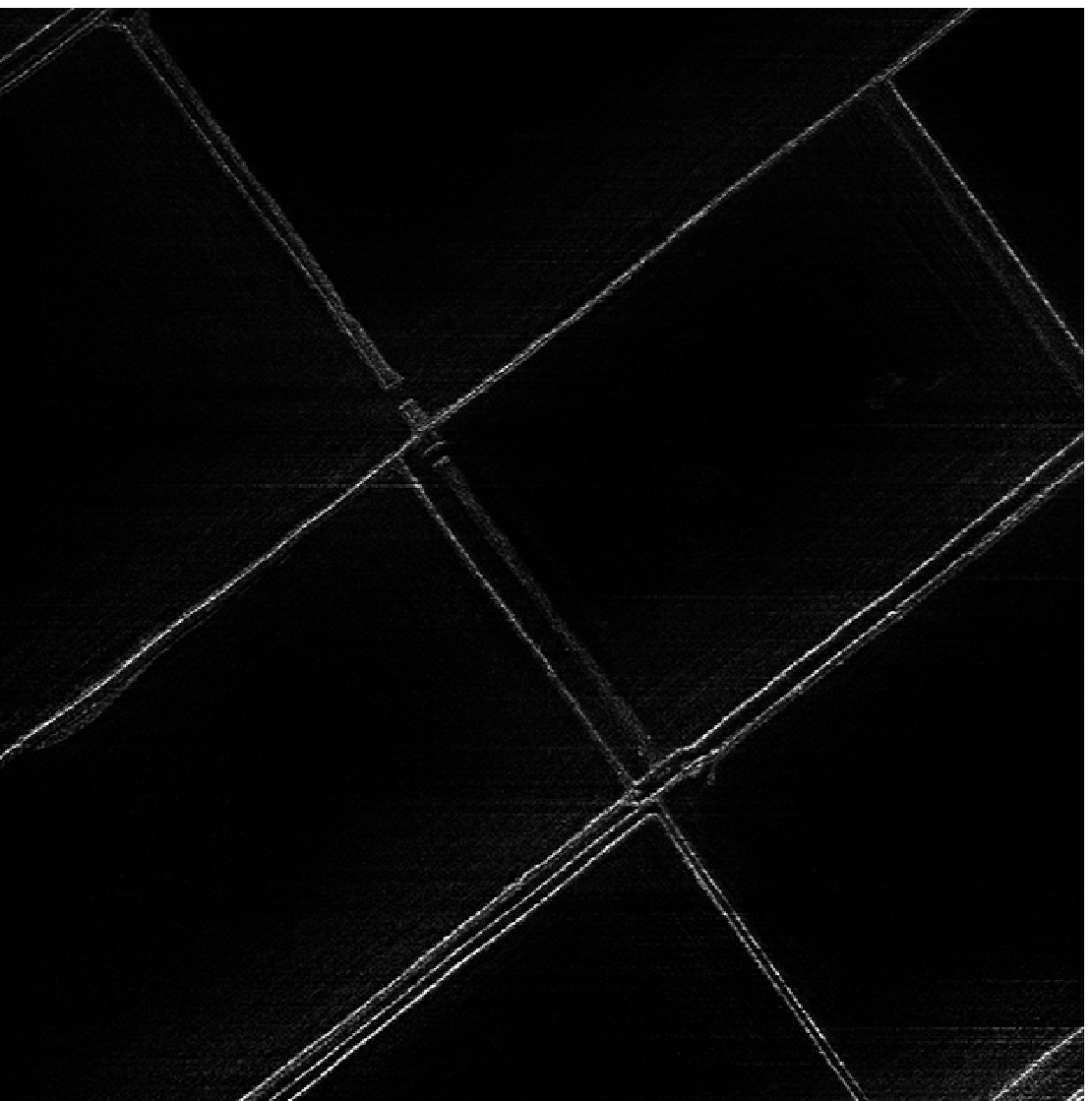}%
		\label{fig:11b}
	}~
	\subfloat[]{
		\includegraphics[width=2in]{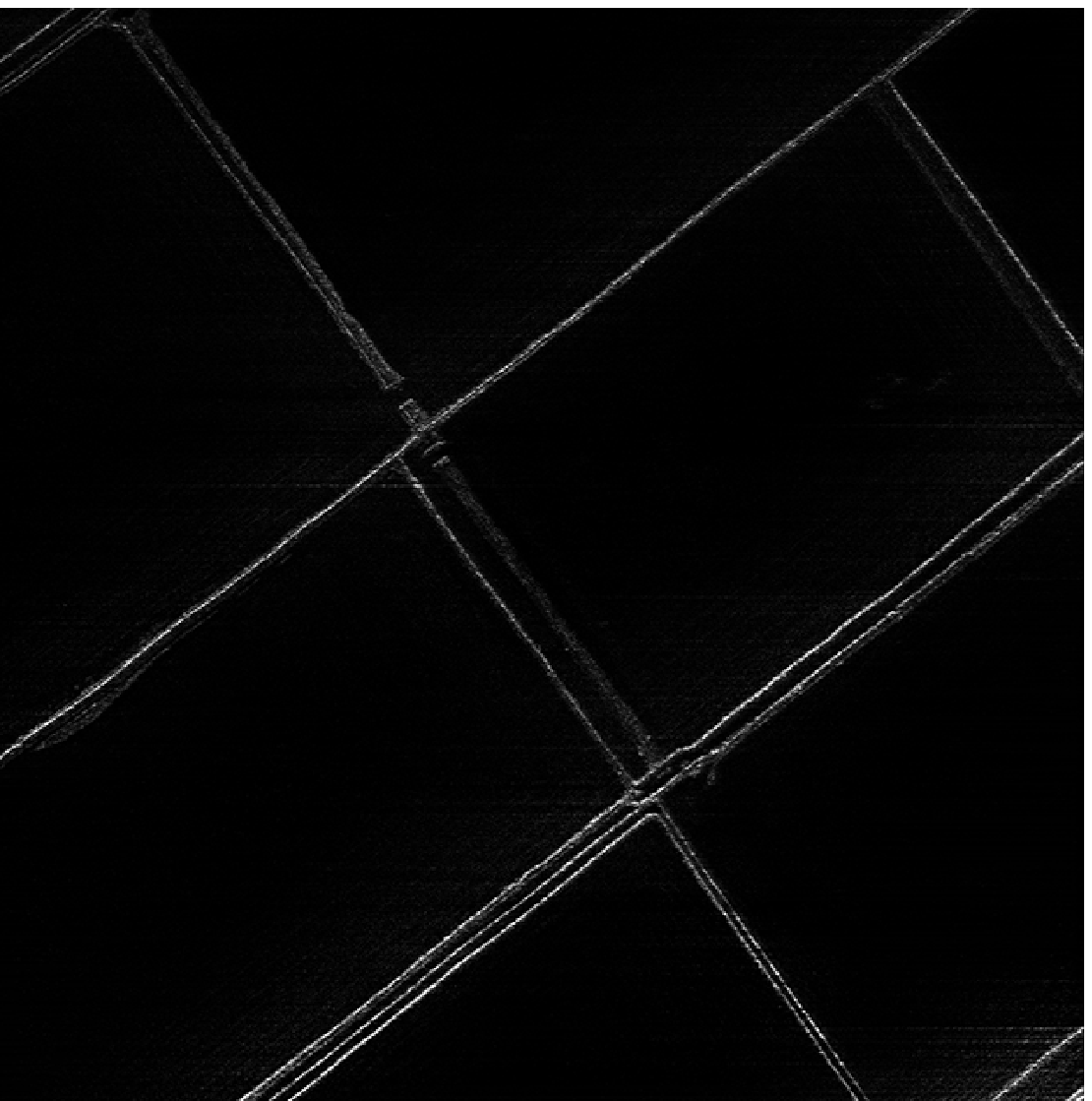}%
		\label{fig:11c}
	}~
	\subfloat[]{
		\includegraphics[width=2in]{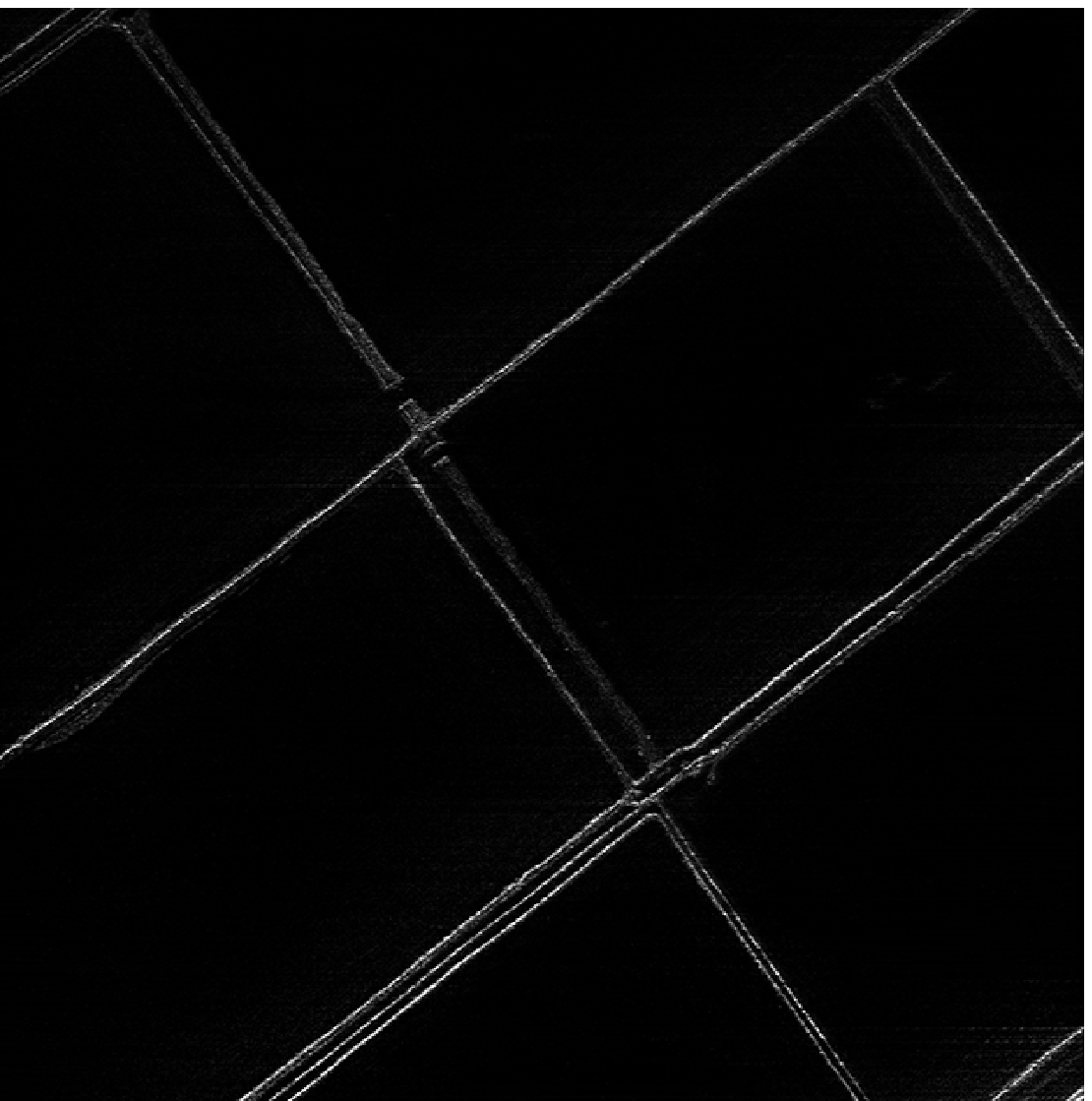}%
		\label{fig:11d}
	}\\
	\subfloat[]{
		\includegraphics[width=2in]{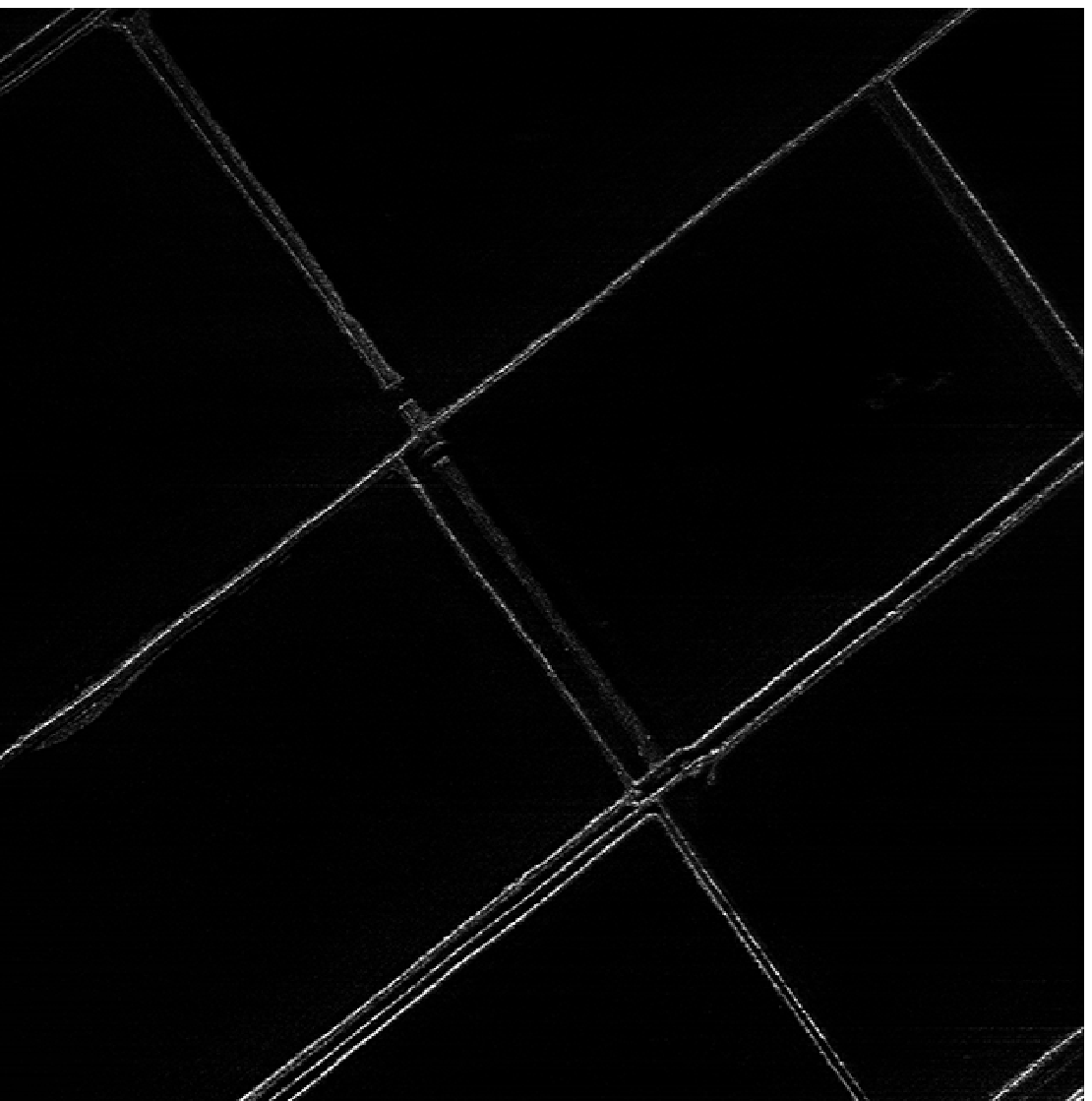}%
		\label{fig:11e}
	}~
	\subfloat[]{
		\includegraphics[width=2in]{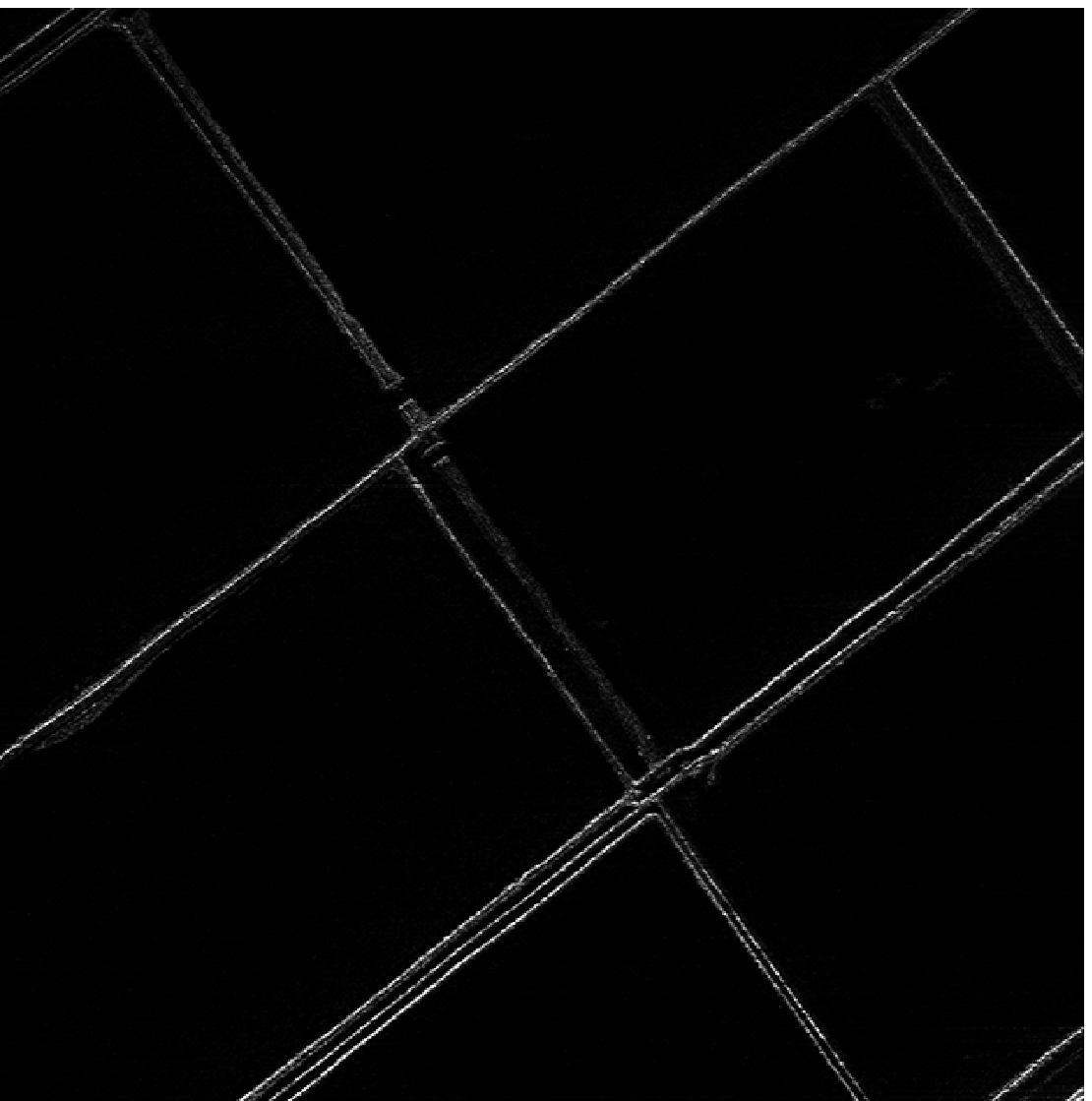}%
		\label{fig:11f}
	}~
	\subfloat[]{
		\includegraphics[width=2in]{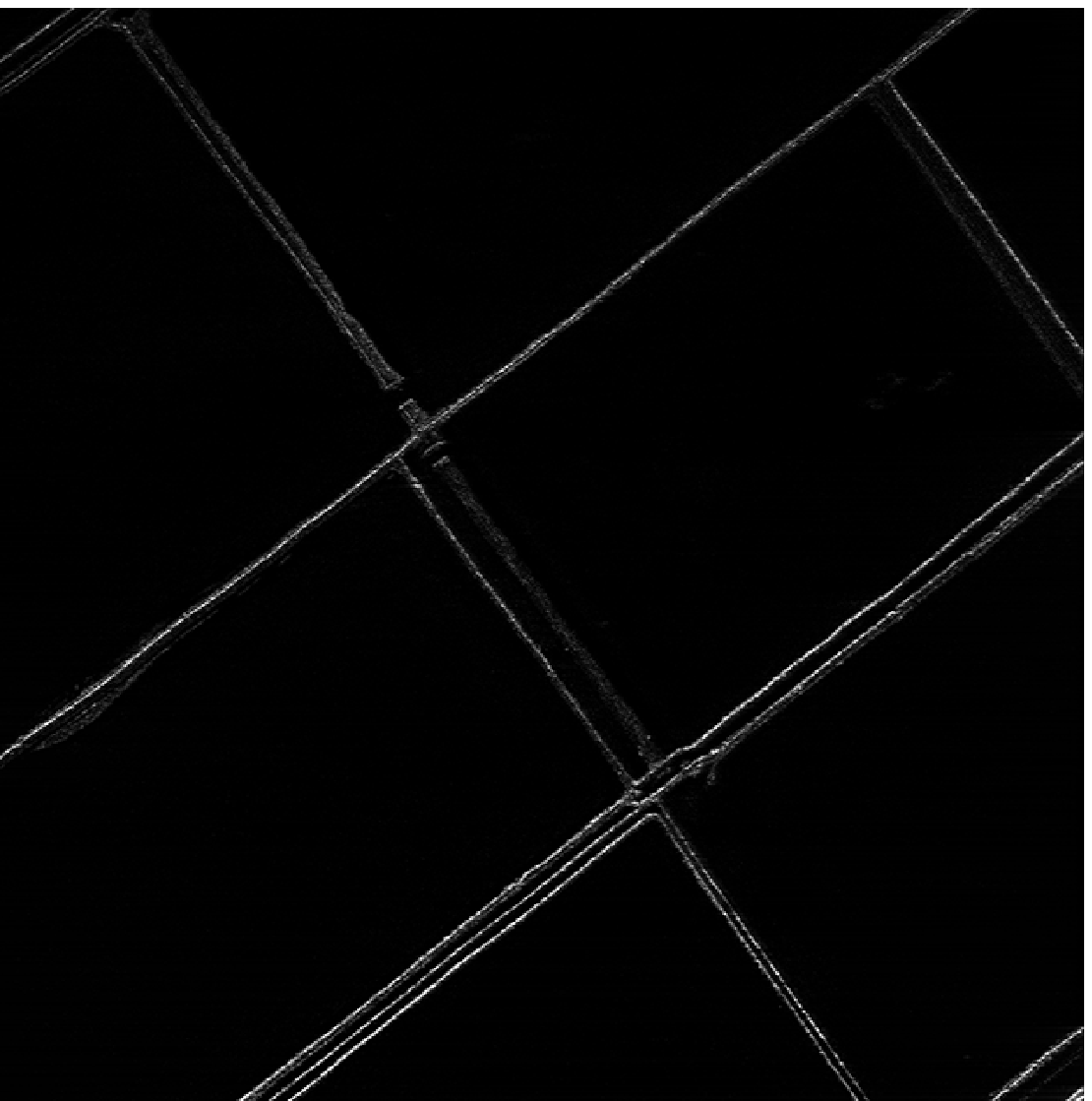}%
		\label{fig:11g}
	}
	\caption{Imaging results of different sampling rates with a scene of sparsity 3\%. (a) Under-sampling rate 40\% of Nyquist rate; (b) under-sampling rate 50\% of Nyquist rate; (c) under-sampling rate 60\% of Nyquist rate; (d) under-sampling rate 70\% of Nyquist rate; (e) under-sampling rate 80\% of Nyquist rate; (f) under-sampling rate 90\% of Nyquist rate and (g) under-sampling rate 100\% of Nyquist rate.}
	\label{fig:11}
\end{figure*}

\begin{figure*}[!t]
	\centering
	\subfloat[]{
		\includegraphics[width=2in]{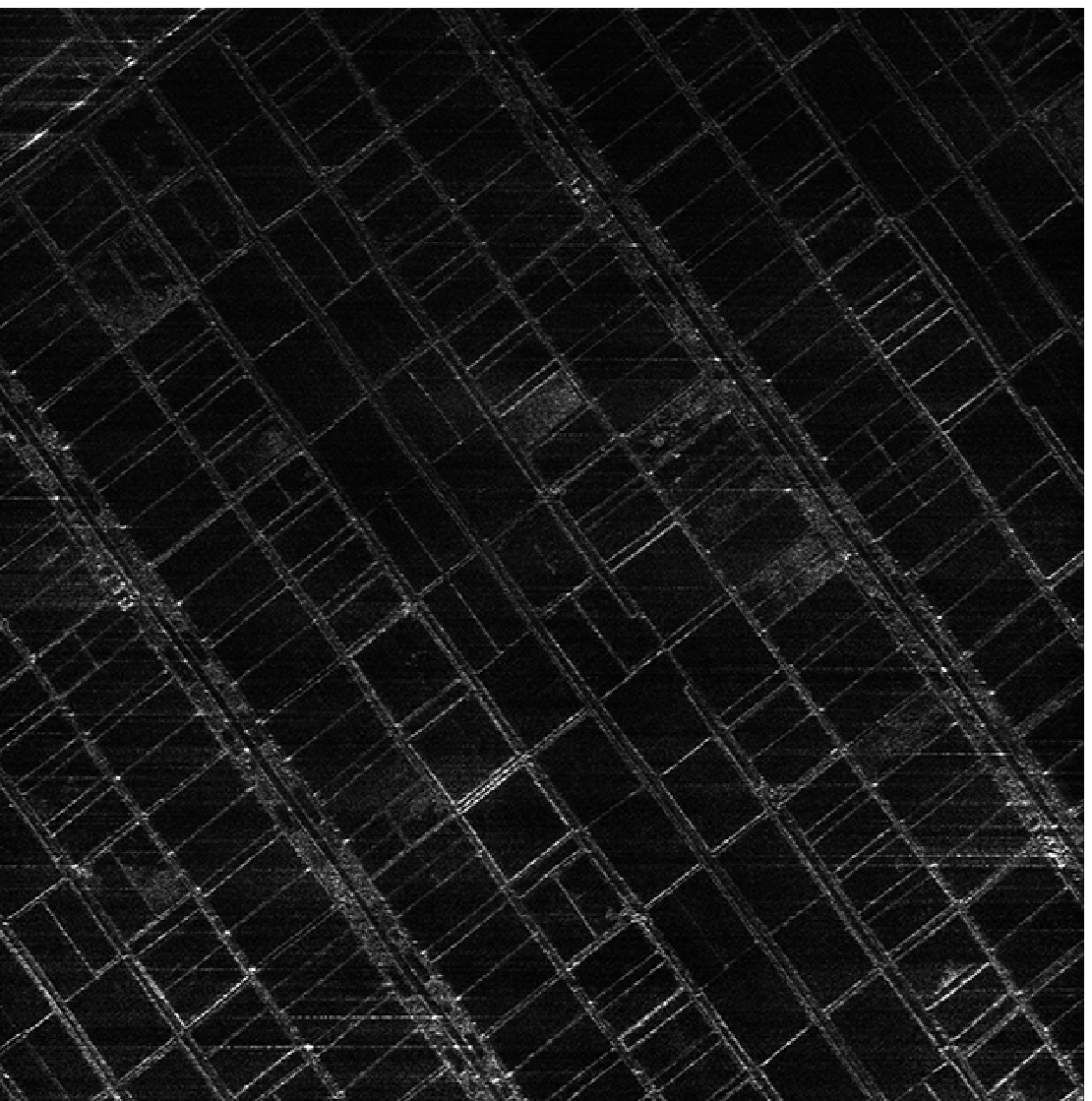}%
		\label{fig:12a}
	}\\
	\subfloat[]{
		\includegraphics[width=2in]{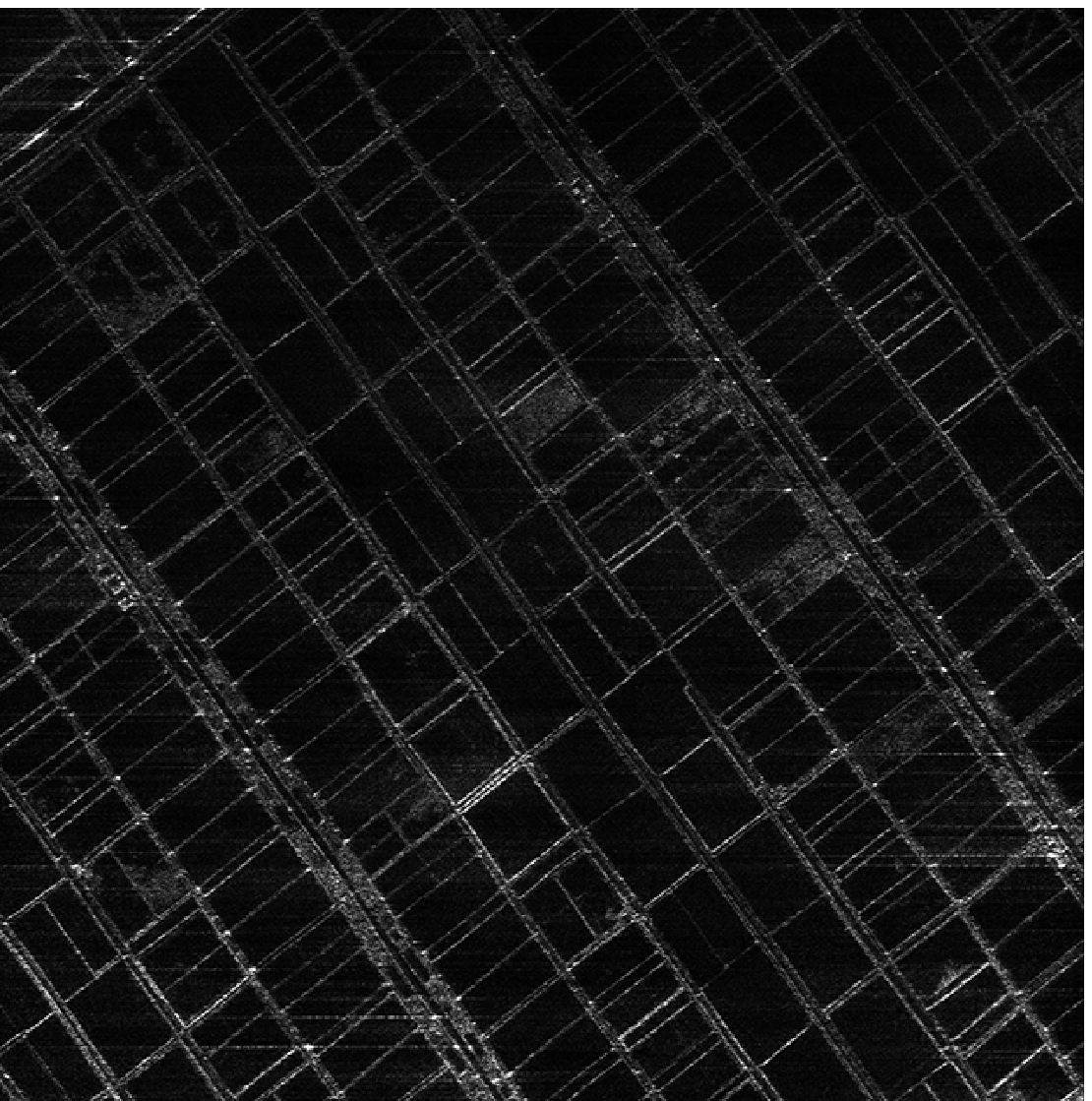}%
		\label{fig:12b}
	}~
	\subfloat[]{
		\includegraphics[width=2in]{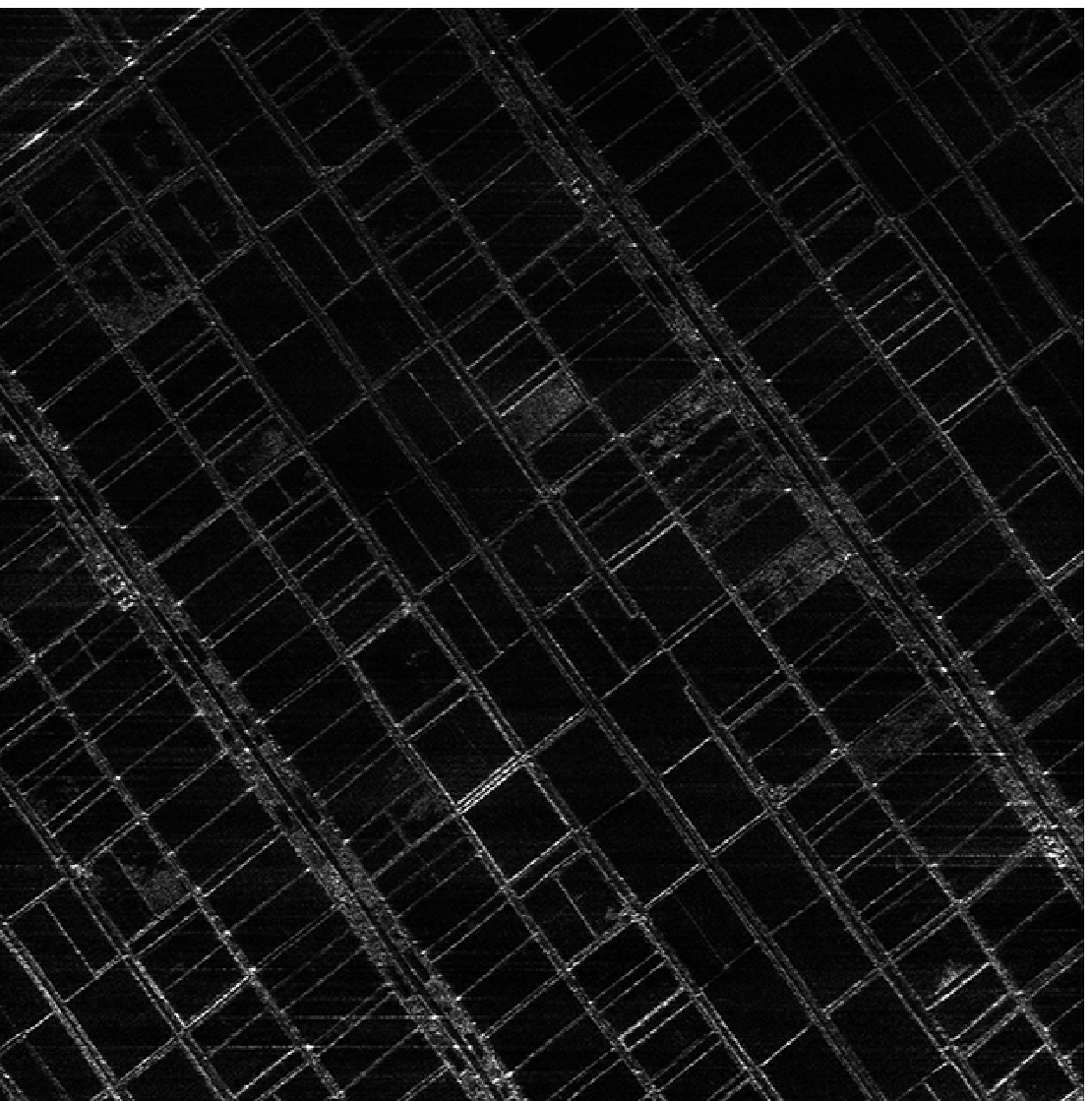}%
		\label{fig:12c}
	}~
	\subfloat[]{
		\includegraphics[width=2in]{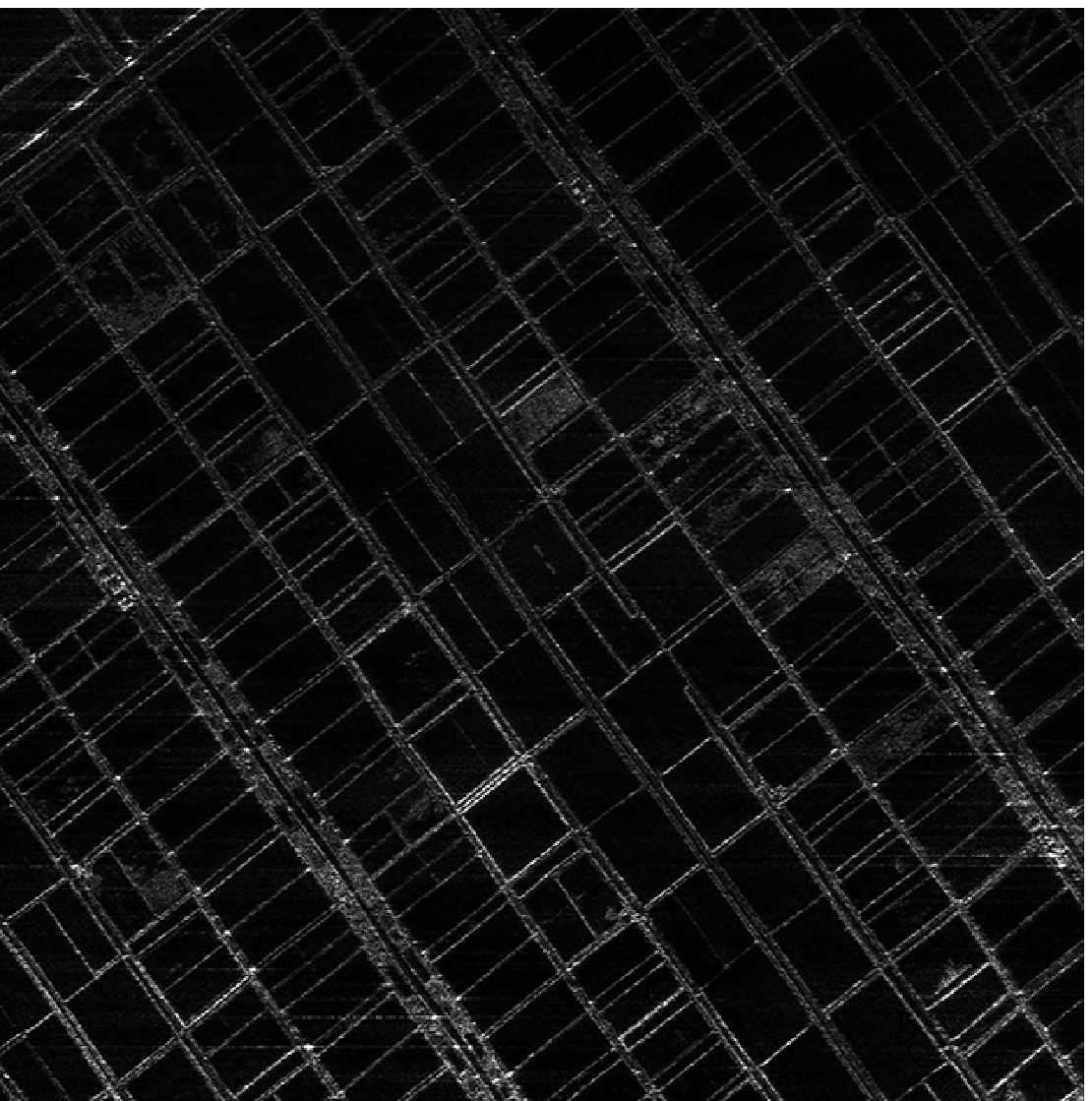}%
		\label{fig:12d}
	}\\
	\subfloat[]{
		\includegraphics[width=2in]{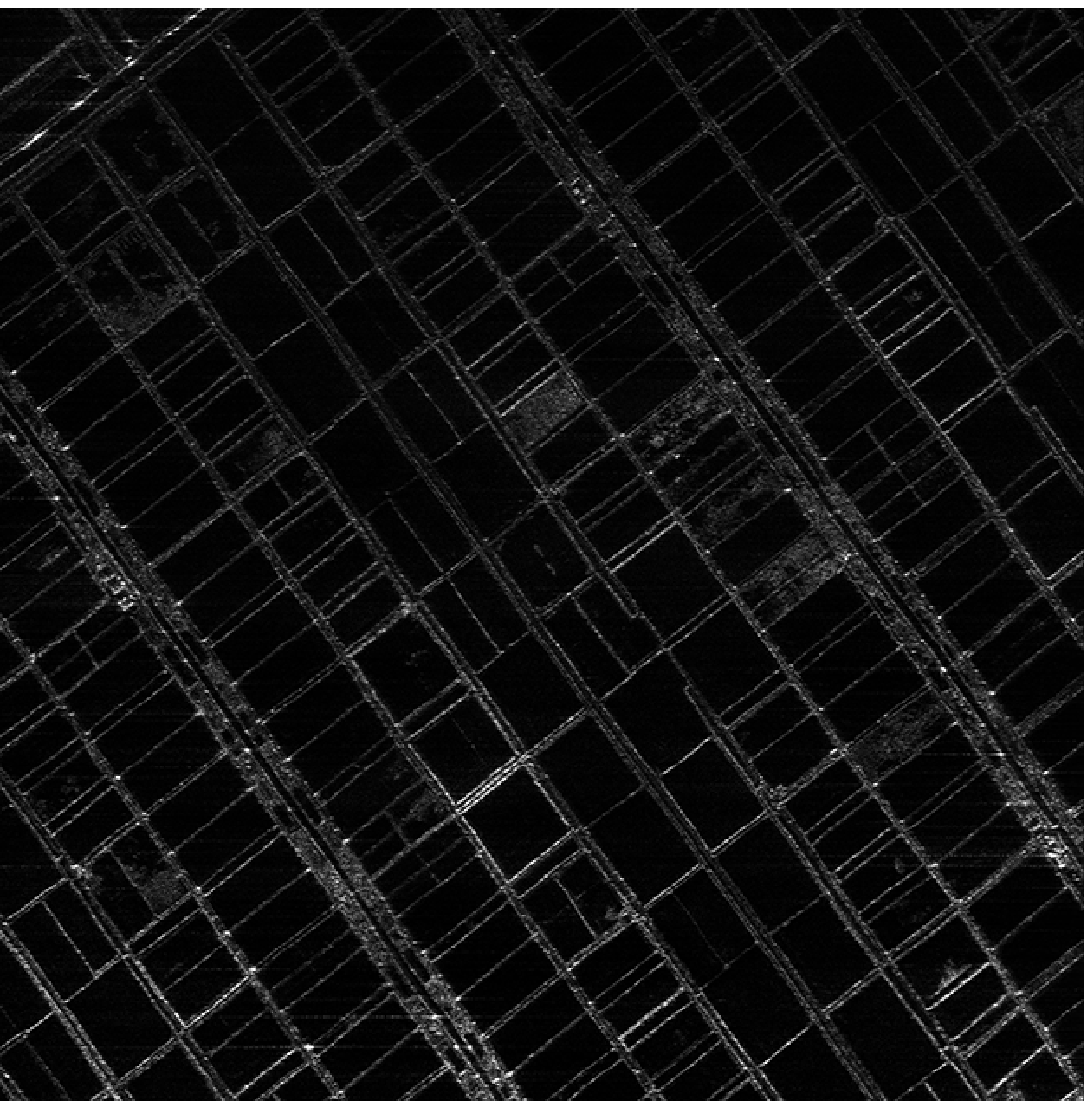}%
		\label{fig:12e}
	}~
	\subfloat[]{
		\includegraphics[width=2in]{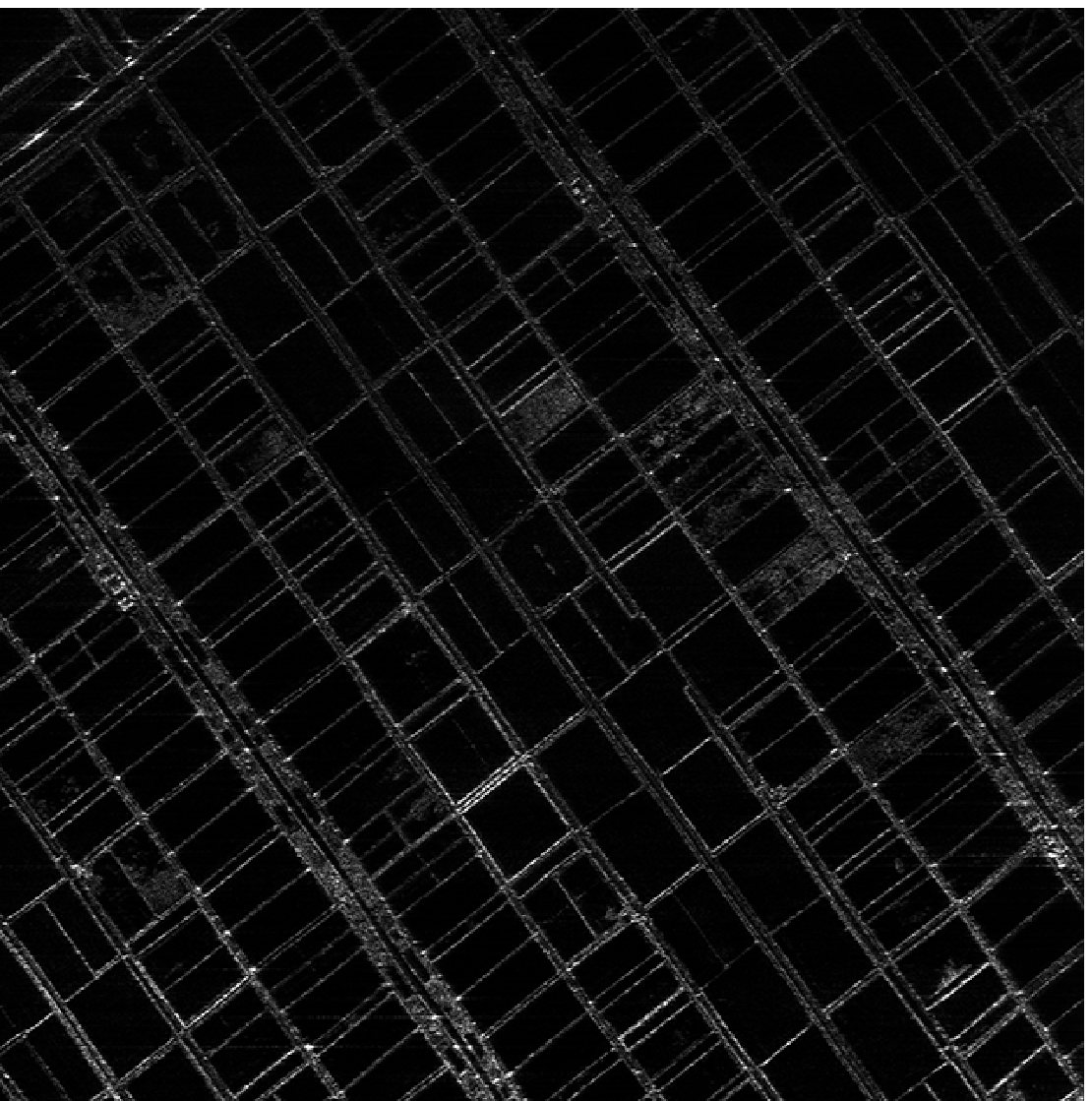}%
		\label{fig:12f}
	}~
	\subfloat[]{
		\includegraphics[width=2in]{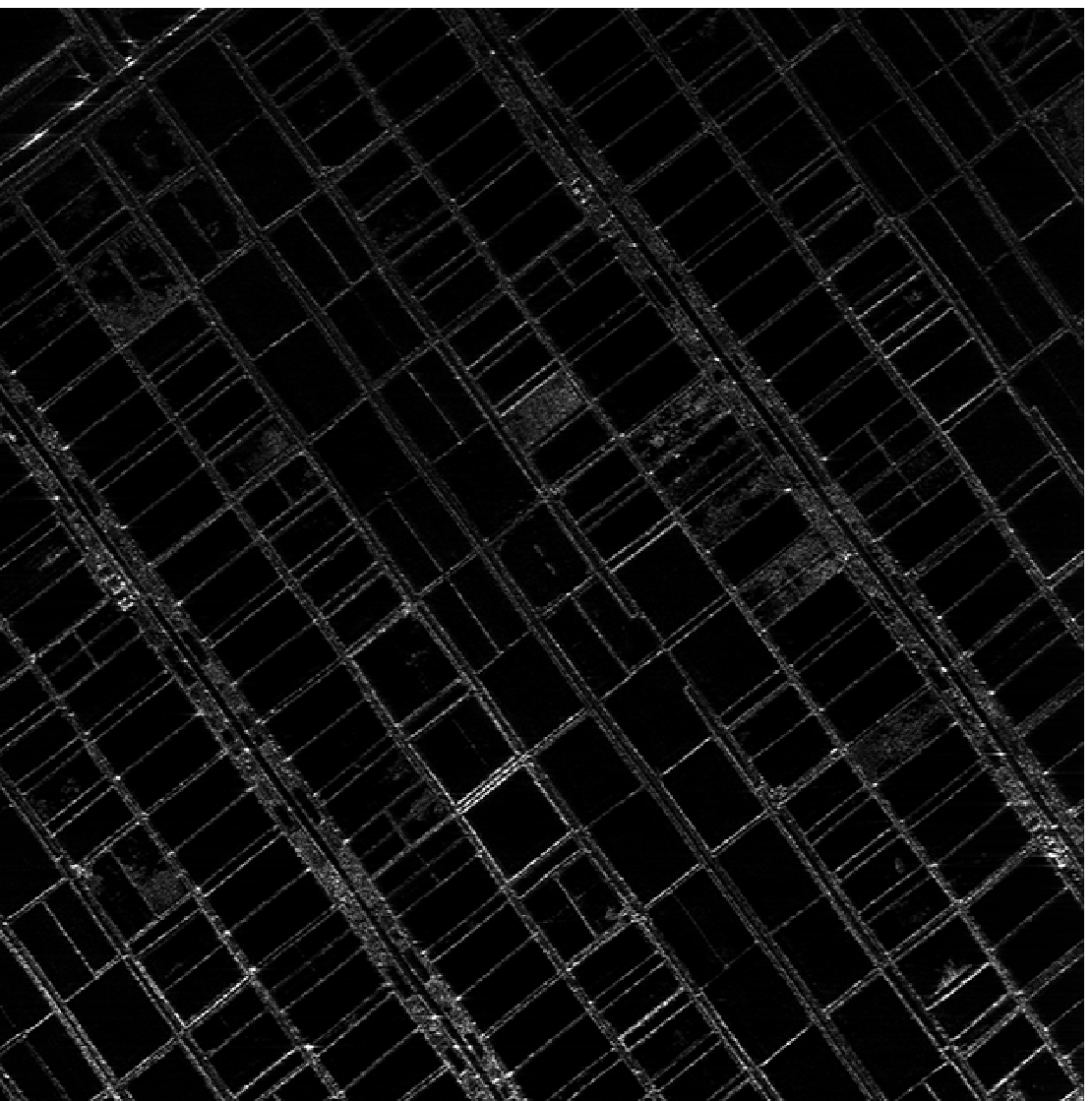}%
		\label{fig:12g}
	}
	\caption{Imaging results of different sampling rates with a scene of sparsity 9.5\%.  (a) Under-sampling rate 40\% of Nyquist rate; (b) under-sampling rate 50\% of Nyquist rate; (c) under-sampling rate 60\% of Nyquist rate; (d) under-sampling rate 70\% of Nyquist rate; (e) under-sampling rate 80\% of Nyquist rate; (f) under-sampling rate 90\% of Nyquist rate and (g) under-sampling rate 100\% of Nyquist rate; }
	\label{fig:12}
\end{figure*}


\begin{figure}[!t]
	\centering
	\includegraphics[width=\columnwidth]{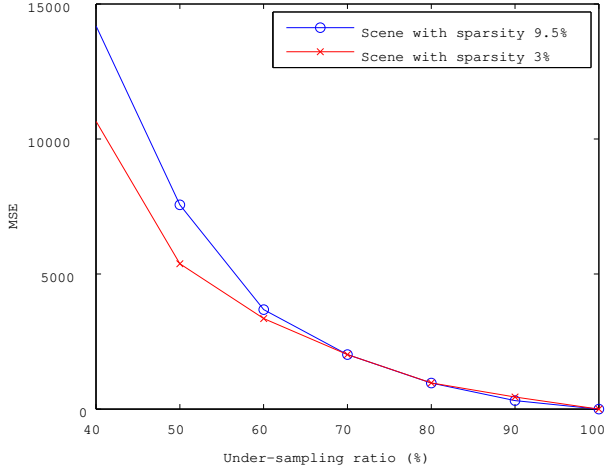}%
	\caption{MSE-sampling curves. Red curve for scene with sparsity 3\% and blue curve for scene with sparsity 9.5\%.}
	\label{fig:13}
\end{figure}

\begin{figure}[!t]
    \centering
  \includegraphics[width=\columnwidth]{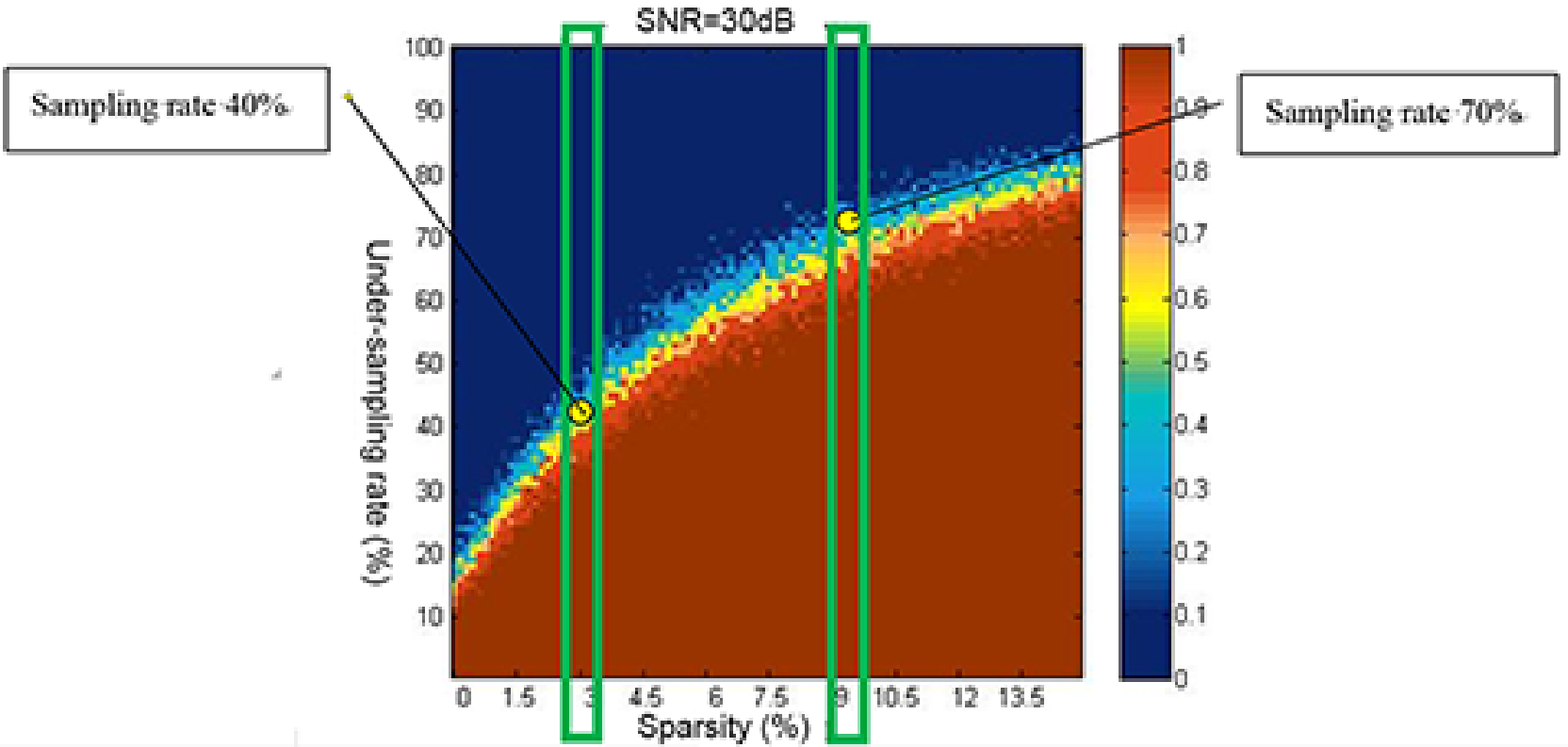}
  \caption{Phase transit diagram of system, red area means the failure area.}
  \centering
  \label{fig:14}
\end{figure}

\begin{figure*}[!t]
	\centering
	\subfloat[]{
		\includegraphics[width=2.5in]{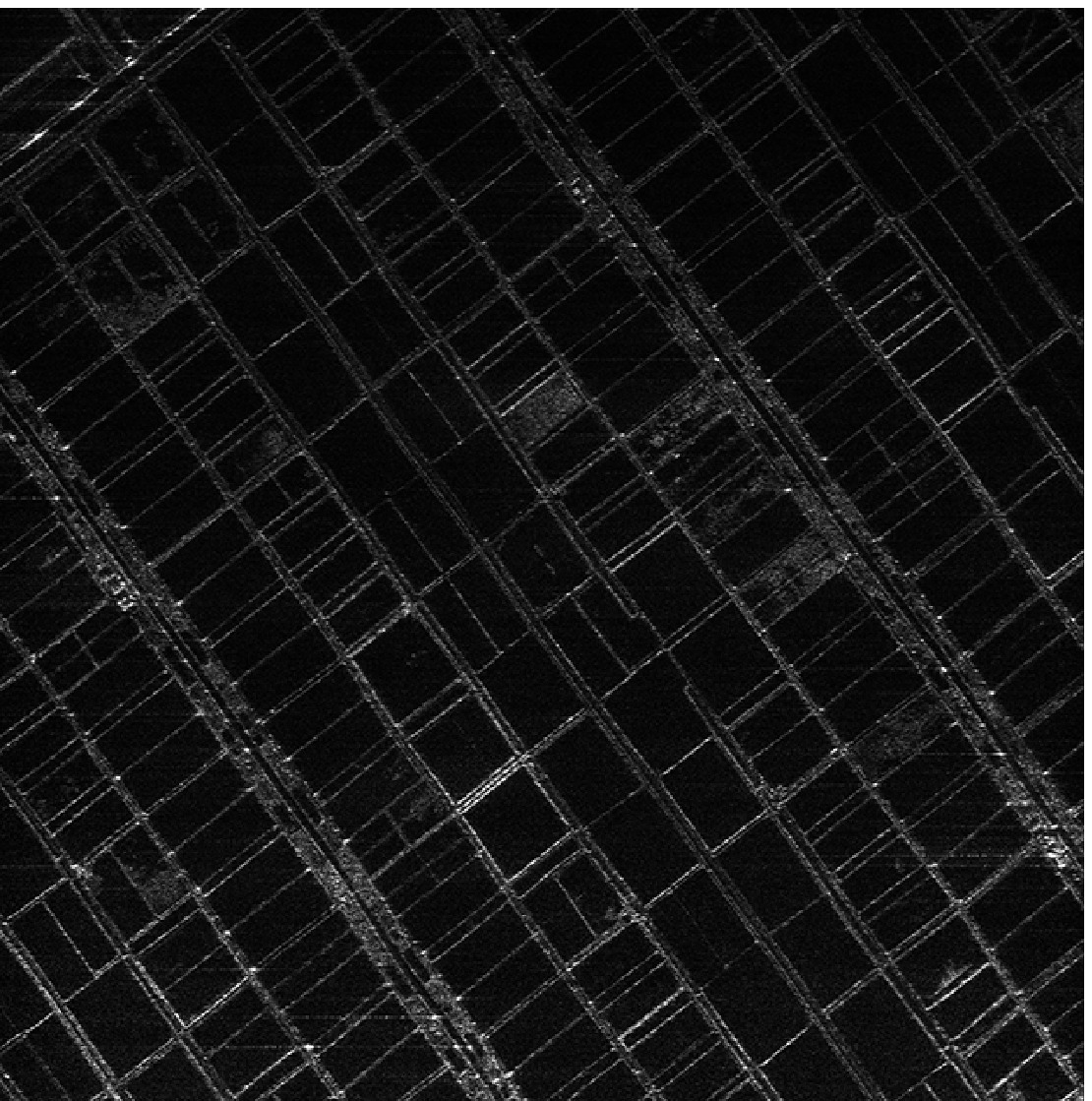}%
		\label{fig:15a}
	}~
	\subfloat[]{
		\includegraphics[width=2.5in]{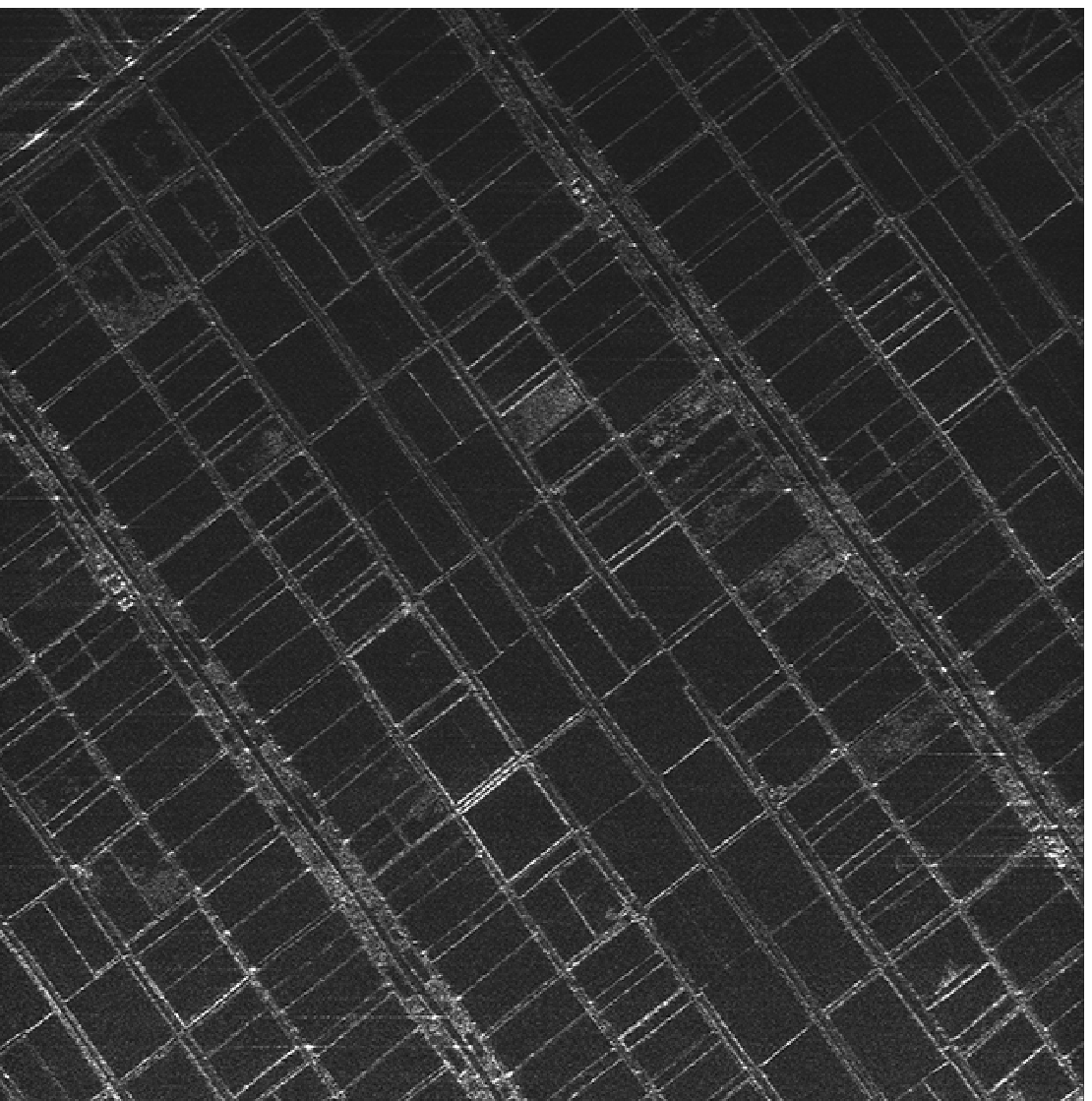}%
		\label{fig:15b}
	}\\
	\subfloat[]{
		\includegraphics[width=2.5in]{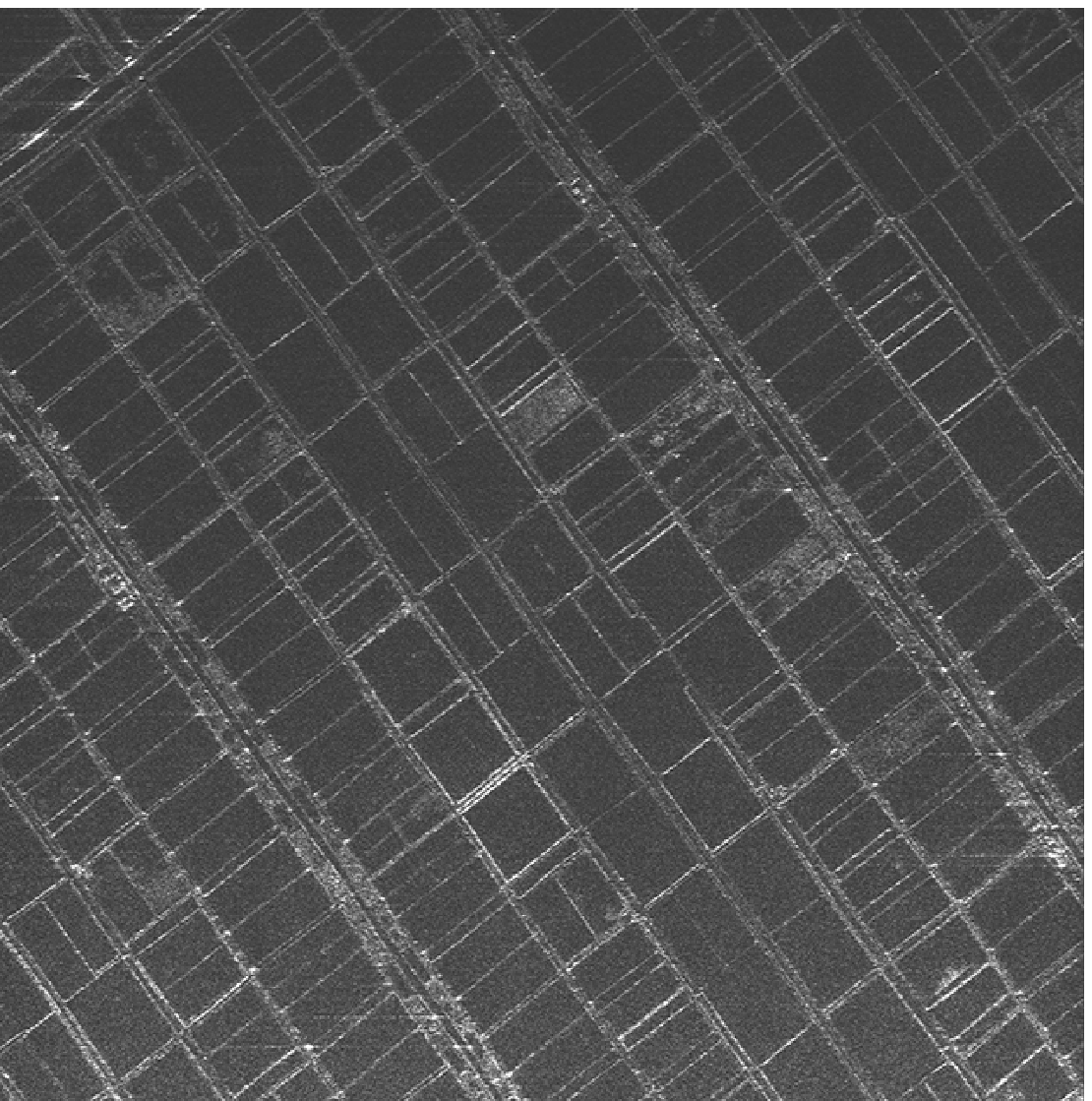}%
		\label{fig:15c}
	}~
	\subfloat[]{
		\includegraphics[width=2.5in]{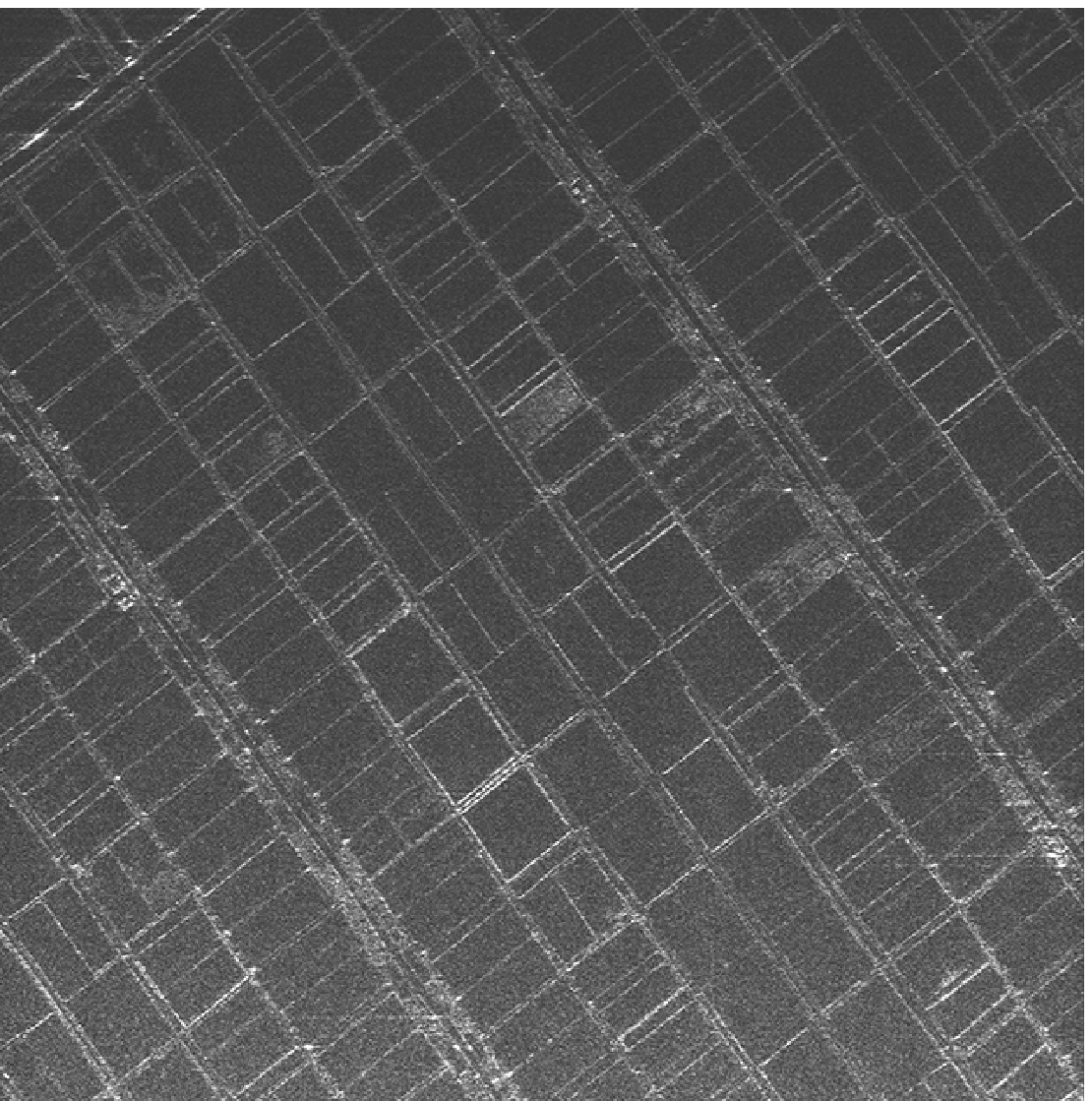}%
		\label{fig:15d}
	}~
	\caption{Imaging results of different SNRs with 70\% down-sampled data and scene of sparsity 9.5\%. (a) SNR 30dB; (b) SNR 27dB; (c) SNR 24dB and (d) SNR 21dB.}
	\label{fig:15}
\end{figure*}

\begin{figure}[!t]
	\centering
	\includegraphics[width=\columnwidth]{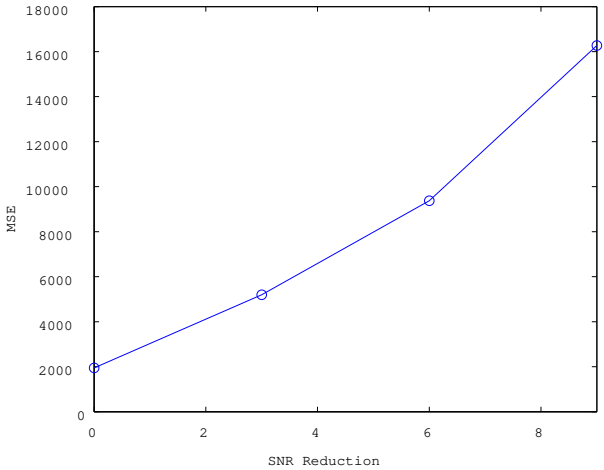}%
	\caption{MSE-SNR curves.}
	\label{fig:16}
\end{figure}

\begin{figure*}[!t]
	\centering
	\subfloat[]{
		\includegraphics[width=2.5in]{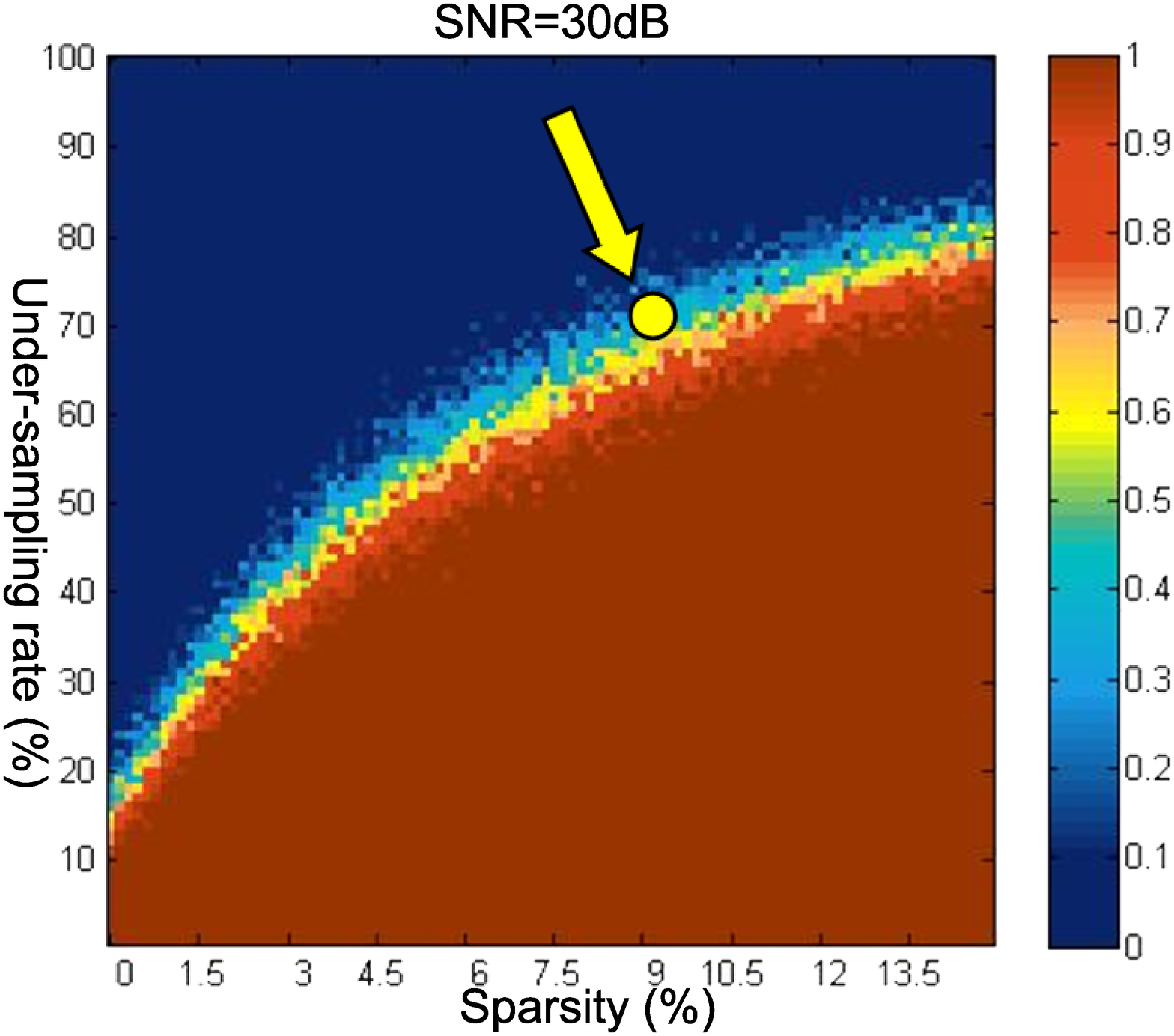}%
		\label{fig:17a}
	}~
	\subfloat[]{
		\includegraphics[width=2.5in]{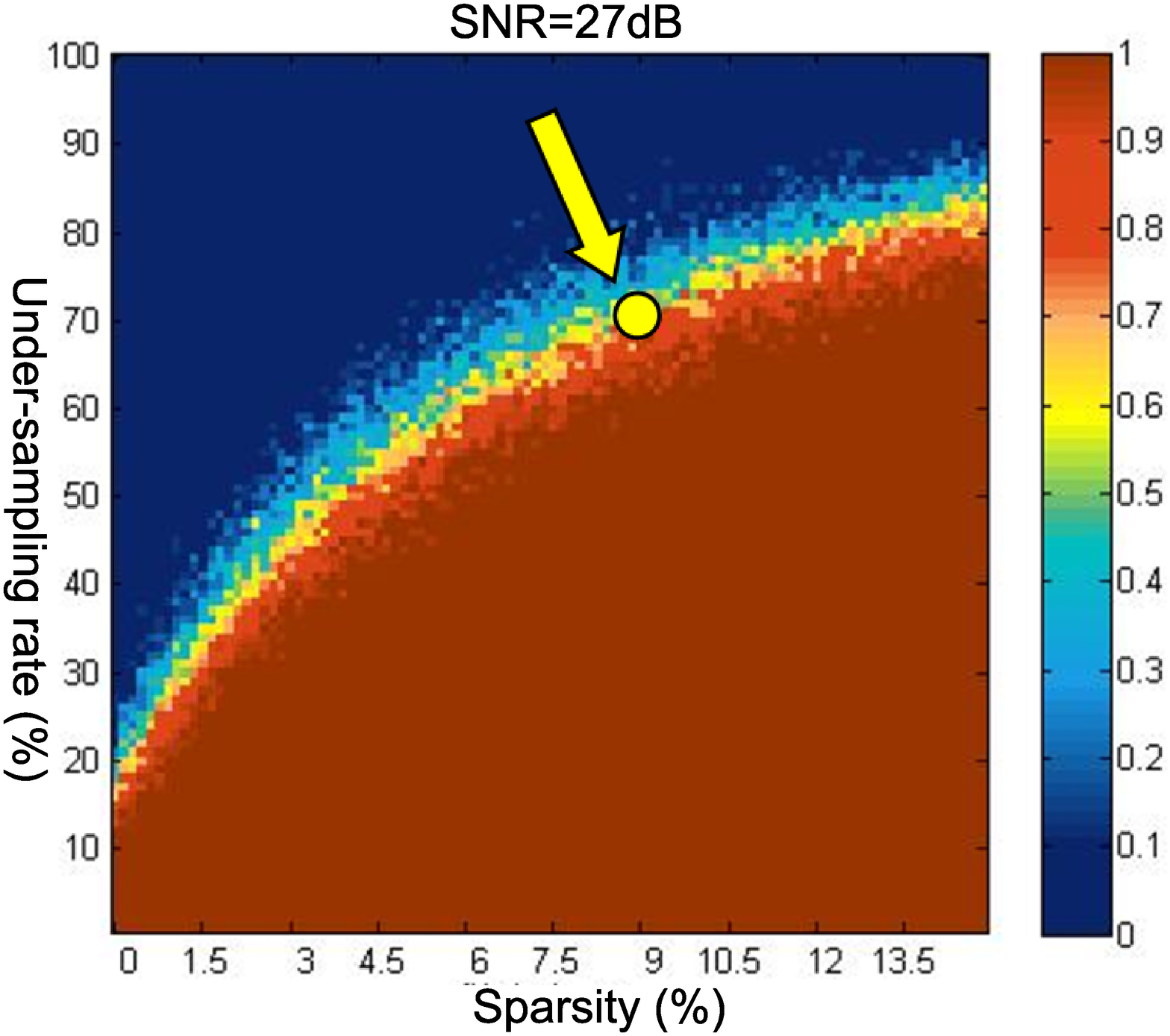}%
		\label{fig:17b}
	}\\
	\subfloat[]{
		\includegraphics[width=2.5in]{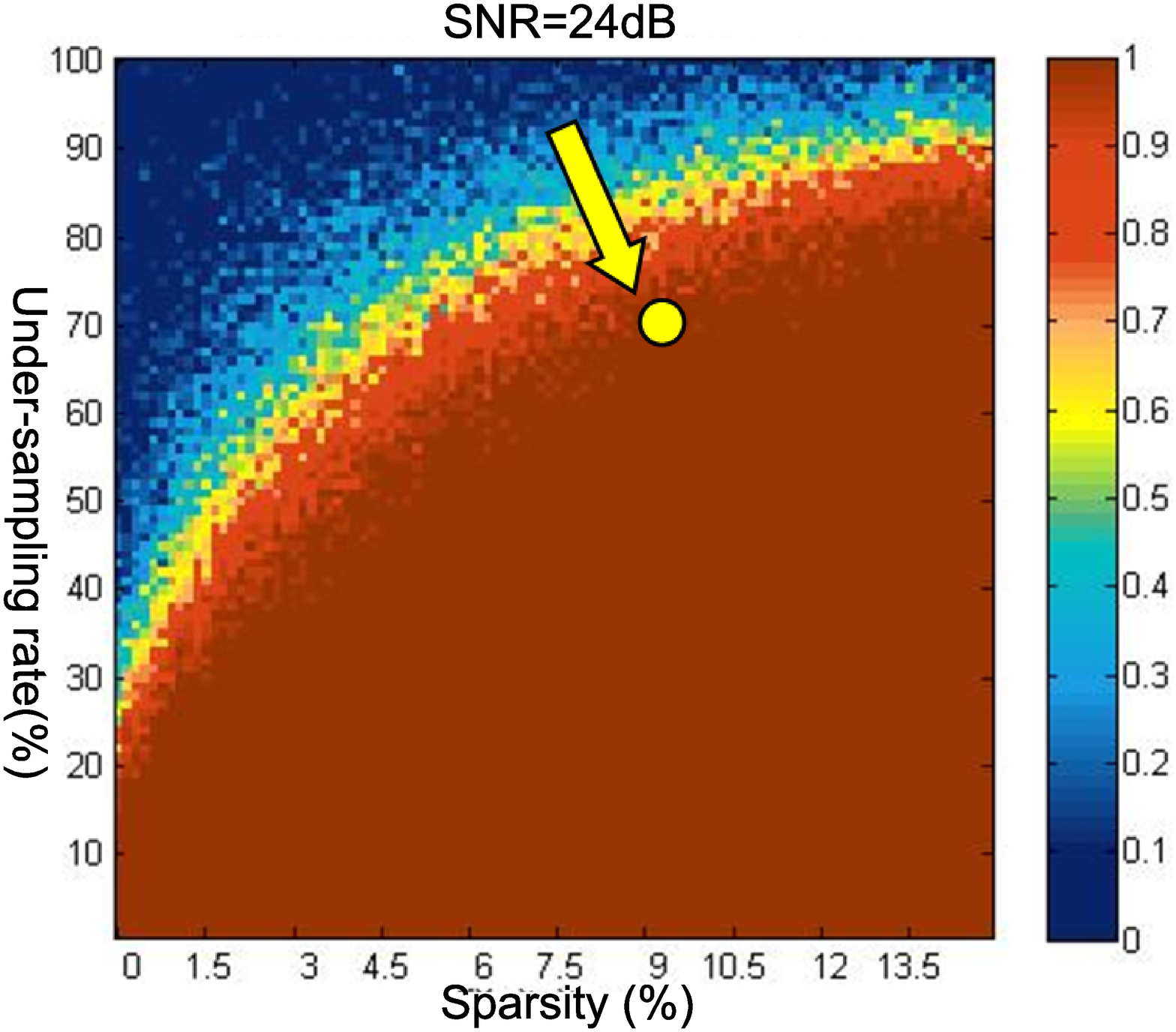}%
		\label{fig:17c}
	}~
	\subfloat[]{
		\includegraphics[width=2.5in]{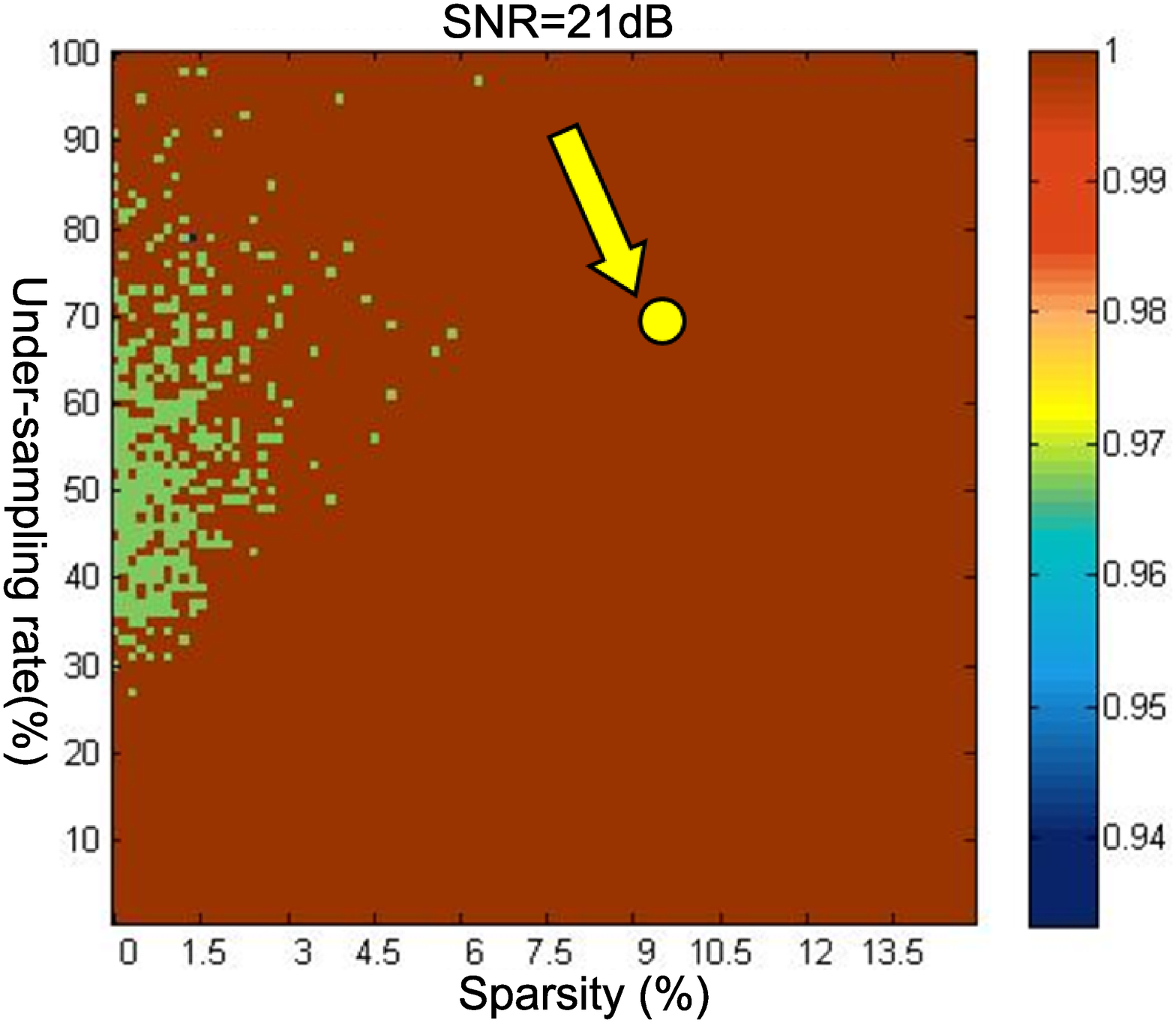}%
		\label{fig:17d}
	}~
	\caption{Phase transit diagrams of different SNRs with 70\% down-sampled data and scene of sparsity 9.5\%. (a) SNR 30dB; (b) SNR 27dB; (c) SNR 24dB and (d) SNR 21dB.}
	\label{fig:17}
\end{figure*}

\begin{figure*}[!t]
	\centering
	\subfloat[]{
		\includegraphics[width=2.5in]{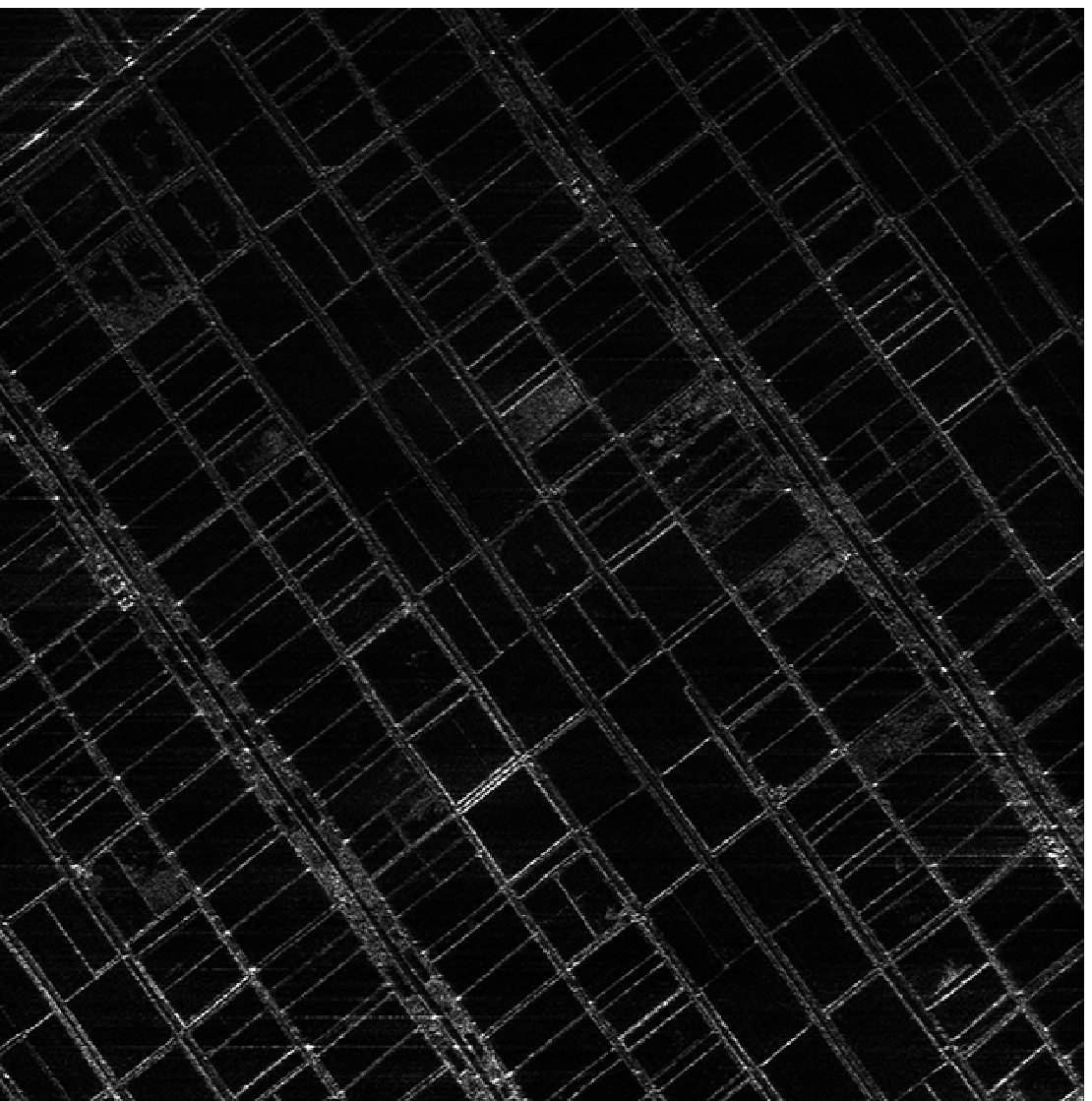}%
		\label{fig:18a}
	}~
	\subfloat[]{
		\includegraphics[width=2.5in]{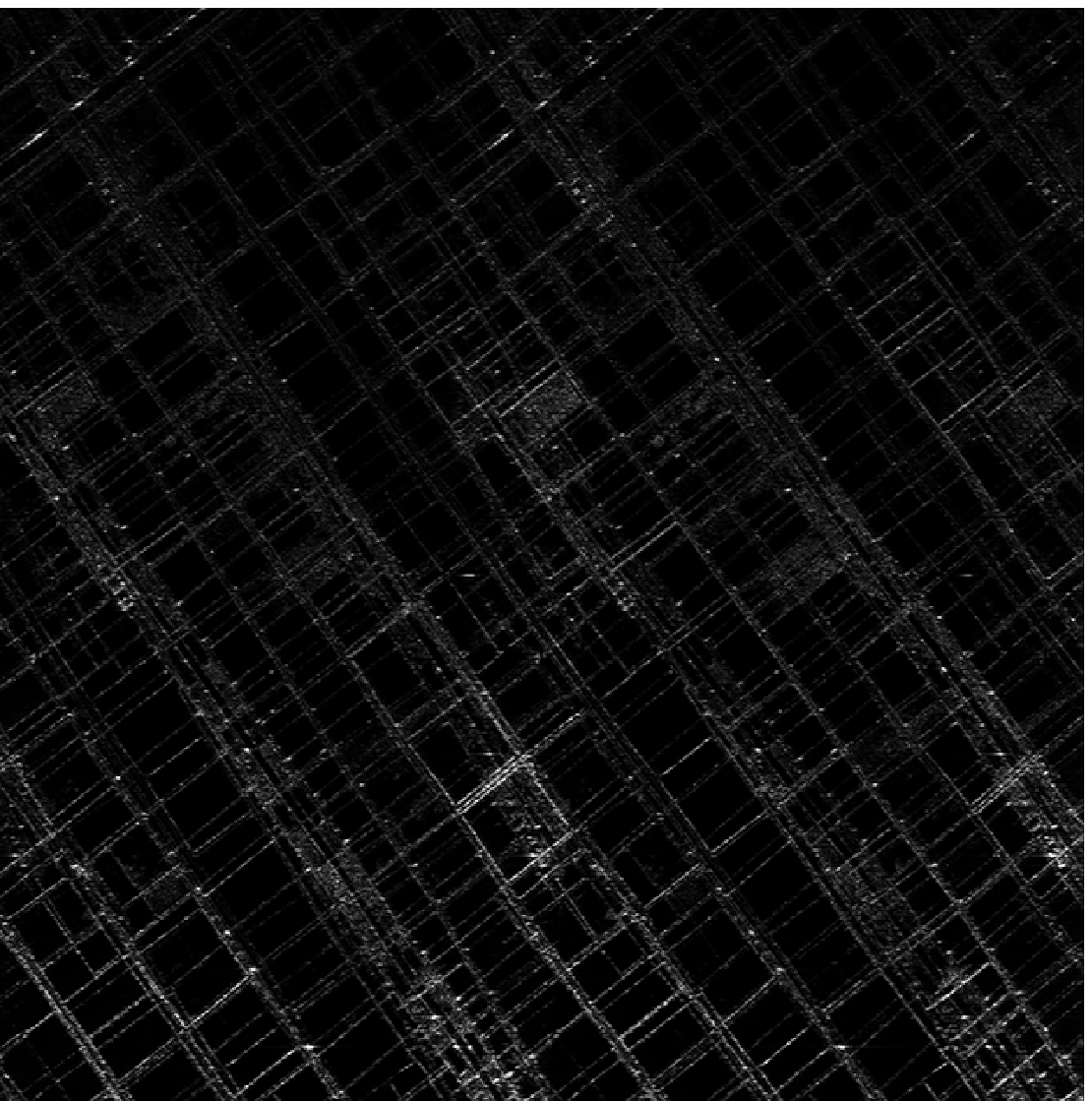}%
		\label{fig:18b}
	}
	\caption{Imaging results of different sampling schemes with 70\% down-sampled data and scene of sparsity 9.5\%. (a) Jittered sampling and (b) uniform sampling.}
	\label{fig:18}
\end{figure*}

\begin{figure*}[!t]
	\centering
	\includegraphics[width=5in]{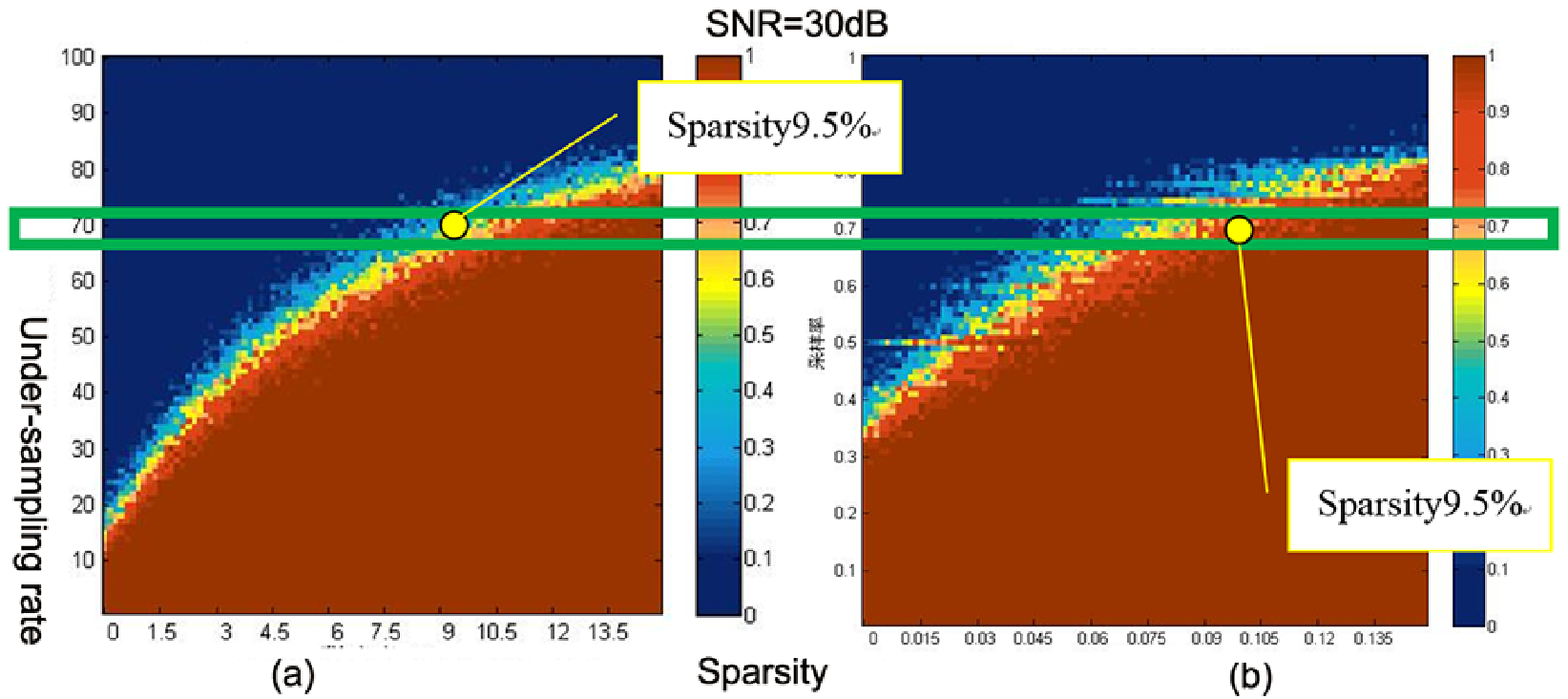}%
	\caption{Phase transit diagrams of different sampling schemes with 70\% down-sampled data and scene of sparsity 9.5\%, (a) Jittered sampling and (b) uniform sampling.}
	\label{fig:19}
\end{figure*}

\subsection{Prototype Radar System}

The design of prototype radar system is based on a existing C-band SAR system, we modified it to a sparse microwave imaging system under the guidance introduced in section III. Basically, the modification includes:
\begin{itemize}
	\item The sampling component.
	
	Sampling circuits in both direction are replaced with random sampling devices. In range direction it is a random sampling and in azimuth direction it is a jittered sampling.
	
	\item The pulse duration.
	
	The pulse duration is relatively extended from 30$\mathrm{\mu}$s to 38$\mathrm{\mu}$s to increase the system SNR.
\end{itemize}

The parameters of radar system are listed in table \ref{table:1}. Most parameters are inherited from the original SAR system. The modified parameters are emphasized in italic.

\begin{table}[!t]
	\renewcommand{\arraystretch}{1.3}
	\caption{The prototype radar system parameters}
	\label{table:1}
	\centering
	\begin{tabular}{|c||c|}
        \hline
        Parameter & Value \\
        \hline
        Bandwidth & 80MHz/500MHz \\
        Carrier Frequency & 5.4~GHz \\
        Antenna Azimuth Size & 0.9m \\
        Platform Velocity & \~108m/s \\
        PRF & \textit{Non-uniform, average 768Hz} \\
        Range Sampling Rate & \textit{Non-uniform, maximal 750MHz} \\
        Pulse Duration & \textit{38$\mathrm{\mu}$s} \\
        \hline
\end{tabular}
\end{table}

In order to evaluate the system performance under different sampling rates, the sampling in both azimuth and range direction are over-sampling. The down-sampling of variety down-sampling ratios is exploited to the raw data in the signal processing period.

\subsection{Description of Experiments}

We exploited two airborne experiments. The experiment setups are listed in table \ref{table:2}.

\begin{table}[!t]
    \renewcommand{\arraystretch}{1.3}
    \caption{Experiment Setup}
    \label{table:2}
    \centering
    \begin{tabular}{|c|c|c|c|c|}
        \hline
        Time & Location & Bandwidth & Height & Platform \\
        \hline
        2013.9 & Tianjin, China & 500MHz & 3900m & Cessna ``Citation'' \\
		2014.5 & Tianjin, China & 80/500MHz & 3900m & Cessna ``Citation'' \\
        \hline
    \end{tabular}
\end{table}

The sparse constraint of scene is required. We selected several typical kinds of sparse scenes, including
\begin{itemize}
	\item Sea and ships;
	\item Habour;
	\item Salt pans;
	\item and Islands.
\end{itemize}
Most sparse scenes have sparsity lower than 30\%, where sparsity if defined as
\begin{equation}
	\mathrm{sparsity}=\frac{\mathrm{Number~of~strong~points~in~discretized~scene}}{\mathrm{Number~of~all~points~in~discretized~scene}}.
\end{equation}
The ``strong target'' is a subjective definition based on the application. In different applications and situations, we have different denfinition of strong target. For example, in the oceanic ship imaging, we treat scatterers of ships as ``strong'' targets, and sea clutters as background. We take figure \ref{fig:51a} as example. This is a salt pan, and we have more interest on the bank area but not the water area. In this case, the ``strong'' targets are marked as blue in figure \ref{fig:51b}.

The scene salt pans is typical sparse scene, where roads are strong targets and water in pans are background. We show an optical image of it in figure \ref{fig:5}. The optical image is acquired from Google Earth.

\subsection{Data Processing}

We use our accelerated $\ell_q$ regularization algorithm to process the raw data. For a typical $8192\times16384$ image, the processing time consuming is about 600s, including some pre-processing operations e.g. error compensation. The data processing platform is listed in table \ref{table:3}. The raw data is full-sampled. We re-sampled it to different under-sampling rates in both directions. To simplify the data processing, the under-sampling rate in range domain is fixed as $85\%$ of Nyquist rate, and the under-sampling rate in azimuth domain varies to achieve different total under-sampling rates.

\begin{table}[!t]
    \renewcommand{\arraystretch}{1.3}
    \caption{Data Processing Platform}
    \label{table:3}
    \centering
    \begin{tabular}{|c|c|}
        \hline
        Workstation & Inspur NF5188 \\
		CPU & Intel Xeon E5620 2.4G 8Core \\
		Memory & 48GB \\
		GPU & nVidia Tesla C2050 \\
		Hard Drive & Seagate 500GB HDD \\
		Programming Language & Matlab and C MEX \\
        \hline
    \end{tabular}
\end{table}

\section{Experiment Results and Analysis}

\subsection{Experiment Results}

We achieved more than 30 scenes of data in the experiment, and the data processing is quite successful. Some images of typical scenes in shown below in figure \ref{fig:6}. The images are achieved with full-sampled data. As a comparison, images that are achieved by conventional means (range-Doppler) are given in figure \ref{fig:61}.

In figure \ref{fig:6} we listed several typical sparse scenes, including sea and ships shown in figure \ref{fig:6a}, salt pans shown in figure \ref{fig:6b}, \ref{fig:6c} and \ref{fig:6d}, habours shown in figure \ref{fig:6e} and urban area shown in figure \ref{fig:6f}. We want to point out that the scene of habour and urban area is not sparse, our signal processing method is still effective to the full-sampled data.

\subsection{Analysis}

In this section, we will analyze the experiment results using the phase transit diagram. As a preliminary result, we use mean square error (MSE) and phase transit diagrams to evaluate the imaging performance. MSE is calculated by amplitude with respect to the full-sampled sparse microwave imaging data, as described in equation (\ref{eq:mse})
\begin{equation}
	\label{eq:mse}
	\mathrm{MSE}=\frac{1}{K}\sum_1^K (S-S_0)^2,
\end{equation}
where $S$ is the evaluated image, $S_0$ is the reference image which is the image acquired with full-sampled sparse microwave imaging data, $K$ is the image size.

We want to emphasize that most analysis are based on down-sampled data, which cannot be processed under traditional SAR framework.

\subsubsection{Analysis of Sparsity}

In figure \ref{fig:9} we provide four images with the same under-sampling rate (70\% of Nyquist rate) but different sparsities. The MSE is shown in table \ref{table:4}. We can find that, if the scene sparsity is below 10\%, the imaging should be successful. The phase transit diagram is also given in figure \ref{fig:10}, which supports the experiment results. Phase transit diagram will drop into the ``failure'' area if the sparsity is above about 10.5\%. 

\begin{table}[!t]
    \renewcommand{\arraystretch}{1.3}
    \caption{Experiment of Different Sparsity}
    \label{table:4}
    \centering
    \begin{tabular}{|c|c|c|}
        \hline
        Scene & Sparsity & MSE \\
        \hline
		Scene 1 & 0.8\% & 943.1 \\
		Scene 2 & 3\% & 2105.1 \\
		Scene 3 & 9.5\% & 2107.8 \\
		Scene 4 & 50\% & 13986.5 \\
		Scene 5 & $>$90\% & 29967.2 \\
        \hline
    \end{tabular}
\end{table}

As a conclusion, we have:
\begin{itemize}
	\item Sparse microwave imaging is effective to sparse scene with under-sampled data;
	\item To non-sparse scene, we need full-sampled data to achieve a successful recovery.
\end{itemize}

\subsubsection{Analysis of Sampling Rate}

Experiment results of different sampling rates are evaluated. We under-sampled the raw data with different under-sampling rates in two scenes. The sampling rate varies from 40\% to 100\% of Nyquist rate. To the first scene with sparsity 3\%, the results are shown in figure \ref{fig:11}. We find that even the sampling rate is as low as 40\% of Nyquist rate, the imaging quantity is still acceptable. To the second scene with sparsity 9.5\%, the results are shown in figure \ref{fig:12}. The imaging quantity dramatically drops when the sampling rate is below 70\% of the Nyquist rate.

The MSE-sampling curve of two scenes are shown in figure \ref{fig:13}, and the phase transit diagram is shown in figure \ref{fig:14}. Generally, the sparser scene has a better MSE performance in the case of low under-sampling rate. The phase transit diagram asserts that for the first scene, all sampling rates more than 40\% of Nyquist rate are in successful area. For the second scene, the phase transition from successful area to failure area takes place in 70\% of Nyquist rate. The assertion follows the experiment results.

As a conclusion, we have:
\begin{itemize}
	\item Sparser scene has more potential ability in under-sampling;
	\item We can use phase transit diagram to select a suitable under-sampling rate during the system designing.
\end{itemize}

\subsubsection{Analysis of SNR}

The original system SNR is about 30dB which is estimated from system $\mathrm{NE\sigma_0}$. We add noises to the raw data to simulate SNR losses in 3dB, 6dB and 9dB to test the system performance under different SNRs. The under-sampling rate is 70\% of Nyquist rate and the scene sparsity is about 9.5\%. The imaging results of different SNRs are shown in figure \ref{fig:15}. We can find that the imaging quantity remarkably drops with SNR loss more than 3dB. The MSE-SNR curve is given in figure \ref{fig:16}, and the phase transit diagrams are given in figure \ref{fig:17}. The evaluation from phase transit diagram also shows that the recovery fails with SNR lower than 27dB.

As a conclusion, we can find that higher SNR will bring significant benefit to the system performance. However, SNR is limited by the system cost, power, complexity and antenna size. We can use phase transit diagram to select an appropriate system SNR.

\subsubsection{Analysis of Sampling Schemes}

As we selected jittered non-uniform sampling in azimuth direction, we also tested its performance compared with uniform down-sampling. We select the scene with sparsity 9.5\% and down-sampling 70\% of Nyquist rate. Imaging results with different sampling schemes are shown in figure \ref{fig:18}. It is obvious that the imaging successes with jittered sampling and fails with uniform sampling, as what is suggested in phase transit diagram figure \ref{fig:19}.

\section{Conclusion}

In this paper, we mainly introduce our work on the first prototype sparse microwave imaging radar system and airborne experiments based on it. The experiment and result analysis leads to the following conclusions. First, the sparse microwave imaging framework is feasible and workable, including its hardware design and signal processing algorithm. It is possible to design an optimized radar system under the guidance of sparse microwave imaging, and we are able to achieve sparse microwave images successfully. Second, the system performance of sparse microwave imaging can be evaluated with the phase transit diagram. Third, the sparse microwave imaging shows its performance advantage compered with tradition SAR framework. We believe that sparse microwave imaging is a quite promising concept in the future development of microwave imaging field, and we are so eager to witness this future.

\section*{Acknowledgment}
This work is supported by National Basic Research Program of China under grant 2010CB731905.

\appendix[Creation of Observation Matrix $\mathbf{H}$]

We take a simple case of stripmap SAR system as example. The baseband echo of a point target with backscattering coefficient $\sigma$ could be written as,
\begin{equation}
    \label{eq:a1}
    \tilde{s}(t, \tau)=\sigma \exp \left \{ -\frac{4j\pi f_c R(t)}{c} \right \} p \left (\tau-\frac{2R(t)}{c} \right ),
\end{equation}
where
\begin{equation}
    \label{eq:a2}
    R(t)=\sqrt{R_{0}^2+v^2 t^2}
\end{equation}
is the instantaneous range distance between point target and the platform, $R_{0}$ is the nearest distance between point target and the platform, $v$ is the relative speed of platform, $\tau$ is the range time, $t$ is the azimuth time, $c$ is the speed of light, $f_c$ is the carrier frequency, and $p(\tau)$ is the transmitting waveform. The additive noise is omitted here.

So, to a target scene $\Omega$, the baseband echo is,
\begin{equation}
    \label{eq:a3}
    \begin{split}
		&\int_\Omega \tilde{s}(t, \tau)\mathrm{d}\Omega=\\
    	&\int_\Omega \sigma(a, b) \exp \left \{ -\frac{4j\pi f_c R(t, a, b)}{c} \right \} p \left (\tau-\frac{2R(t, a, b)}{c} \right )\mathrm{d}\Omega,
    \end{split}
\end{equation}
where
\begin{equation}
    \label{eq:a4}
    R(t, a, b)=\sqrt{R_{0}(a, b)^2+v^2 t^2}
\end{equation}
is the instantaneous range distance between point $(a, b)$ and the platform, $R_{0}(a, b)$ is the nearest distance between point $(a, b)$ and the platform, and $\sigma(a, b)$ is the backscattering coefficient of point $(a, b)$.

Do the sampling to equation (\ref{eq:a3}) in both temporal and spacial domains, we can achieve the model of sparse microwave imaging,
\begin{equation}
    \label{eq:a5}
    y=\mathbf{H}x,
\end{equation}
where
\begin{equation}
    x=[\sigma(a_1, b_1), \sigma(a_1, b_2), \ldots, \sigma(a_M, b_N)]^T,
\end{equation}
is the spacial sampling;
\begin{equation}
    y=[s(t_1, \tau_1), s(t_1, \tau_2), \ldots, s(t_K, \tau_L)]^T,
\end{equation}
is the temporal sampling.

The matrix
\begin{equation} \label{eq:a6}
    \mathbf{H}=\left (
        \begin{array}{ccc}
            h_{1,1}(1,1) &\ldots & h_{M,N}(1,1) \\
            h_{1,1}(1,2) & \ldots & h_{M,N}(1,2) \\
            \vdots & ~ & \vdots \\
            h_{1,1}(1,L) & \ldots & h_{M,N}(1,L) \\
            h_{1,1}(2,1) & \ldots & h_{M,N}(2,1) \\
            \vdots & \ddots &\vdots \\
            h_{1,1}(K,L) &\ldots & h_{M,N}(K,L)
        \end{array}
    \right ),
\end{equation}
is the observation matrix, where
\begin{equation}
	\begin{split}
    	h_{m, n}&(k,l)\\
    	=&\exp \left \{ -\frac{4j\pi f_c R(t_k, a_m, b_n)}{c} \right \} p \left (\tau_l-\frac{2R(t_k, a_m, b_n)}{c} \right ).
    \end{split}
\end{equation} 
It is a $KL$-by-$MN$ matrix.

\ifCLASSOPTIONcaptionsoff
  \newpage
\fi



\bibliographystyle{IEEEtran}
\bibliography{IEEEabrv,IEEEexample}
%
%
%

%

\begin{IEEEbiography}[{\includegraphics[width=1in,height=1.25in,clip,keepaspectratio]{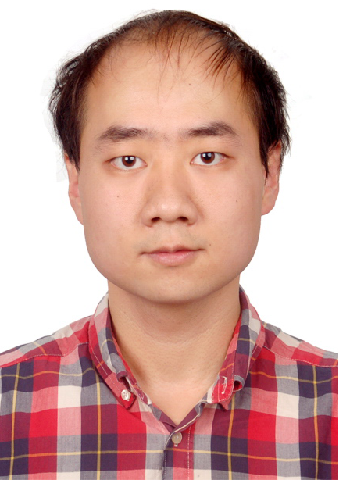}}]{Zhe Zhang}
(S'10) was born in Luoyang, China, in 1988. He received the B.Sc. degree in Information Engineering from Xi'an Jiaotong University, Xi'an, China, in 2008. 

He is currently working towards the Ph.D. degree with the Institute of Electronics, Chinese Academy of Sciences (IECAS) and University of Chinese Academy of Sciences in Beijing, China. His research interests include sparse microwave imaging, sparse signal processing and synthetic aperture radar imaging.
\end{IEEEbiography}

\begin{IEEEbiography}[{\includegraphics[width=1in,height=1.25in,clip,keepaspectratio]{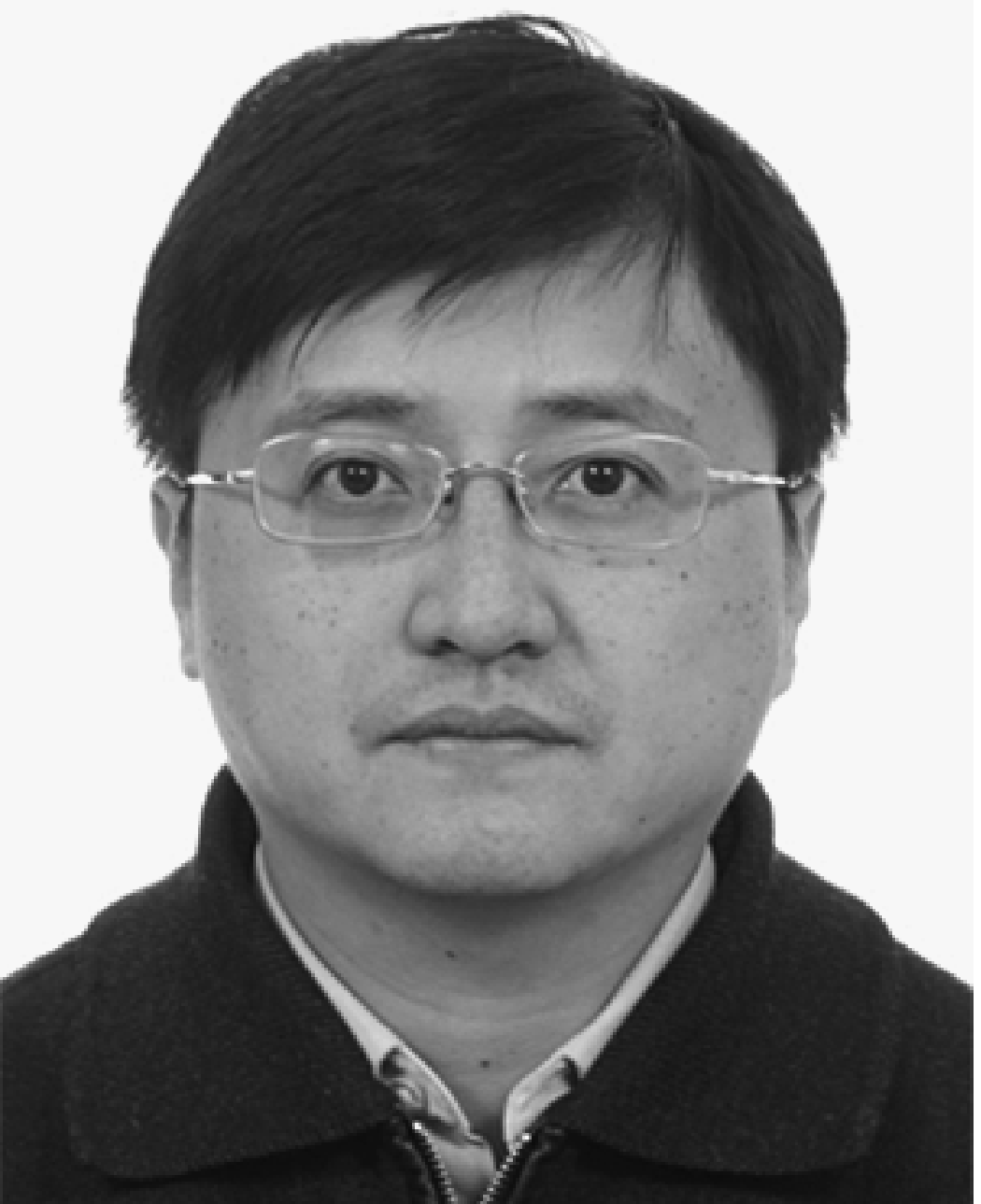}}]{Bingchen Zhang}
received the M.S. degree from the Institute of Electronics of Chinese Academy of Science, Beijing, China, in 1999. 

He currently serves as a scientist in the Institute of Electronics, Chinese Academy of Sciences, Beijing, China. His research interests consist of signal and
information processing, remote sensing technology, and sparse signal processing.
\end{IEEEbiography}

\begin{IEEEbiography}[{\includegraphics[width=1in,height=1.25in,clip,keepaspectratio]{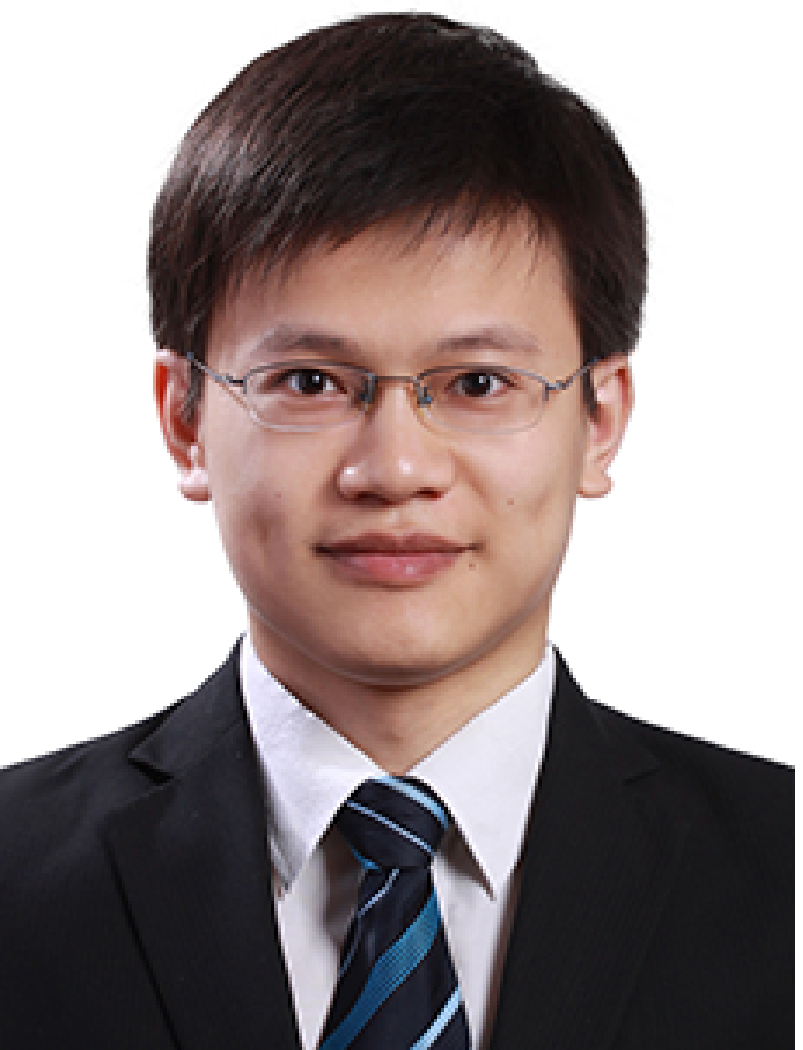}}]{Chenglong Jiang}
(S'14) received the Bachelor degree in electronic engineering and information science (with honors) from University of Science and Technology of China in 2009. 

He is currently pursuiting the Ph.D. degree in signal and information processing at the University of Chinese Academy of Sciences and working in the Science and Technology on Microwave Imaging Laboratory, Institute of Electronics, Chinese Academy of Sciences, Beijing, China. His main research interests include radar imaging, compressive sensing, multi-channel radar signal processing, and spaceborne SAR system design.
\end{IEEEbiography}

\begin{IEEEbiography}[{\includegraphics[width=1in,height=1.25in,clip,keepaspectratio]{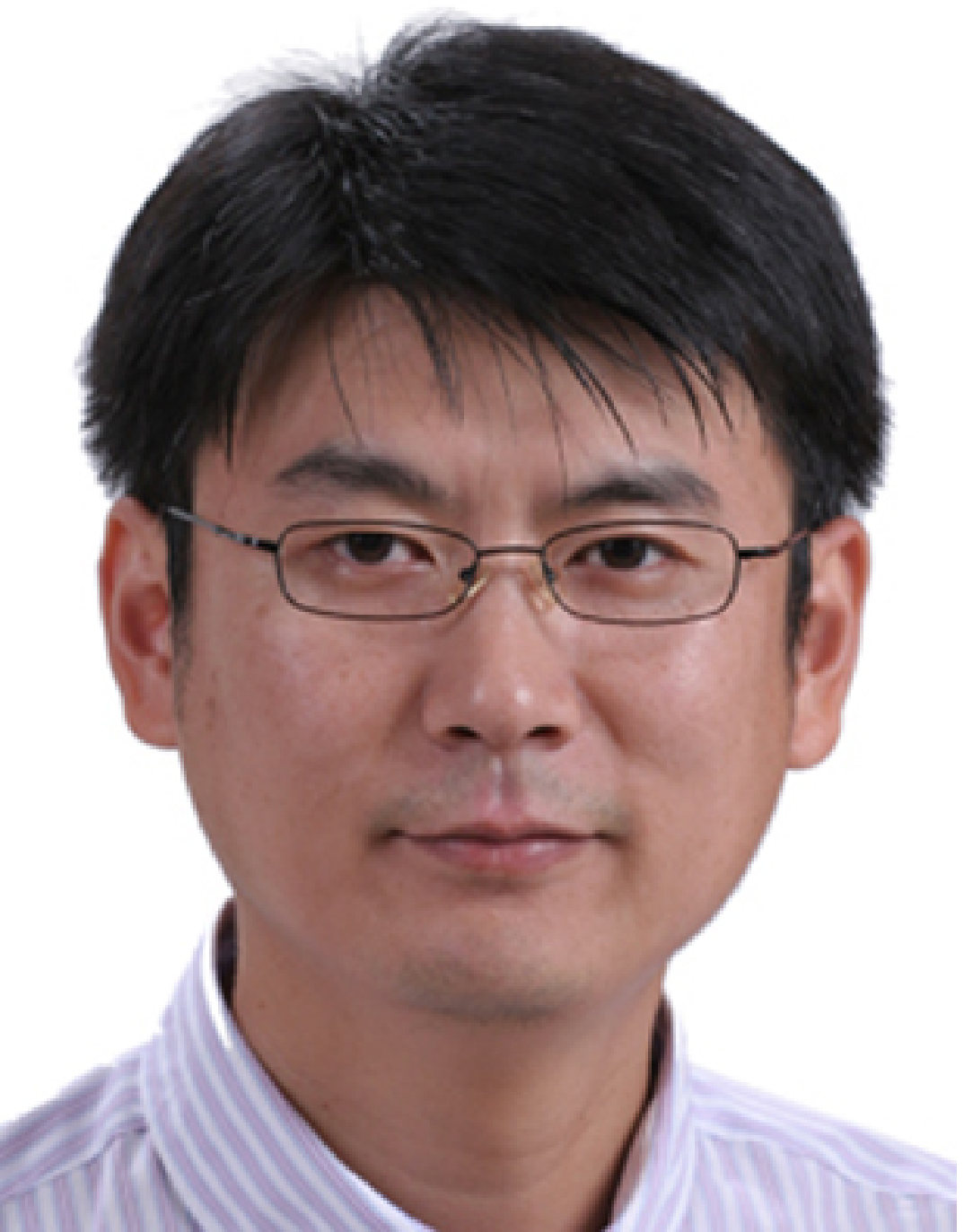}}]{Xingdong Liang}
(M'08) received the Ph.D. degrees from the Beijing Institute of Technology in 2001. Since 2002, he has been with the Institute of Electronics, Chinese Academy of Sciences. During this period, he has product responsibility for many important Chinese national programs, including high resolution SAR systems, polarimetric and interferometric SAR systems, and pre-research for advanced SAR systems. He has been rewarded by the Chinese government for his great contributions to the remote sensing area for several times. 

He is currently a Professor and holds the Director position of the Science and Technology on Microwave Imaging Laboratory, Institute of Electronics, Chinese Academy of Sciences. His research interests include real-time radar signal processing, coherent polarimetric and interferometric SAR systems, and novel radar imaging techniques. 
\end{IEEEbiography}

\begin{IEEEbiography}[{\includegraphics[width=1in,height=1.25in,clip,keepaspectratio]{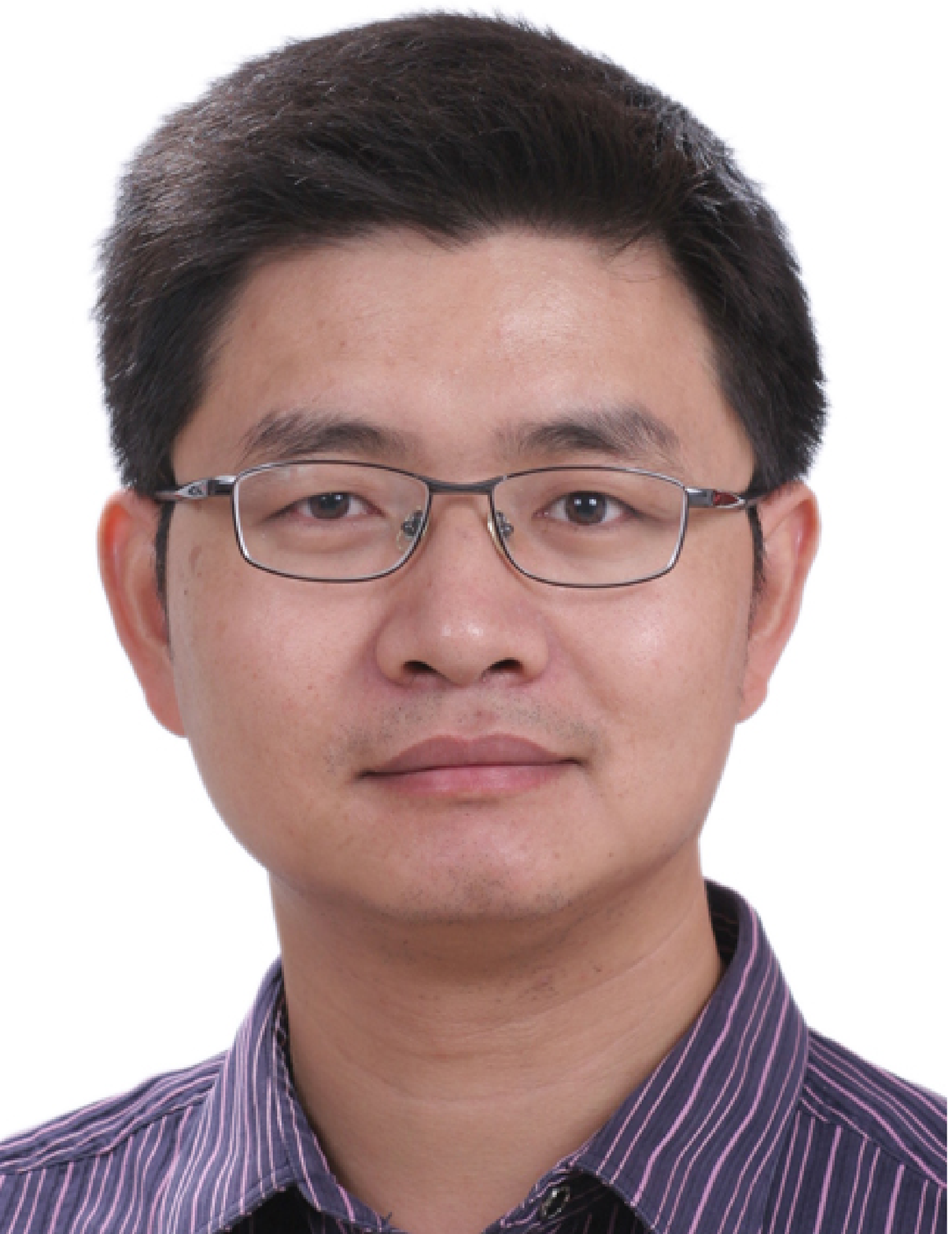}}]{Longyong Chen}
(M'08) received the B.S. degree in electrical engineering from the University of Science and Technology of China, Anhui, China, in 2003, the M.E. and Ph.D. degrees in signal and information processing from the University of Chinese Academy of Sciences, Beijing, China, in 2006 and 2009, respectively. Since July 2009, he has been with the Science and Technology on Microwave Imaging Laboratory, Institute of Electronics, Chinese Academy of Sciences. During this period, he has product responsibility for millimeterwave radars, receivers, polarimetric and interferometric SAR systems, and MIMO SAR systems. 

He is currently an Assistant Professor and holds the Assistant Director position of the Science and Technology on Microwave Imaging Laboratory. His research interests include radar signal processing, coherent polarimetric and interferometric SAR system designing, and novel radar imaging techniques. 
\end{IEEEbiography}

\begin{IEEEbiography}[{\includegraphics[width=1in,height=1.25in,clip,keepaspectratio]{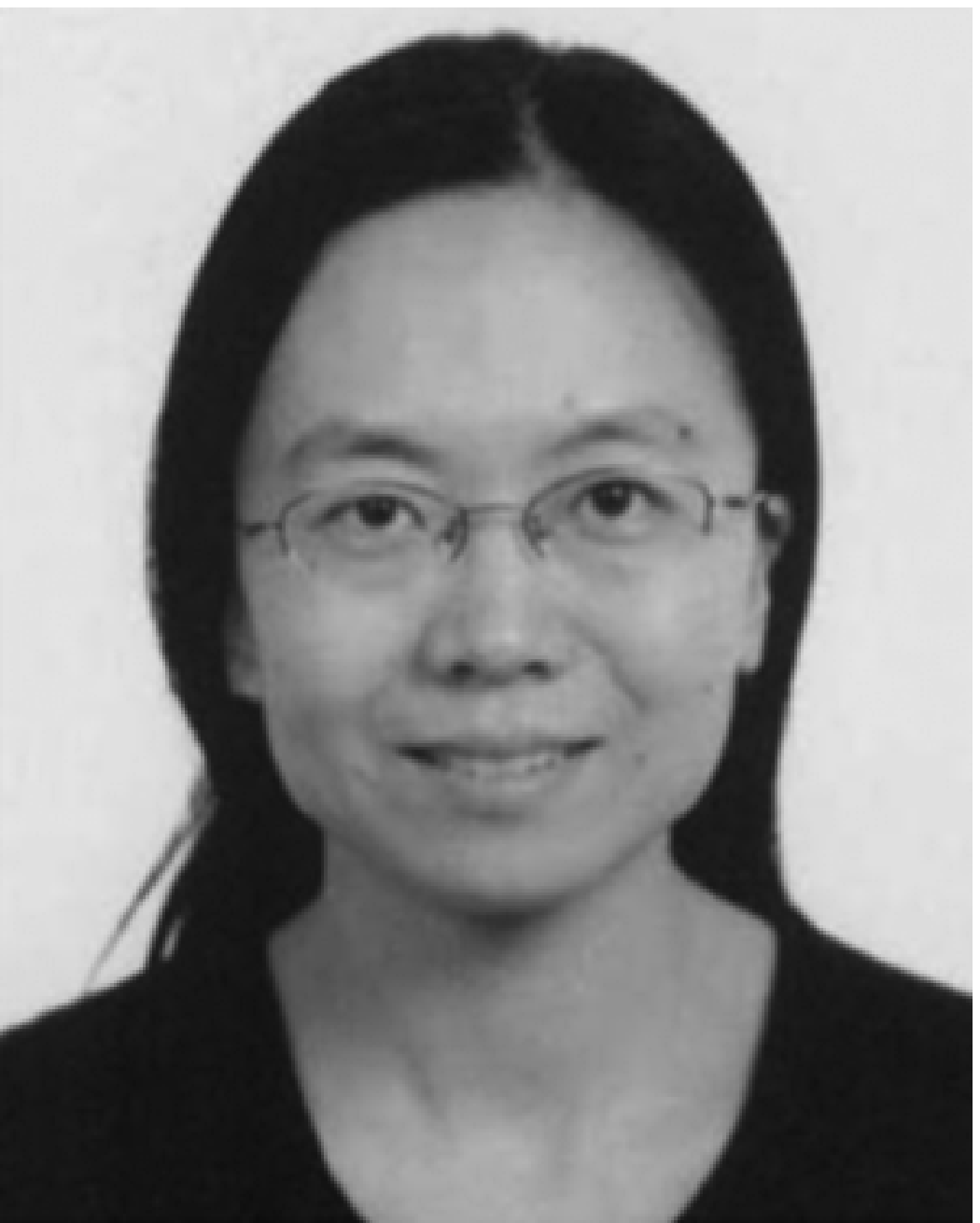}}]{Wen Hong}
(M'03) was born in Shanxi, China, in 1968. She received the M.S. degree from Northwestern Polytechnical University, Xi’an, China, in
1993 and the Ph.D. degree from the Beijing University of Aeronautics and Astronautics (BUAA), Beijing, China, in 1997.

She was formerly a Faculty Member in signal and information processing with the Department of Electrical Engineering, BUAA. She worked as a Guest Scientist for one year with the German Aerospace Center (DLR), Wessling, Germany. Since 2002, she has been with the National Key Laboratory of Microwave Imaging Technology, Institute of Electronics, Chinese Academy of Sciences, Beijing, China, as a Scientist and a Supervisor of the graduate student program. Her research interests include synthetic aperture radar imaging and its applications.
\end{IEEEbiography}

\begin{IEEEbiography}[{\includegraphics[width=1in,height=1.25in,clip,keepaspectratio]{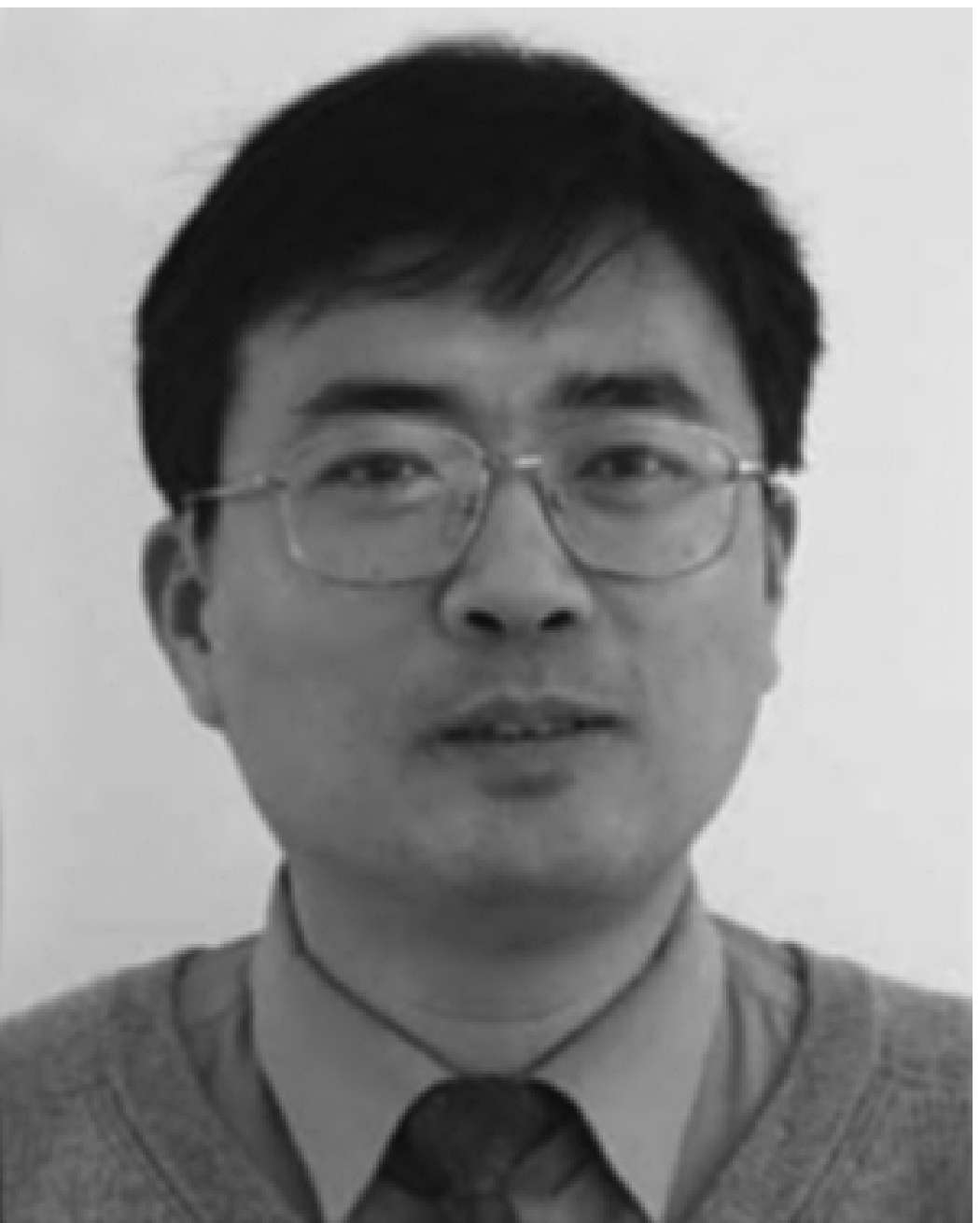}}]{Yirong Wu}
(M'00) received the Ms.D. degree from the Beijing Institute of Technology, Beijing, China, in 1988 and the Ph.D. degree from the Institute of
Electronics, Chinese Academy of Sciences (IECAS), Beijing, China, in 2001.

Since 1988, he has been with IECAS, where he is currently the Director. He has over 20 years of experience in remote sensing processing system design. His current research interests are microwave imaging, signal and information procession, and related applications. Now he is the academician of Chinese Academy of Sciences.
\end{IEEEbiography}



\end{document}